%% file: Article3.tex
\documentclass{aa}

\usepackage{float}
\usepackage{subcaption}
\usepackage{graphicx}
\graphicspath{{Figures/}}
\usepackage[svgnames]{xcolor}
\usepackage{natbib}
\usepackage{amsmath}
\usepackage[varg]{txfonts}
\usepackage{caption}
\usepackage[pdftex,colorlinks,citecolor=MidnightBlue,linkcolor=MidnightBlue]{hyperref}
\usepackage{bm} 
\usepackage{xfrac}
\usepackage[percent]{overpic}
\usepackage{booktabs}
\usepackage{longtable} 

\usepackage{amsfonts}
\newcommand{\tickYes}{\checkmark{}}
\usepackage{pifont}
\newcommand{\tickNo}{\hspace{1pt}\ding{55}}

\usepackage{tikz}
\usepackage{tikzscale} 
\usetikzlibrary{calc,decorations.pathmorphing,decorations.pathreplacing}

\newcommand{\Msun}{\ensuremath{\text{M}_\odot}}

\newcommand{\ramses}{\ensuremath{\mathcal{R}}\textsc{amses}}

\newcommand{\snj}{\text{NF}}
\newcommand{\sj}{\text{PSJo}}
\newcommand{\srh}{\text{HIIRo}}
\newcommand{\sjrh}{\text{PSJ-HIIR}}

\usepackage{siunitx}
\DeclareSIUnit\parsec{pc}
\DeclareSIUnit\lightyear{ly}
\DeclareSIUnit\year{yr}
\DeclareSIUnit{\Msun}{M_\odot}

\usepackage{tikz}
\usepackage{standalone} 
\usepackage{tikzscale} 
\usetikzlibrary{calc,decorations.pathmorphing,decorations.pathreplacing}

\let\pgfimageWithoutPath\pgfimage 
\renewcommand{\pgfimage}[2][]{\pgfimageWithoutPath[#1]{Figures/#2}}

\graphicspath{{Figures/}}

\bibpunct{(}{)}{;}{a}{}{,} 
\begin{document}

  \title{ Influence of protostellar jets and HII regions on the
  formation and evolution of stellar clusters}
  
  \titlerunning{Influence of stellar feedback on stellar clusters}
  
  \author{Antoine Verliat\inst{\ref{inst1}} 
    \and Patrick Hennebelle\inst{\ref{inst1},\ref{inst2}}
    \and Marta González\inst{\ref{inst3}}
    \and Yueh-Ning Lee\inst{\ref{inst5},\ref{inst6},\ref{inst1}}
    \and Sam Geen\inst{\ref{inst7}}
    }

  \institute{Laboratoire AIM, Paris-Saclay, CEA Saclay/IRFU/DAp -- CNRS --
    Universit\'e Paris Diderot, 91191 Gif-sur-Yvette Cedex, France
    \label{inst1}
    \and
    LERMA (UMR CNRS 8112), Ecole Normale Sup\'erieure, 75231 Paris Cedex, France
    \label{inst2}
    \and
    Université Grenoble Alpes, CNRS, IPAG, 38000 Grenoble, France
    \label{inst3}
    \and
     Department of Earth Sciences, National Taiwan Normal University, 116059 Taipei, Taiwan
     \label{inst5}
     \and
     Center of Astronomy and Gravitation, National Taiwan Normal University, 116059 Taipei, Taiwan
     \label{inst6}
     \and
     Anton Pannekoek Institute for Astronomy, Universiteit van Amsterdam, Science Park 904, 1098 XH Amsterdam, The Netherlands
     \label{inst7}
    }

  \date{Received 11 July 2021; accepted 21 January 2022}

  \input{Abstract.tex}


  \maketitle

\input{Introduction}

\input{Numerical_methods}

\input{Results_SFR}

\input{Results_gas}

\input{Results_bound_and_rotation}

\input{Results_alignment}
\input{Results_Q}

\input{Conclusion}

\begin{acknowledgements}
We gratefully acknowledge the anonymous referee for their comments and suggestions that strongly improved the manuscript.
SG acknowledges support from a NOVA grant for the theory of massive star formation.
This work was granted access to HPC
   resources of CINES and CCRT under the allocation x2020047023 made by GENCI (Grand
   Equipement National de Calcul Intensif).
   This research has received funding from the European Research Council
synergy grant ECOGAL (Grant : 855130).
\end{acknowledgements}

\bibliography{References,samgeen_edit}

\begin{appendix}
\input{Annexe_1}
\newpage
\input{Annexe_2}
\newpage
\input{Annexe_3}
\end{appendix}

\end{document}

%% file: Abstract.tex
  \abstract
    {Understanding the conditions in which stars and stellar clusters form is of great
    importance. In particular the role that stellar feedback may have is still hampered by large uncertainties.
    }
    {We investigate the role played by ionising radiation and protostellar outflows during the formation and evolution of a stellar cluster. To self-consistently take into account gas accretion, we start with clumps of tens of parsecs in size.
    }
    {Using an adaptive mesh refinement code, we run magneto-hydrodynamical numerical simulations aiming at describing the collapse of  massive clumps with either no stellar feedback or taking into account ionising radiation and/or protostellar jets. 
    }
    {Stellar feedback substantially modifies the protostellar cluster properties, in several ways. We confirm that protostellar outflows reduce the star formation rate by a factor of a few, although the outflows do not stop accretion and likely enough do not modify the final cluster mass. On the other hand, ionising radiation, once sufficiently massive stars have formed, efficiently expels the remaining gas and reduces the final cluster mass by a factor of several. We found that while HII radiation and jets barely change the distribution of high density gas, the latter increases, at a few places, the dense gas velocity dispersion again by a factor of several. As we are starting from a relatively large scale, we found that the clusters whose mass and size are respectively on the order of a few 1000 M$_\odot$ and a fraction of parsec, present a significant level of rotation. Moreover we found that the sink particles which mimic
    the stars themselves, tend to have rotation axis aligned with the cluster large scale rotation. Finally, computing the classical $Q$ parameter used to quantify stellar cluster structure, we infer that when jets are included in the calculation, the $Q$ values  are typical of observations, while when protostellar jets are not included, the $Q$ values tend to be significantly lower. This is due to the presence of sub-clustering that is considerably reduced by the jets. 
    }
    {Both large scale gas accretion  and stellar feedback, namely HII regions and protostellar jets, appear to significantly influence the formation and evolution of stellar clusters. 
        }

%% file: Introduction.tex
\section{Introduction}
\label{section: introduction}

It is largely established that a large fraction, and likely the majority, of stars forms in stellar clusters \citep{ladalada2003,bressert2010}.
As such, stellar clusters are certainly 
amongst the most important structures to understand 
in galaxies. This is particularly important to know the 
physical conditions that prevail as a star and its surrounding 
planets form and evolve. 

How exactly stellar clusters form remains largely unknown and 
has been the subject of several recent reviews
\citep{longmore2014,krumholz2019,krause2020,adamo2020}. 
While it is well established that their births take place through 
the collapse of massive gas proto-clusters, 
which likely have been observed \citep{urquhart2014,elia2017}, 
the exact conditions under which the collapse proceeds are still 
a matter of debate. 
Not only the initial conditions are still hampered by large uncertainties, 
in particular because gas clouds do not seem to present gas densities 
sufficiently high to reproduce the ones needed to explain the stellar densities
in clusters \citep{krumholz2019} but also because the feedback 
effects and efficiency have still only been partially explored and quantified. 

Starting from dense clumps, several authors have performed  
collapse calculations with various spatial resolutions ignoring 
stellar feedback
\citep[e.g.][]{bonnell2008, bate2012,kuznetsova2015}. In all 
these works, compact concentration of both gas and stars are naturally 
produced as a consequence of gravitational collapse. \citet{Lee2016-1,Lee2016-2}
closely investigated the radius of the gas proto-cluster as well as 
the stellar clusters and concluded that their mass-radius relation present 
similar scaling laws that the ones inferred from observations \citep[e.g.][]{pfalzner2016}. In this process, accretion-driven 
turbulence \citep{klessen2010} is playing a fundamental role. As gas falls in the 
gas proto-cluster, it sustains turbulent motion that get virialised and 
leads to the observed mass-size relation. 

However it is firmly established that stellar feedback plays a 
fundamental role during the course of stellar cluster formation and evolution. 
The main stellar feedback processes believed to be significant by the time 
of the cluster formation are protostellar outflows, ionising radiation and stellar 
winds. The supernovae themselves arrive at least \SI{4}{Myr} after the 
formation of the most massive stars and therefore 
likely do not play a significant role. Numerous studies have considered 
one or several feedback processes.
This is particularly the case for ionising feedback which has been extensively 
studied \citep[e.g.][]{Dale2012,dale2014,walch2013,Geen2015,geen2017,gavagnin2017,grudic2019}. There is
a general agreement that the ionising radiation can efficiently disperse
molecular clouds provide they are not bound, meaning that typically the 
escape velocity should not be larger than the sound speed of the ionised 
gas that is around \SI{5}{km.s^{-1}}. More precisely for molecular clouds typical of the 
solar neighbourhood, it has been inferred that the typical efficiency, that is to say the gas mass fraction which is converted into 
stars, is on the
order of a few percents, although there are considerable variations. These variations are due
on one hand to the stochastic nature of the gravo-turbulent fragmentation and 
on the other hand to the 
strong dependence of the ionising flux onto the stellar mass \citep{geen2018}.

While several studies have considered the role of jets in 
cores \citep[e.g.][]{offner2017} or low mass clusters \citep[e.g.][]{cunningham2018}, a somewhat more restricted numbers of studies have considered the 
role of jets at the scale of more massive clusters, that is to say a
few parsecs across. For instance \citet{nakamura2007} and \citet{wang2010} considered
star forming clumps of about \SI{1000}{\Msun} and show that turbulence can 
be maintained through the driving of jets but also that the star formation 
rate (SFR) can be reduced by a factor of a few compared to the case 
without jet driving leading to SFR that are typical of observed values. 
\citet{guszejnov2021} explored the influence of jets
on a large variety of clouds and also found that 
they have a significant influence on their 
evolution, particularly  on the mass spectrum of the stellar objects. 
Similar results regarding the star formation efficiency (SFE) have been obtained using periodic boxes which in total also 
contain about \SI{1000}{\Msun}
\citep{carroll2009,Federrath2014,murray2018}. 

So far to our knowledge, no work have 
thus investigated the influence of both 
ionising radiation and protostellar outflows
simultaneously. Moreover, most works that 
have included protostellar jets considered 
computational domain on the order of a few parsecs 
which makes the time both for accretion to proceed 
and jets to reach the computational domain a little short. Both aspects however are essential to assess the respective role of accretion driven turbulence and feedback driven turbulence. Indeed, protostellar outflows from stars in HII regions have been observed in our Galaxy, such as jets detected at the tip of dense gas pillars similar to the ``Pillars of Creation'' in M16 \citep{McLeod2016}.

In the present work, we aim at simulating the formation of a stellar cluster by describing self-consistently spatial scales from a few tens of pc down to about \SI{1000}{au}. This allows us to describe the proto-cluster environment while getting a 
reasonable description of the gaseous proto-cluster whose size ranges between 0.1 and \SI{1}{pc}. We treat 
both ionising radiation and stellar jet feedback. 
By performing a series of simulations that include
none, one or both feedback processes, we can 
induce their respective influence, in conjunction with large scale infall,  on the gas and the stars.

The paper is organised as follows. 
The \hyperref[section: numerical methods]{second part} of the paper describes the numerical methods and 
the initial conditions of the runs performed. It is complemented 
by an \hyperref[annexe: implementation jets]{appendix} which describes in details our jet
implementation. The \hyperref[section: article2-Description,SFE,SFE]{third part} of the paper presents the general description of the simulation
results as well as a detailed analysis of the 
gas properties. It also discuss the star formation 
rate and efficiency obtained in the various simulations.
The \hyperref[section: article2-gas]{fourth part} focuses on the gas properties 
with particular emphasis on the kinematic 
and the influence of the various types of feedback.
The \hyperref[section: article2-sinks properties]{fifth part} investigates the star/sink
particle properties, namely the global rotation of the cluster,
the stellar rotation axis orientation and the cluster 
structure, that we quantify using the $Q$ parameter. \hyperref[section: article2-conclusion]{Section six} concludes the paper.

%% file: Numerical_methods.tex
\section{Numerical methods and initial conditions}
\label{section: numerical methods}

\subsection{Code and numerical parameters}
\label{section-sub: code and numerical parameters}
To investigate the joint influences of HII regions and protostellar jets on the formation of a stellar cluster, we carry out a set of four simulations starting from \SI{e4}{\Msun} of gas in a cubic box of side \SI{30.4}{pc}, differing only by the included type of stellar feedbacks. Thus, one simulation is including no feedback at all, one is including HII regions only, one is including protostellar jets only, and the last one is including both HII regions and protostellar jets.

These simulations have been made using \ramses{} \citep{Teyssier2002}. This numerical Eulerian code uses Adaptative Mesh Refinement (AMR) technique to enhance resolution locally, where it is needed, on a Cartesian mesh. We use five levels of AMR, from 7 to 12. 
This gives a cell size of at least \SI{0.24}{pc} (\SI{5e4}{au}) everywhere, and \SI{7.4e-3}{pc} (\SI{1.5e3}{au}) in the most refined cells. Our refinement criterion is based on Jeans length such that each local Jeans length is described by at least 40 cells. 

We use open boundary conditions for the hydro solver to allow the matter to flow out of the box, and we use periodic boundaries for gravity. The refinement is not allowed in the outer 5\% of the simulation box, to avoid the appearance of numerical instabilities in the matter flowing out of the box.

When the density in a cell rises above \SI{e7}{cm^{-3}}, we create a sink particle \citep{Krumholz2004,Bleuler2014} which interacts gravitationally with the surrounding gas and by the way of accretion and ejection processes. The radius and luminosity of the sinks are provided through evolutionary tables of \citet{Kuiper2013}.

Our four simulations will include magnetic field. This latter is treated through the ideal magnetohydrodynamics approximation \citep{Fromang2006}.  The cooling and heating processes are as described in \citet{Audit2005}. It includes the most important atomic cooling and the heating  photo-electric effect on PaH due to an external standard galactic UV field. Typically the resulting cooling curve is almost identical to the one obtained for instance in \citet{koyama2000}. This is in good agreement with earlier conclusions by \citet{levrier2012} and \citet{glover2012} where comparisons between atomic and molecular cooling have been performed.

\subsection{Stellar objects and ionising radiation}
\label{section-sub: stellar objects}
The resulting IMF in those simulations does not compare to the observed ones, partly due to a lack of numerical resolution but also to the 
fact that we do not treat here the infrared radiation that leads to a strong heating of 
the high density gas \citep{krumholz2007,hennebelle2020}.
Indeed \citet{leeh2018b} found that when using an explicit barotropic equation of state to mimic the temperature at high density, a numerical resolution typically below 10 AU is needed to get numerical convergence on the stellar mass spectrum, in particular to describe the peak. This condition becomes even more severe, typically at least few AU of resolution are requested, when infrared radiation is treated \citep{hennebelle2020}. While one may argue that not properly resolving the formation of low mass stars does not preclude to resolve the formation of high mass stars, the issue is that too many of the latter will then be produced because the total mass of gas converted in stars does not strongly depend on resolution. This issue may even be more severe when infrared radiation is treated since the resulting gas heating may then be overestimated and the number of massive stars further amplified.

Notably, the lack of massive stars in our simulations would be problematic for studying the influence of HII regions which are mainly triggered by massive stars. 
More generally the ionising flux emitted 
by massive stars strongly depend on their 
mass \citep[e.g.][]{Vacca1996}.
Even for a thousand solar masses in stars, the most massive star in our simulation hardly exceeds \SI{10}{\Msun} while observations suggest that statistically, for a pool of \SI{120}{\Msun} of stars, typically one massive star lies in them. To overcome this issue, we proceed as in \citet{colling2018}, where 
a simple sub-grid model is presented. Every time that \SI{120}{\Msun} of gas have been accreted into stars, we consider that one massive stars should have formed, so we create a stellar object whose mass is randomly chosen between 8 and \SI{120}{\Msun} assuming that the mass function
is a power-law of index -2.35. 

We do not use the approach in, e.g., \cite{He2019}, where each sink particle is considered to be a core containing one massive star. This is a) because it is still unclear how much these cores should fragment into close binaries, which would change the distribution of massive stars in the cloud, and b) in order to allow a consistent set of stellar masses to be implemented in comparisons between simulations. 
 The influence of the particular prescriptions made to 
introduce massive stars has been analysed in details 
by \citet{grudic2019}, who concluded that it indeed results in large uncertainties, which can be as important as a factor of 3. Typically quantifying the influence of the prescriptions such as the random choices of the massive star masses \citep{geen2018} would require to perform several runs
which is outside the scope of the present work.
Along the same line, a statistical sample should also be run to quantify the influence of different random seeds of the initial turbulence, although this would increase the cost of the work even further.

Each stellar object emits 
an ionising radiation as given by \citet{Vacca1996}.
This  ionises the surrounding gas, which then expands into a HII regions provided the strength of the radiative field  compared to the density of the environment is sufficient, as described in \citet{Geen2015}. 
The ionising radiation is treated using 
the RT method developed in \citet{rosdahl2013} which treats the propagation of light using the 
reduced speed approximation (a factor of \SI{e-4}{} is employed in this work as it is sufficient to capture the expansion of the ionisation front, provided it does not expand faster than \SI{30}{km.s^{-1}} --- a few times the speed of sound). 
As in \citet{Geen2015}, we used 3 groups of photons
to describe the ionisation of hydrogen and helium.

\subsection{Implementation of protostellar jets}
\label{section-sub: implementation of protostellar jets}
As our maximum resolution is of the order of a thousand astronomical units, we do not resolve all the physics responsible for the ejection of matter from young accreting stars known as protostellar jets. To still be able to take into account the effect of these jets on the large scale evolution, we implemented a sub-grid model based on the properties of the sink particles. Once a sink grows a mass higher than \SI{0.15}{\Msun}, at each time step it expels $1/3$ of the mass accreted during this time step in the form of a circular biconic jet. The matter is expelled with a velocity equal to 24\% of the escape velocity at the surface of the star. The direction of the ejection is given by the angular momentum of the sink. Each circular cone has an opening angle of $20 ^\circ$. The expelled material has the same specific thermal energy than the direct surrounding of the sink particle. Technical details of the implementation and references for the values above can be find in Appendix \ref{annexe: implementation jets}.

\subsection{Neglect of supernovae and stellar winds}

Our work omits the influence of stellar winds and supernovae from massive stars. Simulations using a similar setup have explored the interaction between photoionisation feedback and winds \citep{dale2014,Geen2020,Lancaster2021b,Lancaster2021c} or supernovae \citep{Geen2016,Kimm2021}. The reason for this is both scientific and technical. Except for large, long-lived cloud complexes, supernovae either occur too late or do not have sufficient compressive power to strongly influence either their host cloud or other nearby clouds \citep{Seifried2018}, although they are thought to be a trigger for new cloud formation \citep{Inutsuka2015,Fujii2021}.

Winds are more complex, since they are produced by stars during their main sequence, similarly to ionising radiation. Stellar winds shock the cloud material to very high temperatures \citep[$>10^6~$K, sufficient to emit x-rays, e.g.][]{Guedel2008}, which create hot bubbles that store large quantities of energy. Their low density means the bubbles themselves do not typically radiate energy strongly \cite{MacLow1988}. However, strong cooling channels exist either through evaporation of dense structures in the cloud overrun by the wind bubble \citep{Arthur2012}, or by turbulent mixing with the cloud material (\citealp{Rosen2014,Lancaster2021a}, see also \citealp{Tan2021} for an analysis of turbulent mixing on shockfronts). In this case, wind bubbles become momentum-driven rather than pressure-driven \citep{Silich2013}, and their dyanmical influence drops considerably. In this mode, winds only significantly drive the expansion of HII regions at small radii \citep{Geen2019,Oliver2021}, similar to the influence of radiation pressure. However, they can still play a strong role in structuring the photoionised region around the star \citep{Pellegrini2007,Rahner2017} or even trapping ionising radiation at early times \citep{Geen2021}. This has a non-linear influence on the evolution of HII regions that should be explored in more detail.

A more technical reason for omitting stellar winds is the relatively high computational cost in simulating them. The sound speed in photoionised gas is $\sim10~$km/s, compared to speeds sometimes exceeding 3000$~$km/s for stellar winds \citep{Vink2011,Ekstrom2012}. The Courant condition for gas flows on the simulation grid thus requires around 100 times more timesteps if winds or other sources of hot gas such as supernovae are included. We thus justify omitting them both because their influence is more subtle and requires additional research to determine when they are important, but also because to do so would drastically increase our computational costs and thus limit our ability to explore the core physics in this work.

\subsection{Initial conditions}
\label{section-sub: initial conditions}
Our four simulations have the same initial conditions as they only differ by the types of stellar feedbacks included. Those initial conditions are very similar to those used by \citet{Lee2016-1}. 
To set the density, we divide the box into 3 concentric regions. The inner one is a Bonnor-Ebert-like spherical cloud whose diameter is \SI{15.2}{pc}, half of the box length. In this inner cloud the number density $n$ is distributed according to a top hat function:
\begin{equation}
n \left( r \right) = \frac{n_0}{1+ \left( \frac{r}{r_0} \right) ^2}
\end{equation}
with $r$ the distance of the cell from the centre of the box, $n_0 = \SI{8e2}{cm^{-3}}$, and $r_0 = \SI{2.5}{pc}$. This lead to a mass of \SI{e4}{\Msun} in this central region. The second region, beginning at the cloud edge, is a ring whose outer diameter is equal to the box length. The density in this ring is constant and equal to about \SI{8}{cm^{-3}}, a tenth of the cloud edge density. The rest of the box, the eight corners, are filled with a constant density of \SI{1}{cm^{-3}}.

The initial temperature set by the cooling function is about \SI{10}{K} in the dense gas. We initialised the magnetic field with a uniform mass-to-flux ratio of about 8 in the $x$ direction.

To initialise the velocity we do not consider those three regions separately. We mimic roughly the turbulence of the interstellar medium by constructing a velocity field according to a probability distribution of the turbulence \citep[see for example section 3 of][for a review on turbulence in interstellar clouds]{Hennebelle2012}: 
in Fourier space, the power spectrum of the velocity field follows a Kolmogorov turbulence law\footnote{$\mathcal{P}_v (k) \propto k^{-11/3}$.}, and the phases are randomly chosen. This turbulent field is normalised to have a Mach number of 6.7.

Those initial conditions are set by specifying ratios of characteristic timescales in the inner regions of size $r_0$. We then define the free-fall time to sound-crossing time ratio $\tfrac{t_\text{ff}}{t_\text{sc}} = 0.15$, the free-fall time to Alfvén-crossing time ratio $\tfrac{t_\text{ff}}{t_\text{ac}} = 0.2$, and the free-fall time to turbulent-crossing time $\tfrac{t_\text{ff}}{t_\text{tc}} = 1$. The two first ratios show that the thermal and magnetic support against gravity are low, while the turbulent support is the dominant one.

\subsection{Runs performed}
We used this numerical setup to run four simulations which differ only by the stellar feedback that is included. The first simulation -- \snj{} -- is run without feedback at all. One simulation includes only protostellar jets as stellar feedback -- \sj{} -- and another includes only HII regions -- \srh{}. The last one -- \sjrh{} -- includes both protostellar jets and HII regions. The acronyms used to refer to those simulations are summarised in Table \ref{table: acronyms table}.

\begin{table}[h]
\resizebox{\linewidth}{!}{
\begin{tabular}{ccccc}
  \toprule
  Simulation & No feedback & Jets only & HII regions only & Jets \& HII regions \\
  \midrule
Protostellar jets & \tickNo & \tickYes & \tickNo & \tickYes \\
HII regions & \tickNo & \tickNo & \tickYes & \tickYes \\
\midrule
Acronym & \snj & \sj & \srh & \sjrh \\
  \bottomrule
\end{tabular}
}
\caption{Summary of the four runs performed. The check marks and crosses indicate whether or not the protostellar jets or the HII regions are included in the simulations. The acronym for each simulation is given on the last line. These acronyms will be used all across the article.}
\label{table: acronyms table}
\end{table}

We also ran two simulations with protostellar jets as the unique source of feedback but with different properties for the jets than in the \sj{} simulation. The jets in the simulation ``jets - wider'' exhibit a wider ejection angle of $30 ^\circ$ while it is $20 ^\circ$ in \sj{}. The jets in the simulation ``jets - faster'' are expelled with a velocity equal to 48\% of the escape velocity of the sink, while it is 24\% in \sj{}.

%% file: Results_SFR.tex
\section{General description, star formation rate and star formation efficiency}
\label{section: article2-Description,SFE,SFE}

\subsection{Global appearance of the emerging structures}
\label{section-sub: Global appearance of the emerging structures}

\begin{figure*}[hbtp]
\begin{overpic}[width=0.333\textwidth]{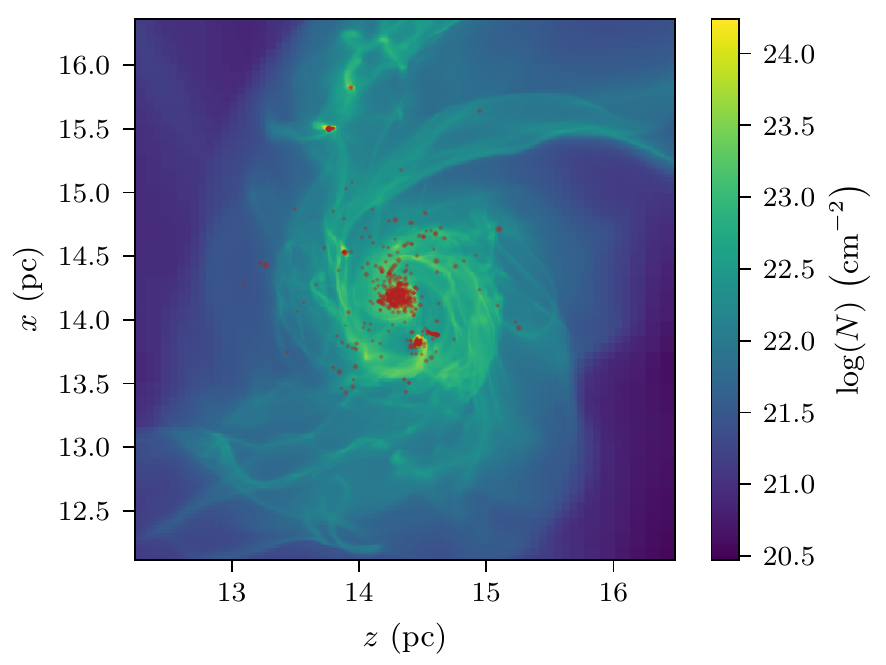} \put (20,65) {\textcolor{white}{\snj}} \end{overpic}%
\includegraphics[width=0.333\textwidth]{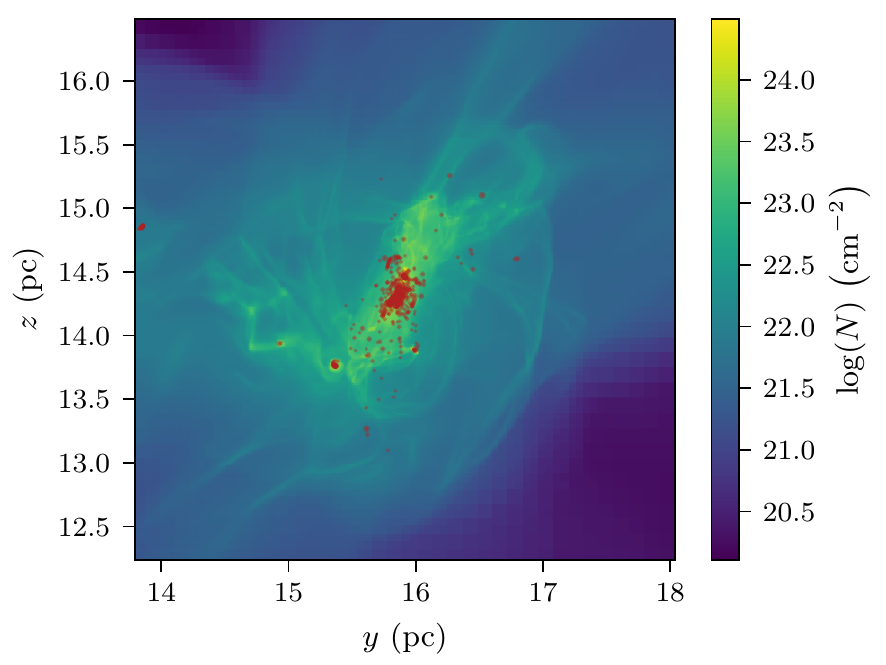}%
\includegraphics[width=0.333\textwidth]{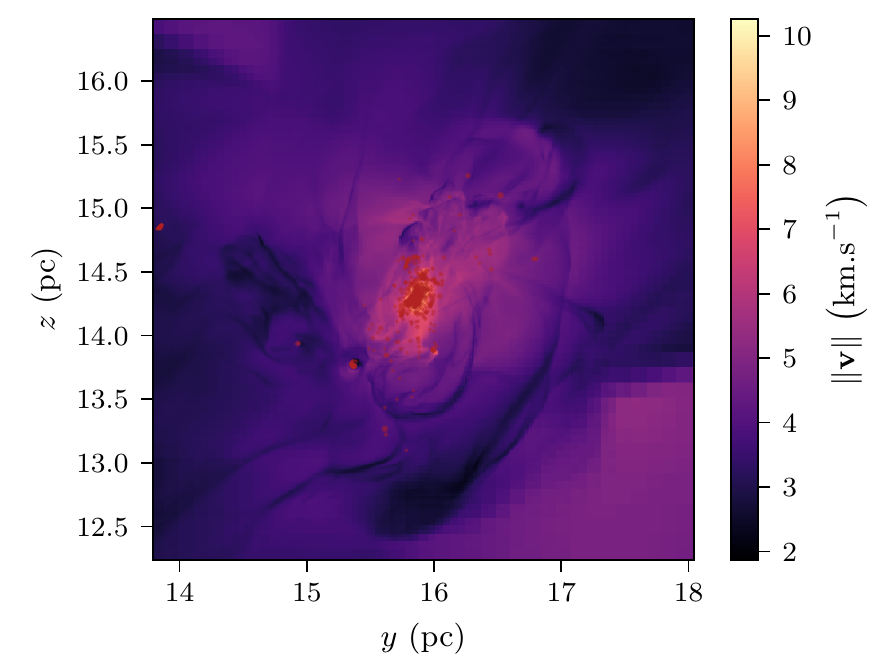}
\begin{overpic}[width=0.333\textwidth]{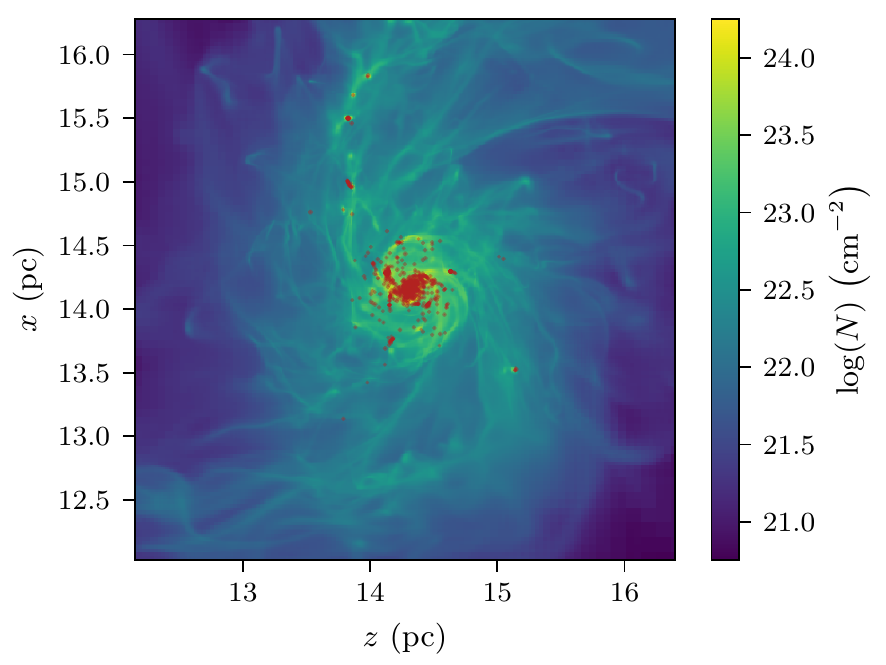} \put (20,65) {\textcolor{white}{\sj}} \end{overpic}%
\includegraphics[width=0.333\textwidth]{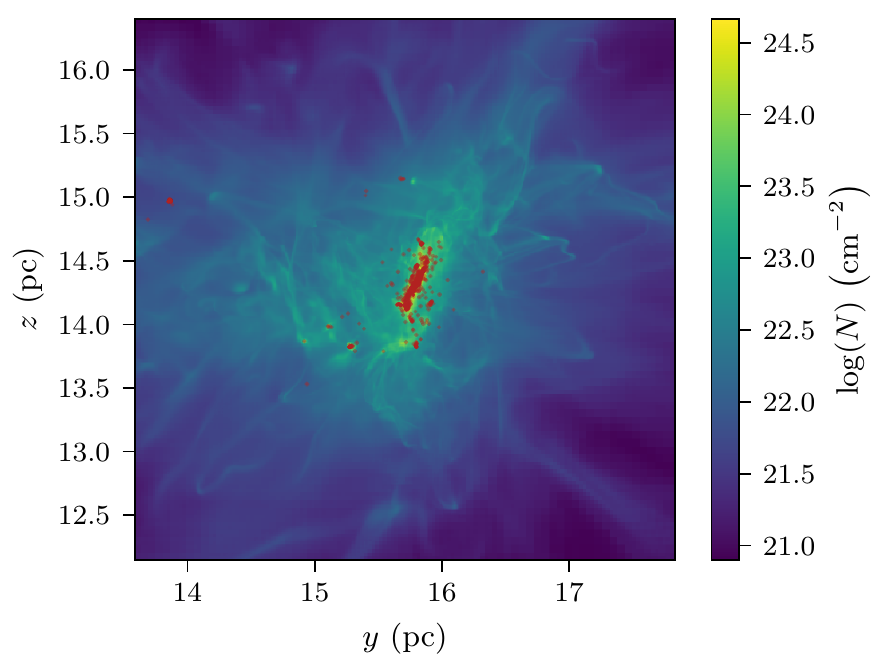}%
\includegraphics[width=0.333\textwidth]{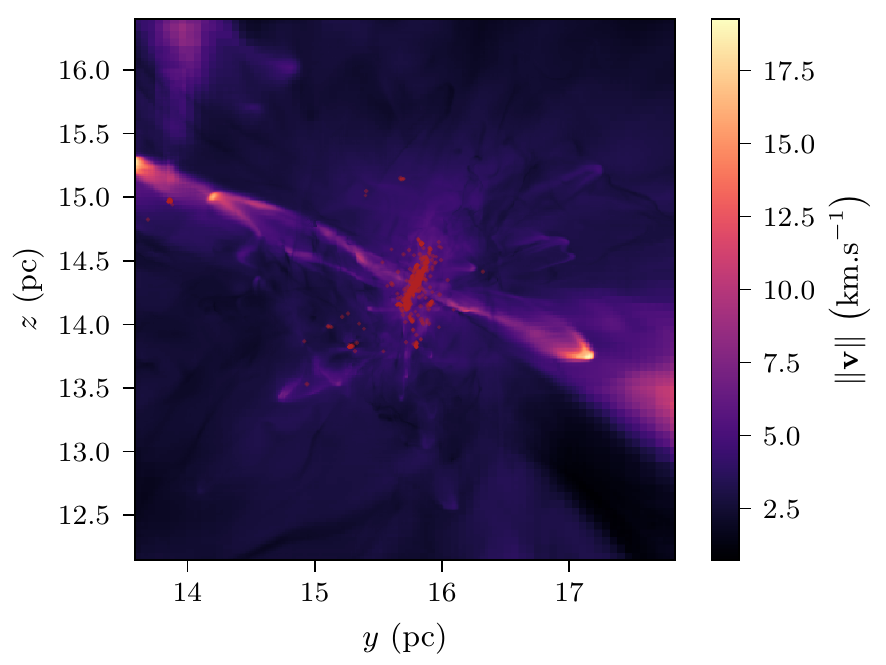}
\begin{overpic}[width=0.333\textwidth]{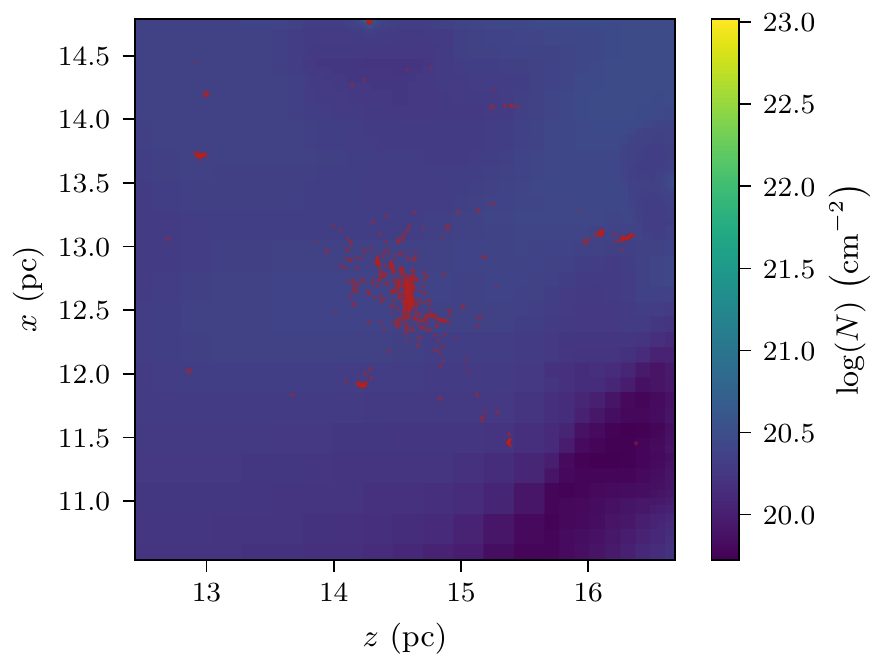} \put (20,65) {\textcolor{white}{\srh}} \end{overpic}%
\includegraphics[width=0.333\textwidth]{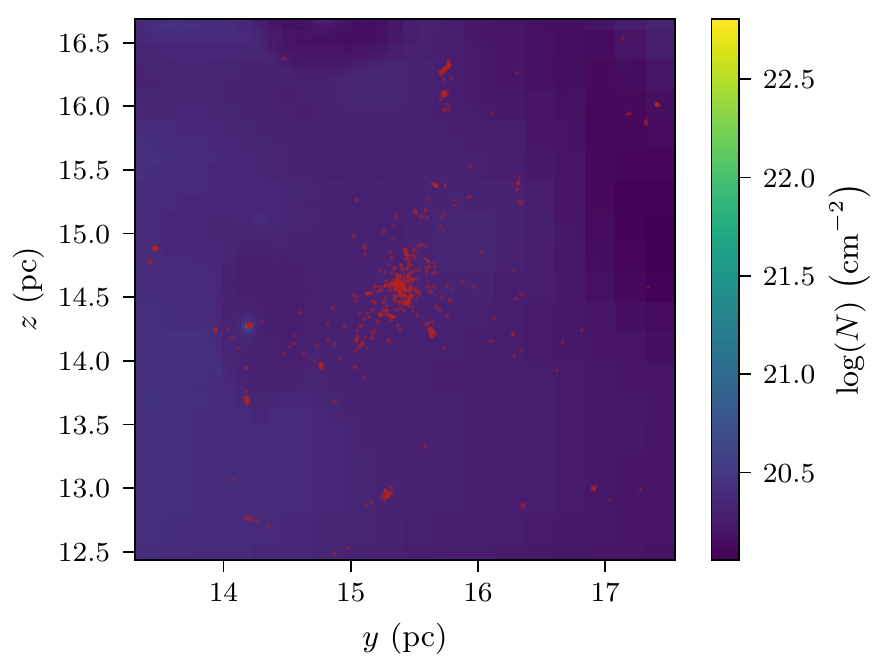}%
\includegraphics[width=0.333\textwidth]{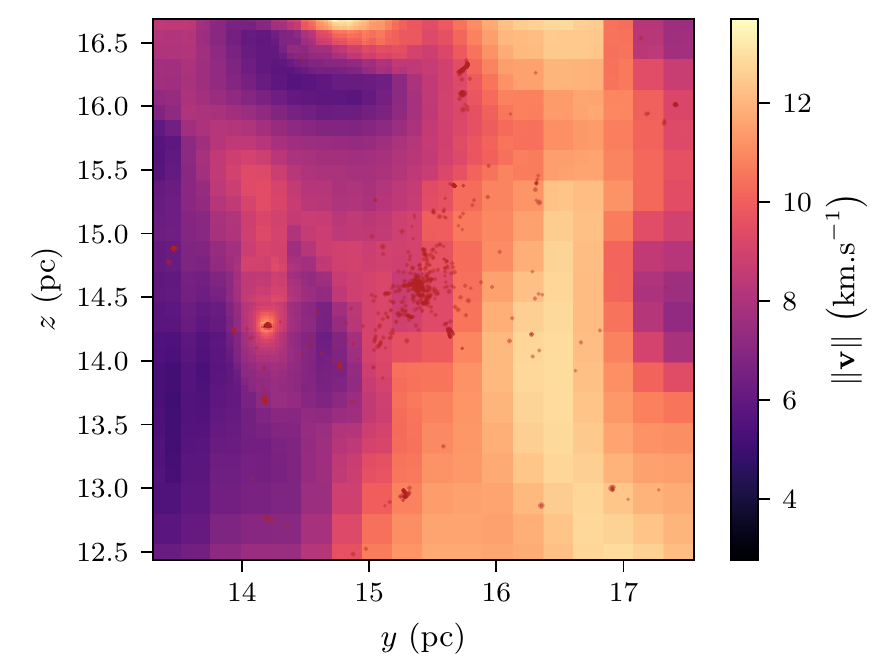}
\begin{overpic}[width=0.333\textwidth]{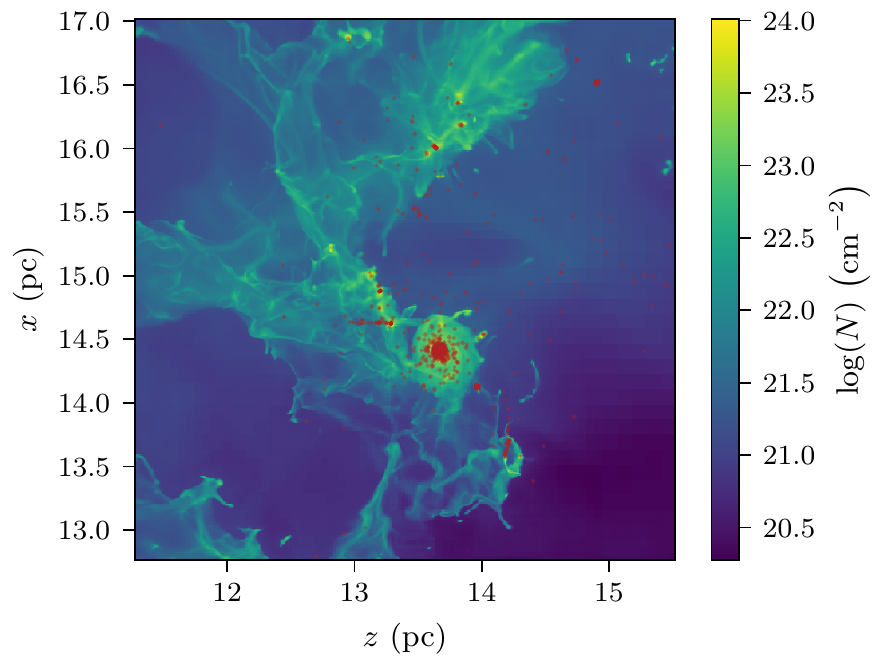} \put (20,65) {\textcolor{white}{\sjrh}} \end{overpic}%
\includegraphics[width=0.333\textwidth]{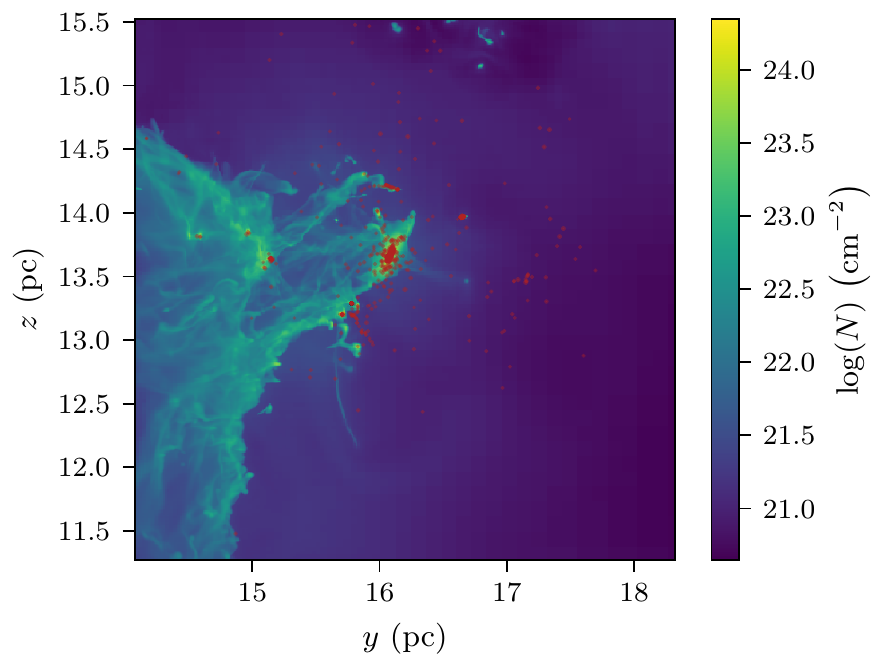}%
\includegraphics[width=0.333\textwidth]{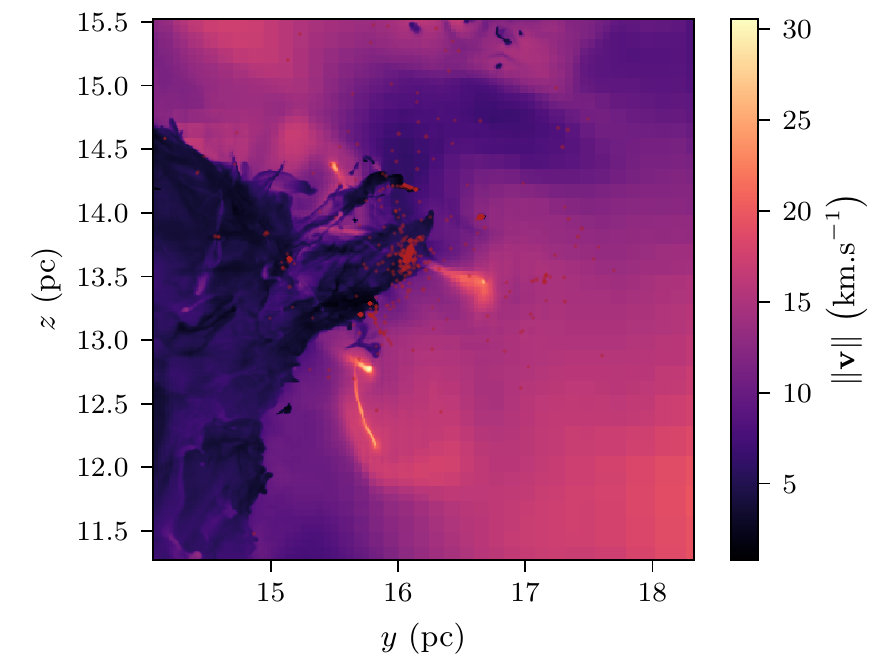}
\caption{Global appearance of the star cluster surrounded by its gas environment. Each row corresponds to a different simulation. The four simulations are visualised at the same time of \SI{3.5}{Myr}. From top to bottom are the simulations without feedback, with protostellar jets only, with HII regions only, and with both jets and HII regions. The two first columns are column density along the $y$ and $x$ axis of the simulation, and the third column is the mean of the velocity norm integrated along the line of sight, weighted by the density. The colour scales are not common and depend on each map. The overplotted red circles represent the sink particles. As the view on the two last rows, corresponding to the simulations with HII regions, is a bit narrow, we presented on Fig. \ref{figure: global appearance of simulations - large scale} the same maps with a spatial scale four times larger. We also present zooms on the central star cluster on Fig. \ref{figure: global appearance of simulations - small scale}.}
\label{figure: global appearance of simulations}
\end{figure*}

\begin{figure*}[hbtp]
\begin{overpic}[width=0.333\textwidth]{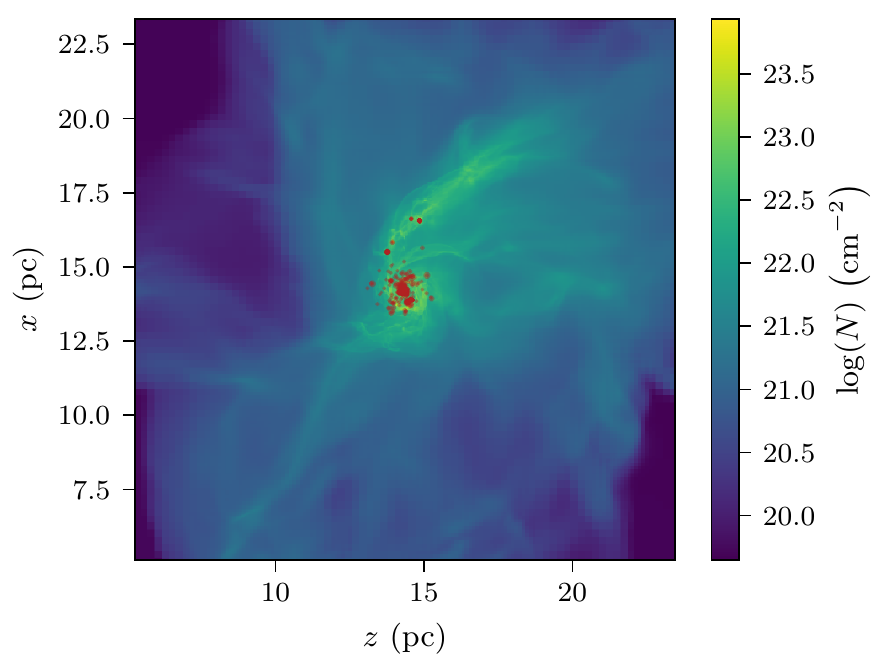} \put (20,65) {\textcolor{white}{\snj}} \end{overpic}%
\includegraphics[width=0.333\textwidth]{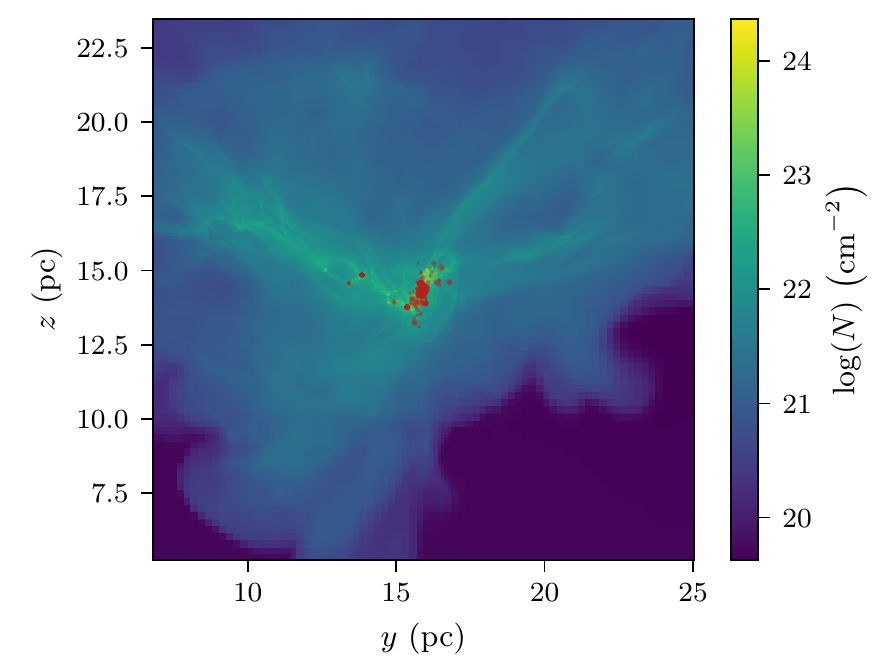}%
\includegraphics[width=0.333\textwidth]{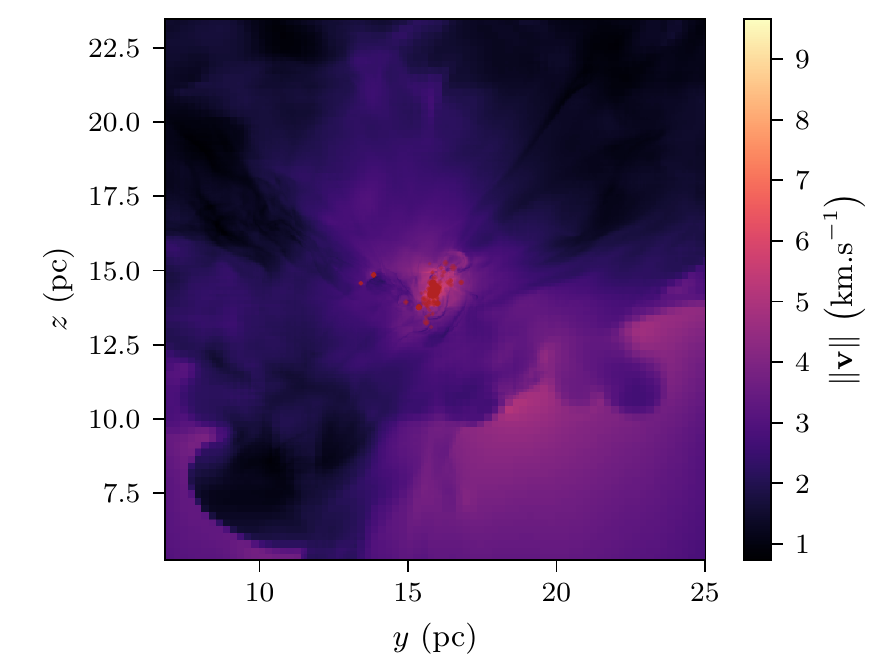}
\begin{overpic}[width=0.333\textwidth]{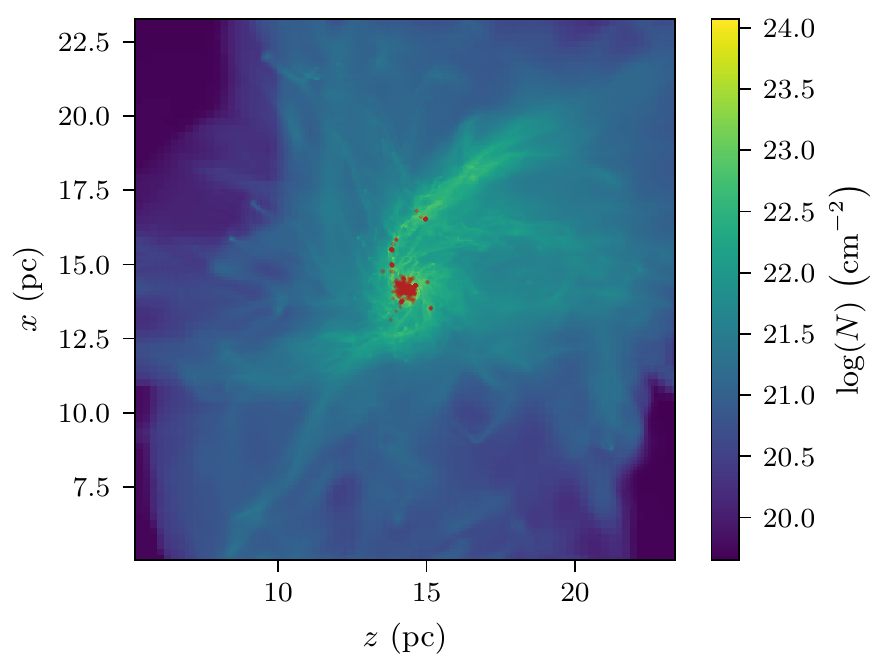} \put (20,65) {\textcolor{white}{\sj}} \end{overpic}%
\includegraphics[width=0.333\textwidth]{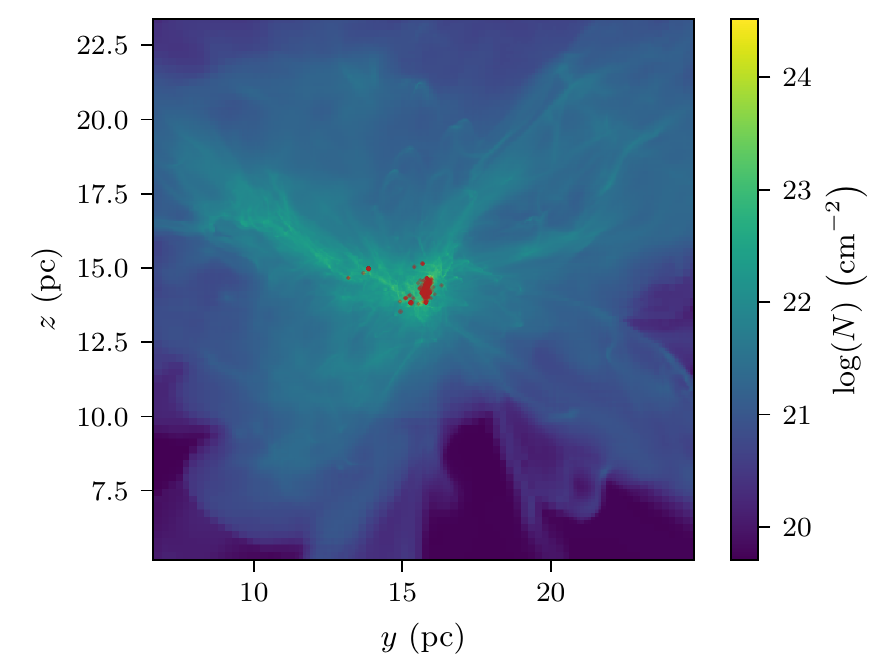}%
\includegraphics[width=0.333\textwidth]{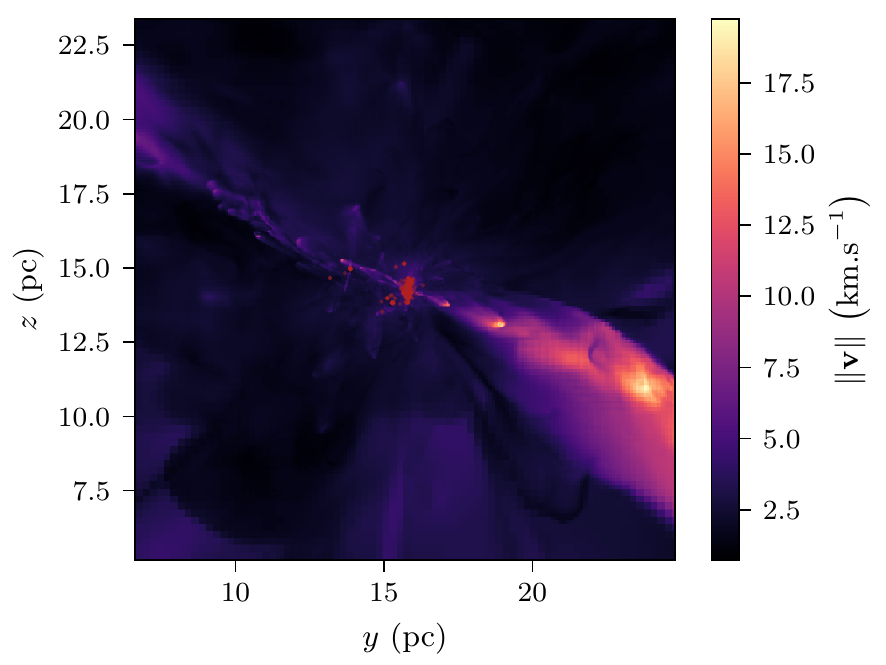}
\begin{overpic}[width=0.333\textwidth]{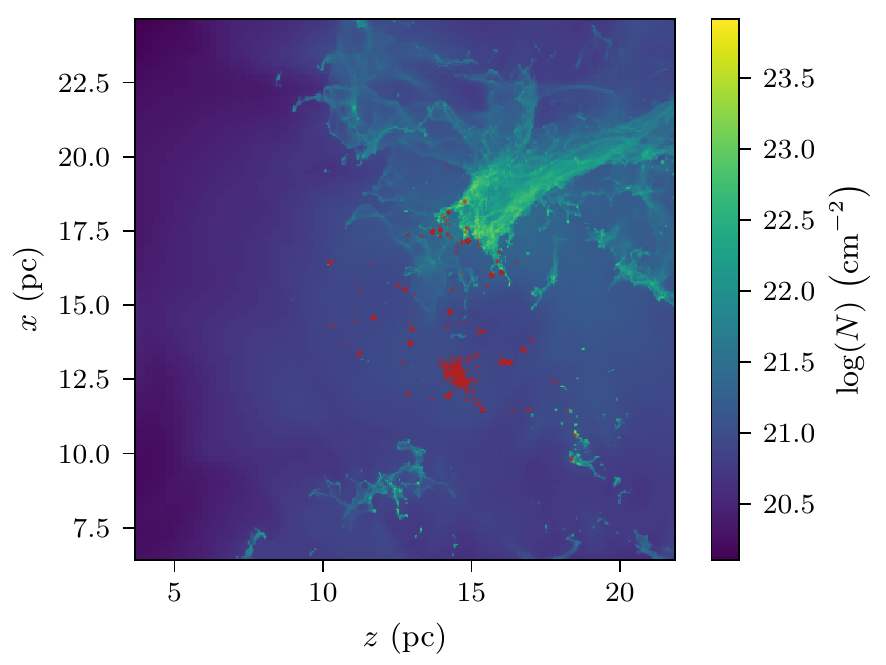} \put (20,65) {\textcolor{white}{\srh}} \end{overpic}%
\includegraphics[width=0.333\textwidth]{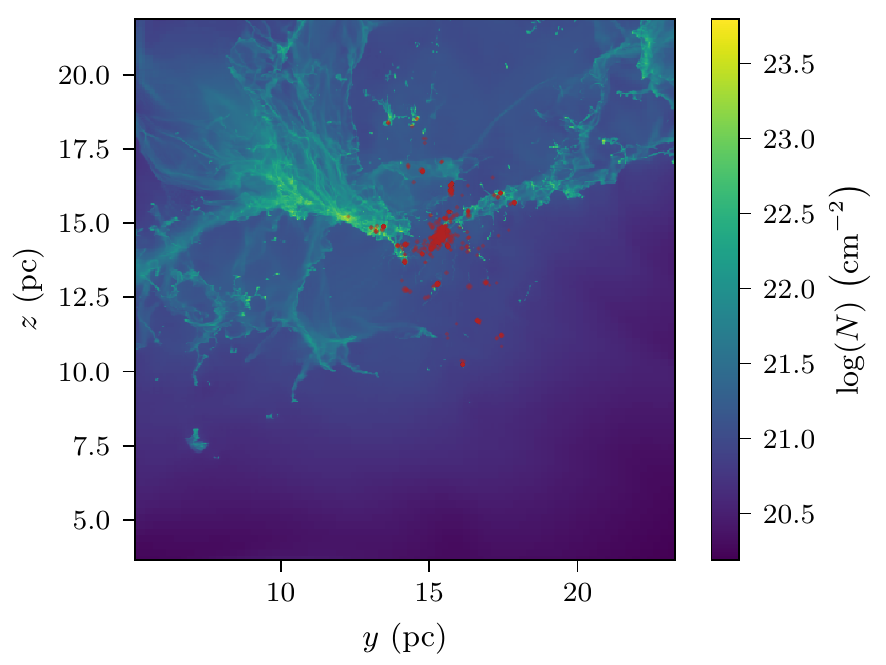}%
\includegraphics[width=0.333\textwidth]{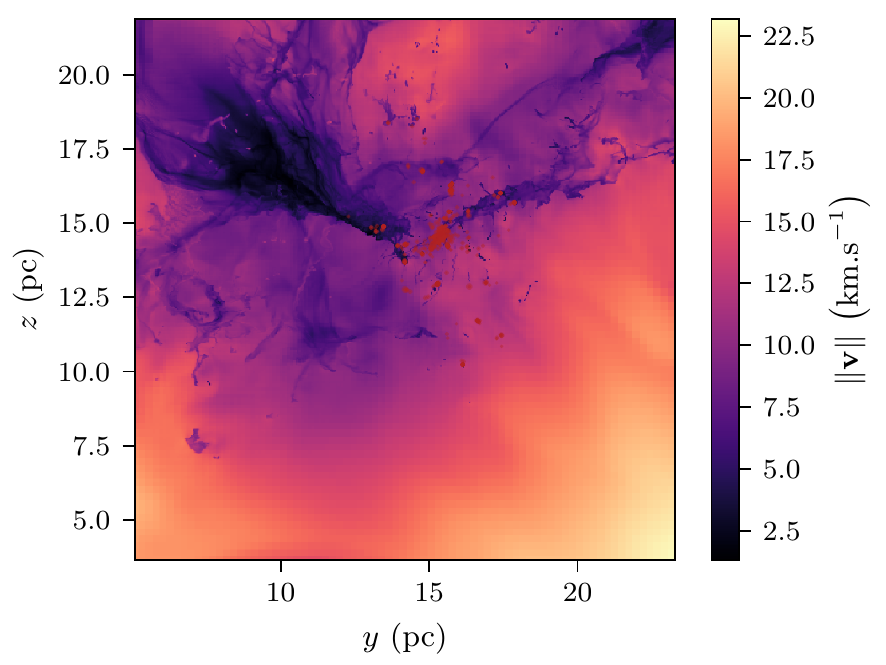}
\begin{overpic}[width=0.333\textwidth]{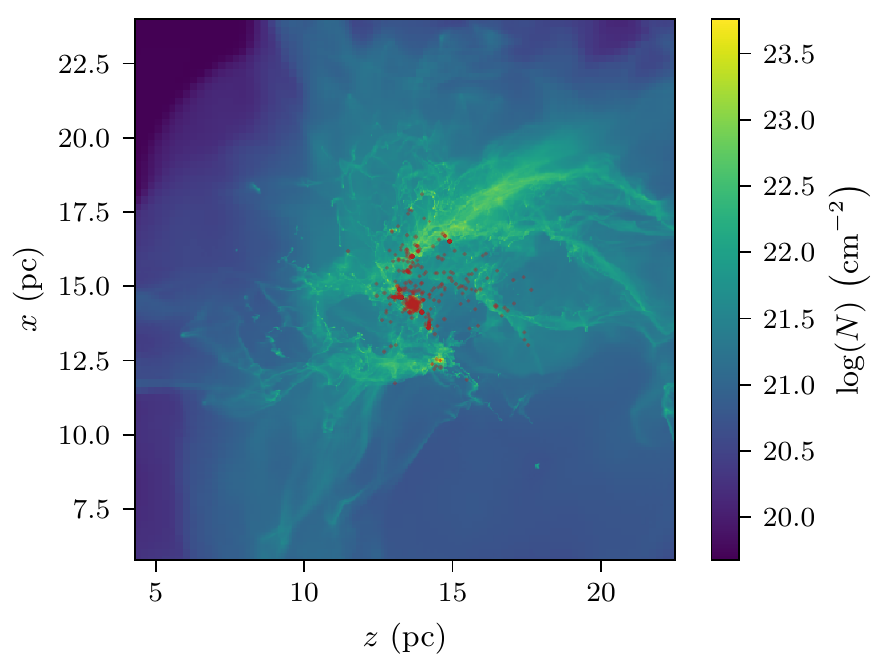} \put (20,65) {\textcolor{white}{\sjrh}} \end{overpic}%
\includegraphics[width=0.333\textwidth]{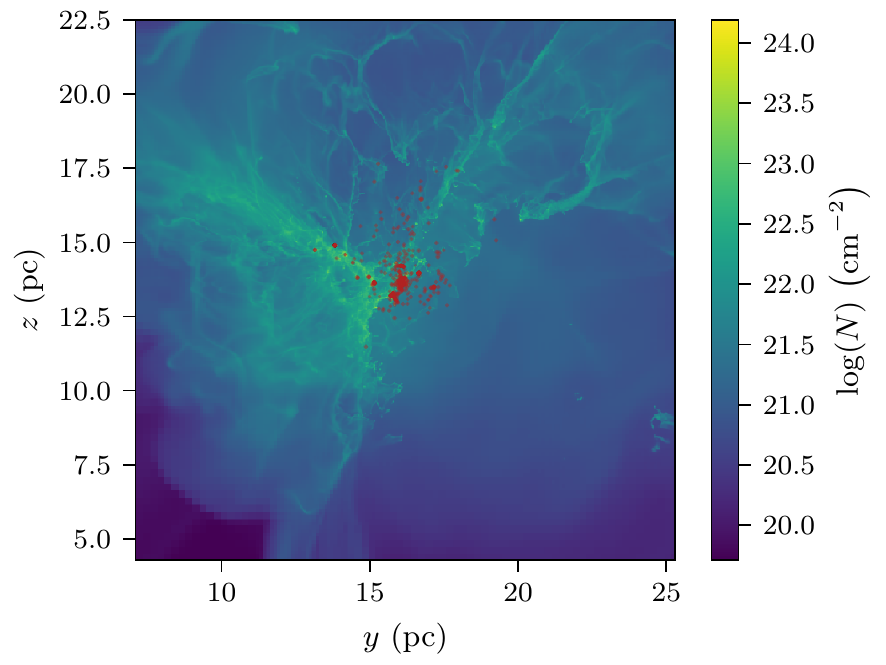}%
\includegraphics[width=0.333\textwidth]{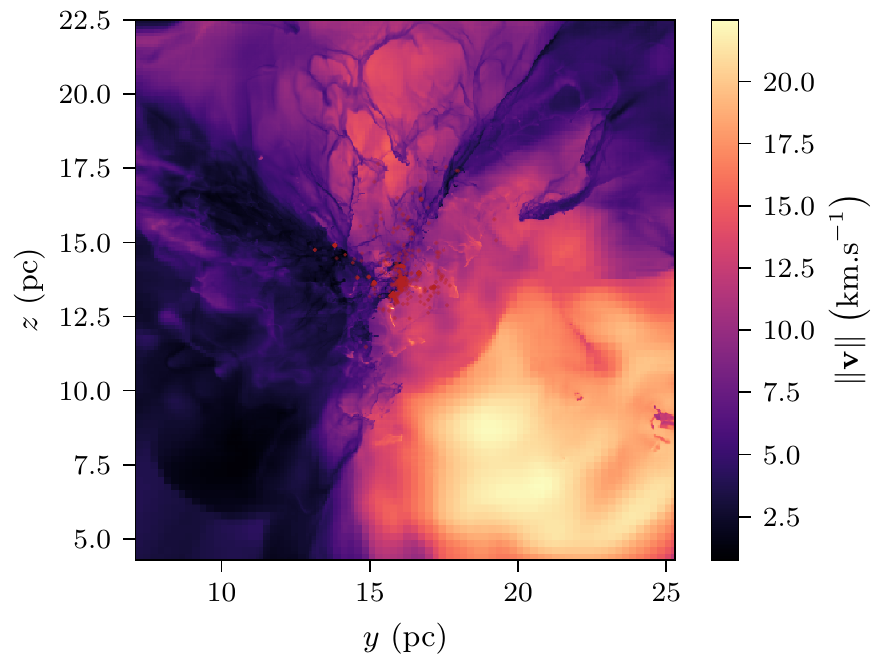}
\caption{Global appearance of the star cluster surrounded by its gas environment, at a spatial scale four times larger than Fig. \ref{figure: global appearance of simulations}. The four simulations are visualised at the same time of \SI{3.5}{Myr}. Each row corresponds to a different simulation. From top to bottom are the simulations without feedback, with protostellar jets only, with HII regions only, and with both jets and HII regions. The two first columns are column density along the $y$ and $x$ axis of the simulation, and the third column is the mean of the velocity norm integrated along the line of sight, weighted by the density. The colour scales are not common and depend on each map. The overplotted red circles represent the sink particles.}
\label{figure: global appearance of simulations - large scale}
\end{figure*}

We first look at the appearance of the emerging star clusters and surrounding gas on Fig. \ref{figure: global appearance of simulations} and \ref{figure: global appearance of simulations - large scale} for the four simulations. Figure~\ref{figure: global appearance of simulations} shows the emerging clusters at intermediate scales, whereas Fig.~\ref{figure: global appearance of simulations - large scale} shows them at large scales. The two first columns show column density along two different lines of sight, and the third column shows the mean velocity norm along the second line of sight, weighted by density. The overplotted red circles represent the sink particles. The four simulations are visualised at the same time of \SI{3.5}{Myr}. The first row exhibits the column densities of the simulation without feedback, \snj{}, the second one those of the simulation with jets only, \sj{}, the third one those of the simulation with HII regions only, \srh{}, and the last row exhibits the column density of the simulation with both protostellar jets and HII regions, \sjrh{}. The simulations that do not include HII regions are pretty similar, exhibiting a flattened cluster of stars, surrounding by spiralling gas with a disk-like shape. In the case of simulations with jets, the distribution of the gas is a bit more messy, due to the presence of jets. Those jets are visible in the third panel of the second row as high velocity components from either side of the cluster. On the other hand, the simulations that include HII regions have a qualitatively different appearance. The two simulations exhibit a distribution of the gas which is much more shredded, due to the expansion of HII regions, which is clearly visible from the velocity maps displayed in 
the third column of Fig. \ref{figure: global appearance of simulations} and \ref{figure: global appearance of simulations - large scale}.
The disk-like shape of the gas which was visible in the simulations without HII regions is no longer visible. However the star cluster also seems to be a bit flattened in these simulations.

In the case of the simulations without HII regions, this disk-like shape of the gas correlated to the flattened aspect of the star cluster seems to indicate that a preferred angular direction emerges as the cluster forms. This is not surprising, as \citet{Verliat2020} exposed a mechanism through which a protostellar disk naturally form around a protostar even in the absence of initial rotation.
Angular momentum with respect to the cluster center, which is not the mass center of the system,  
is produced by inertial forces.
The same mechanism can be invoked here to explain the formation of this rotating structure emerging from the gravitational collapse. 
Moreover \citet{Lee2016-1} inferred that the clusters formed in their 
simulations without feedback indeed  significantly rotate due to angular 
momentum inherited from the large scale collapse. 

\begin{figure*}[hbtp]
\begin{overpic}[width=0.333\textwidth]{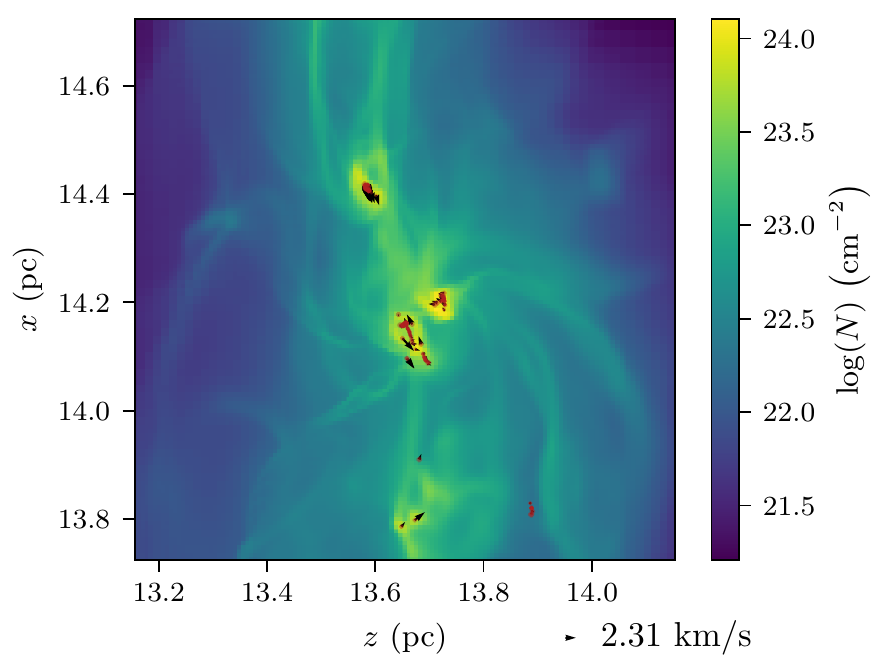} \put(20,65){\textcolor{white}{\snj}} \put(33,75){\textcolor{black}{$t=\SI{2}{Myr}$}}\end{overpic}%
\begin{overpic}[width=0.333\textwidth]{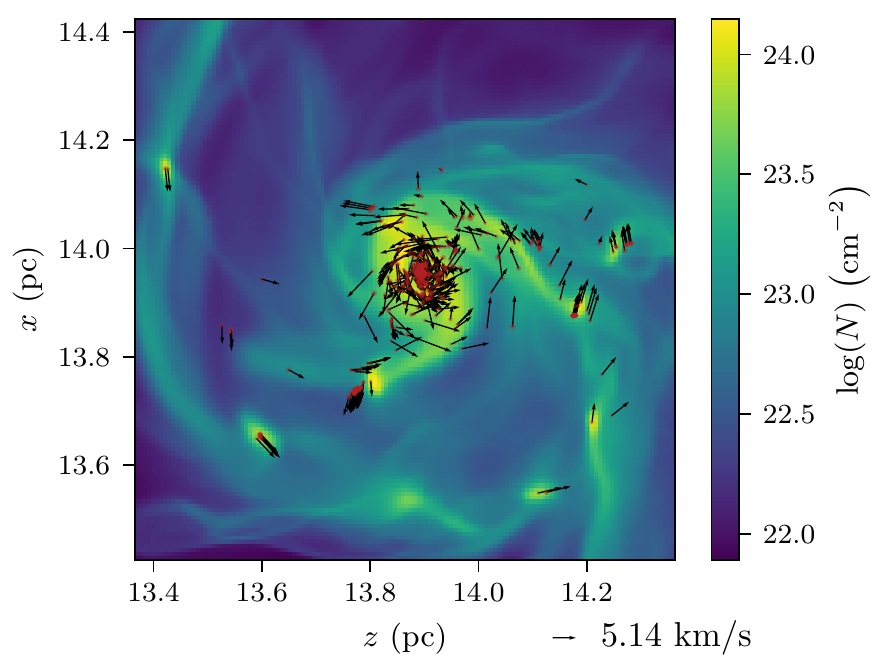} \put(33,75){\textcolor{black}{$t=\SI{2.75}{Myr}$}}\end{overpic}%
\begin{overpic}[width=0.333\textwidth]{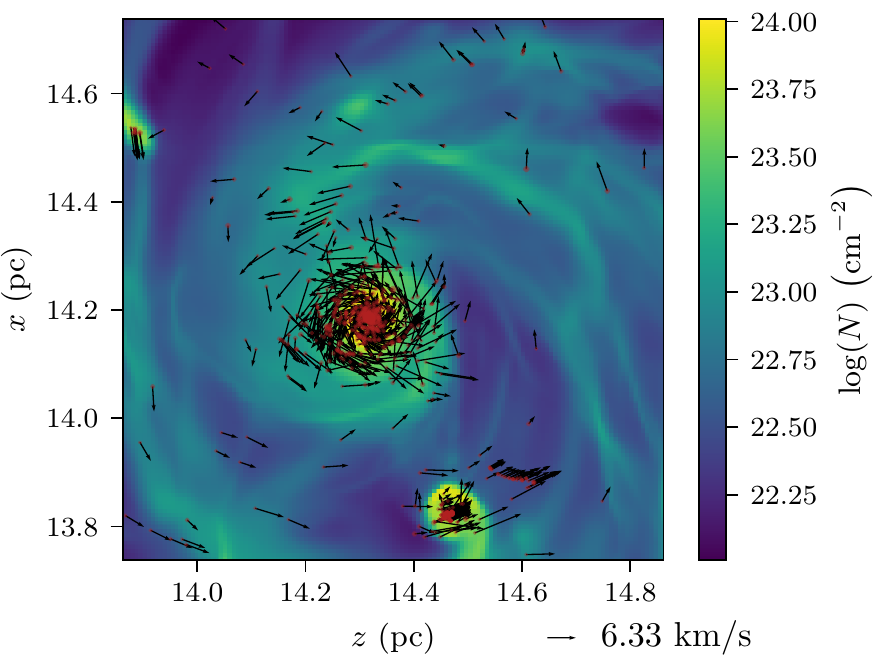} \put(33,75){\textcolor{black}{$t=\SI{3.5}{Myr}$}}\end{overpic}
\begin{overpic}[width=0.333\textwidth]{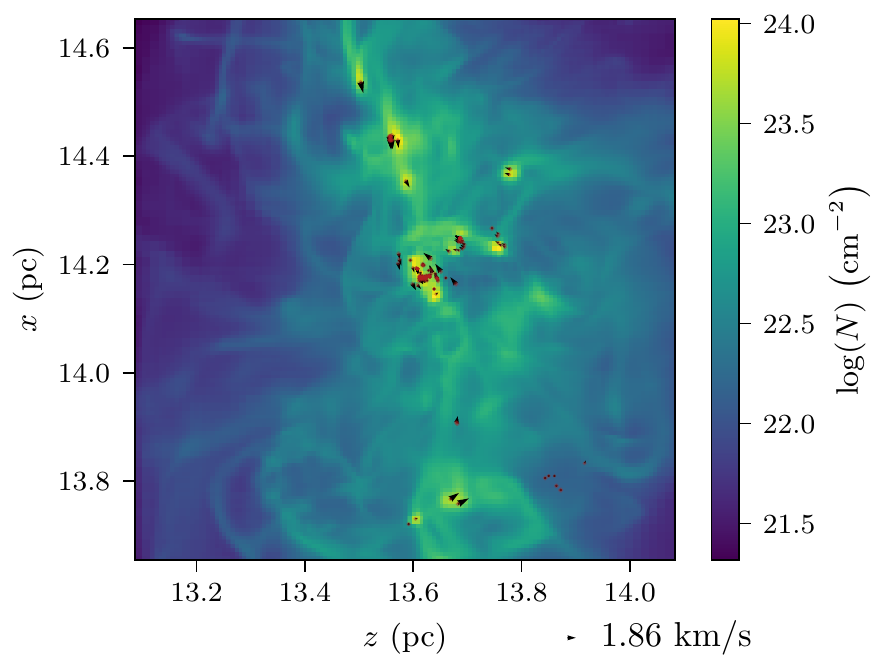} \put (20,65) {\textcolor{white}{\sj}} \end{overpic}%
\includegraphics[width=0.333\textwidth]{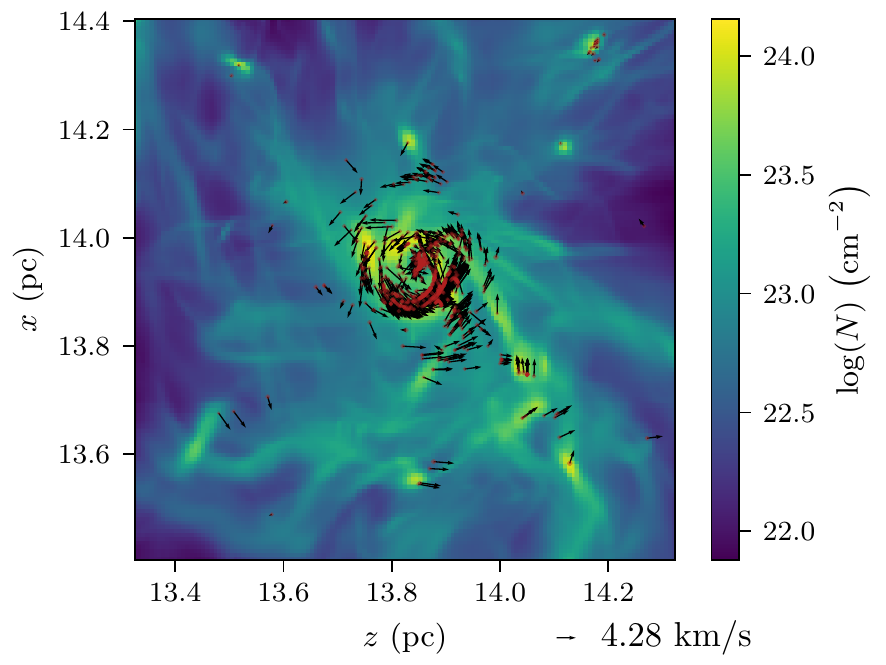}%
\includegraphics[width=0.333\textwidth]{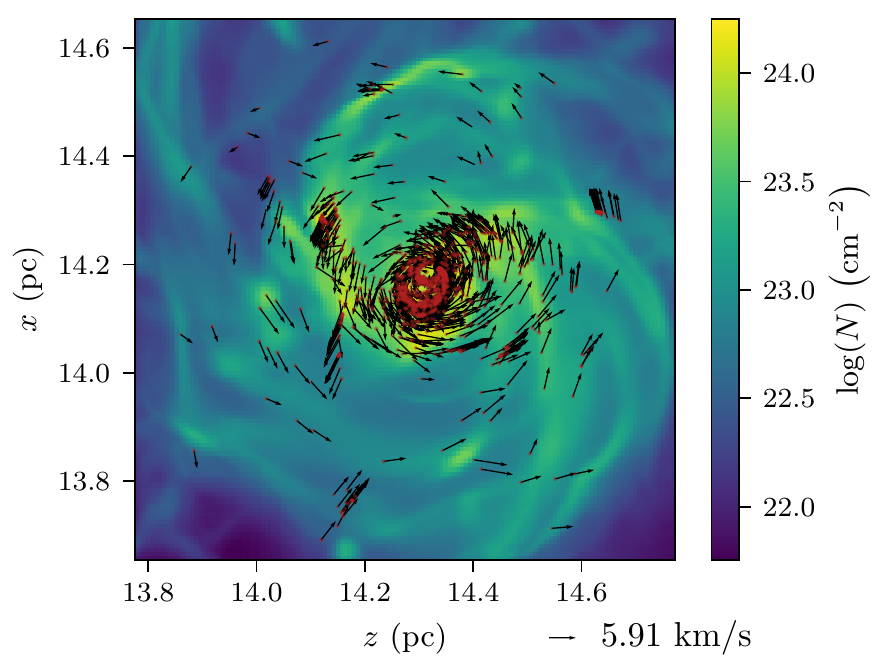}
\begin{overpic}[width=0.333\textwidth]{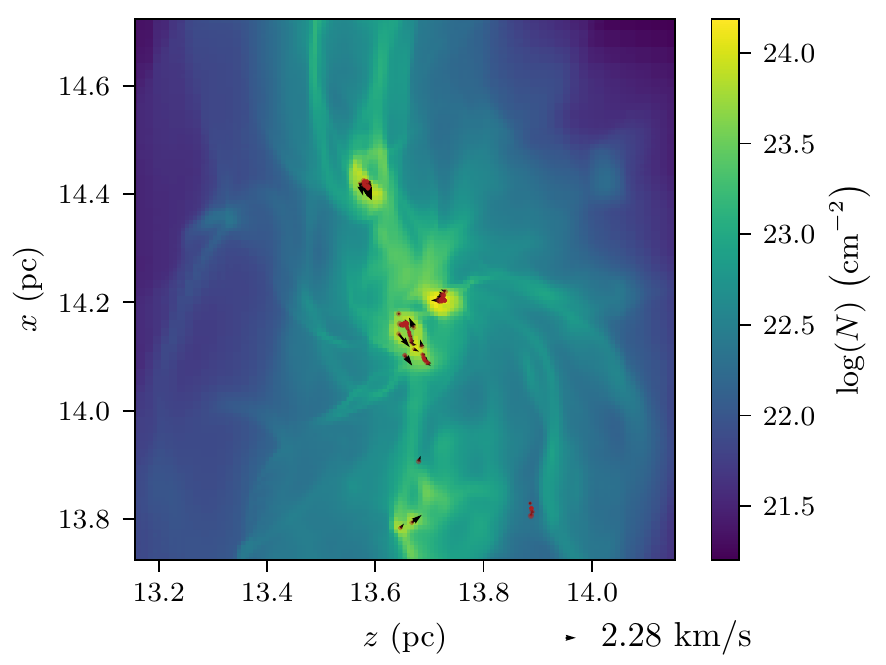} \put (20,65) {\textcolor{white}{\srh}} \end{overpic}%
\includegraphics[width=0.333\textwidth]{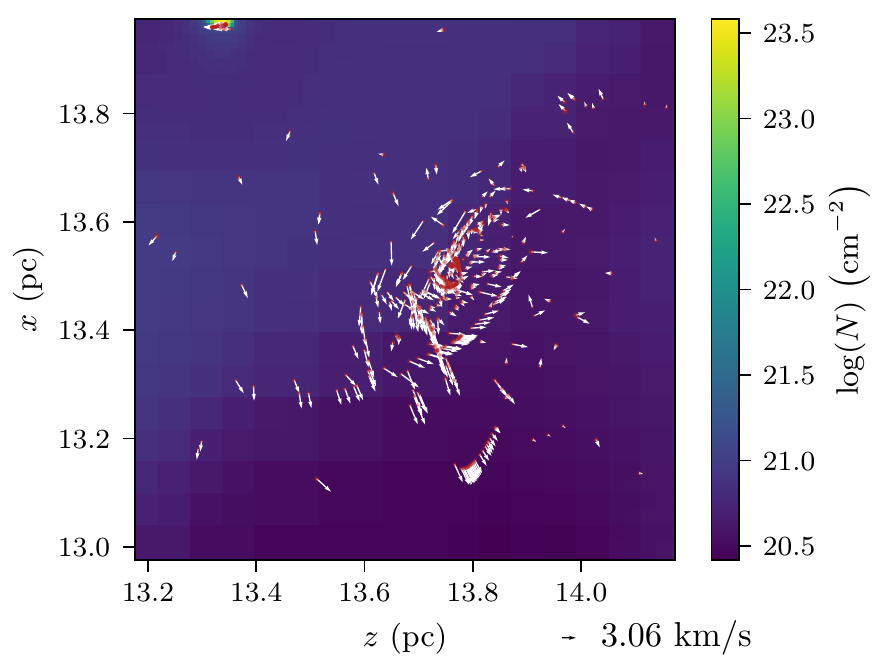}%
\includegraphics[width=0.333\textwidth]{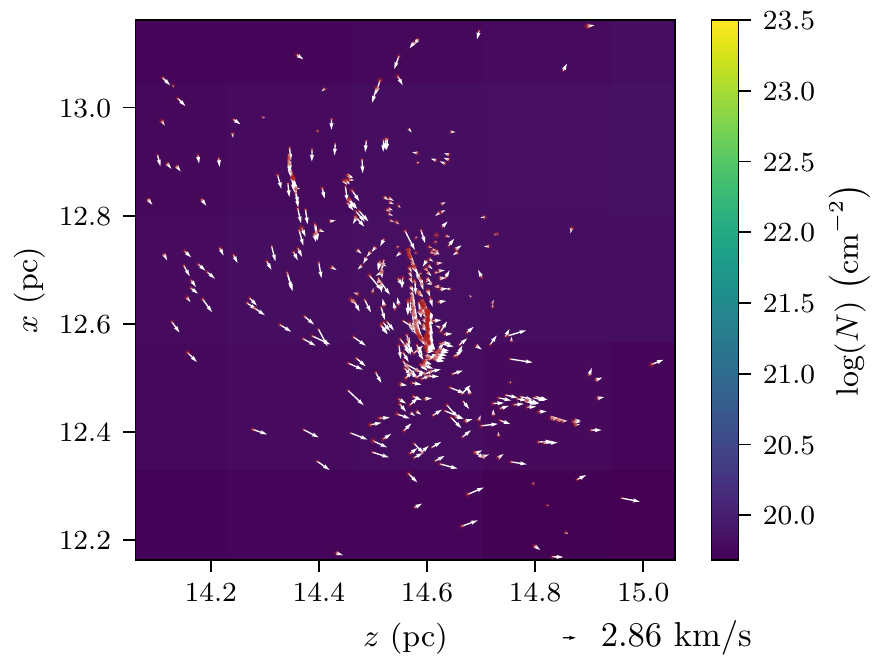}
\begin{overpic}[width=0.333\textwidth]{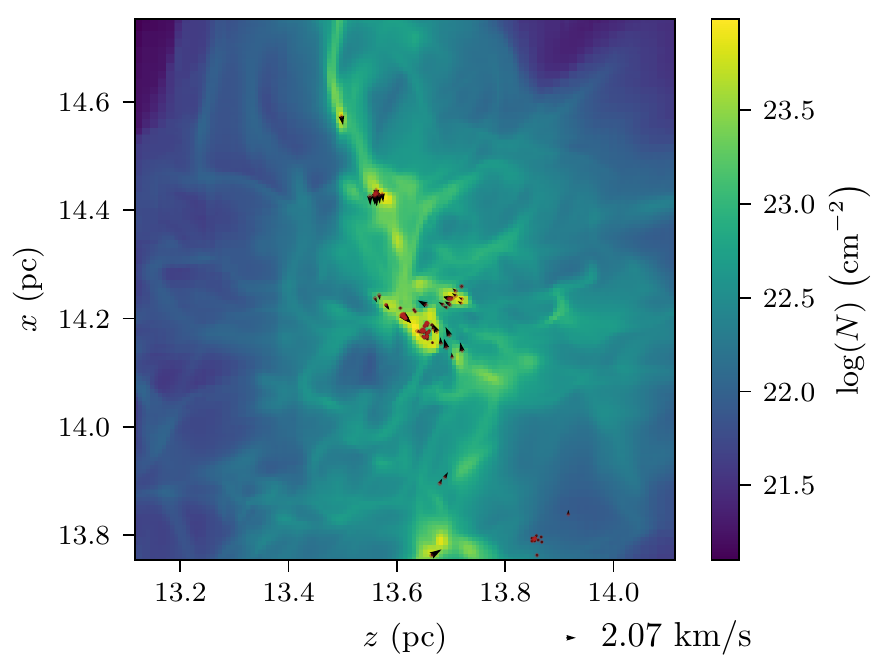} \put (20,65) {\textcolor{white}{\sjrh}} \end{overpic}%
\includegraphics[width=0.333\textwidth]{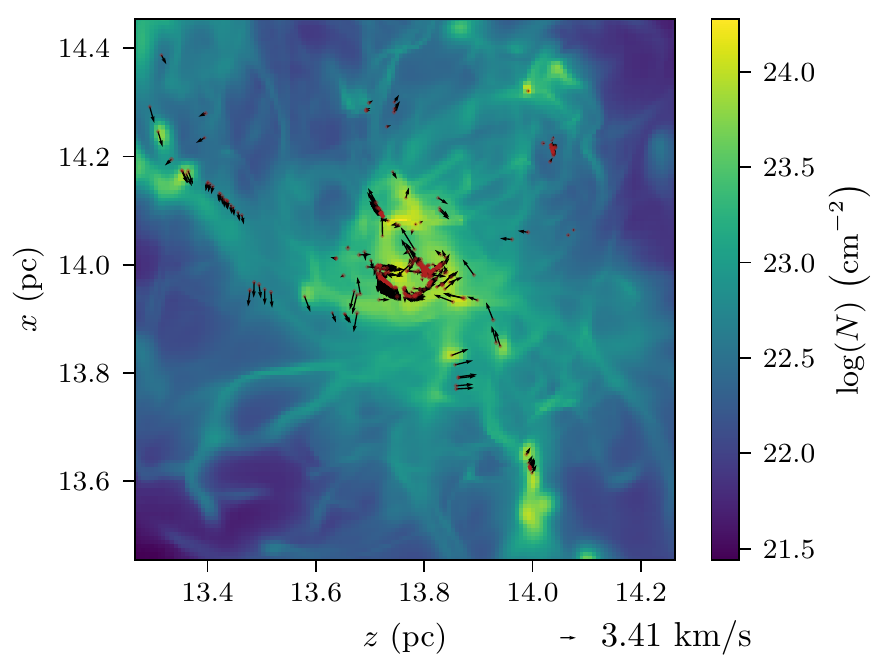}%
\includegraphics[width=0.333\textwidth]{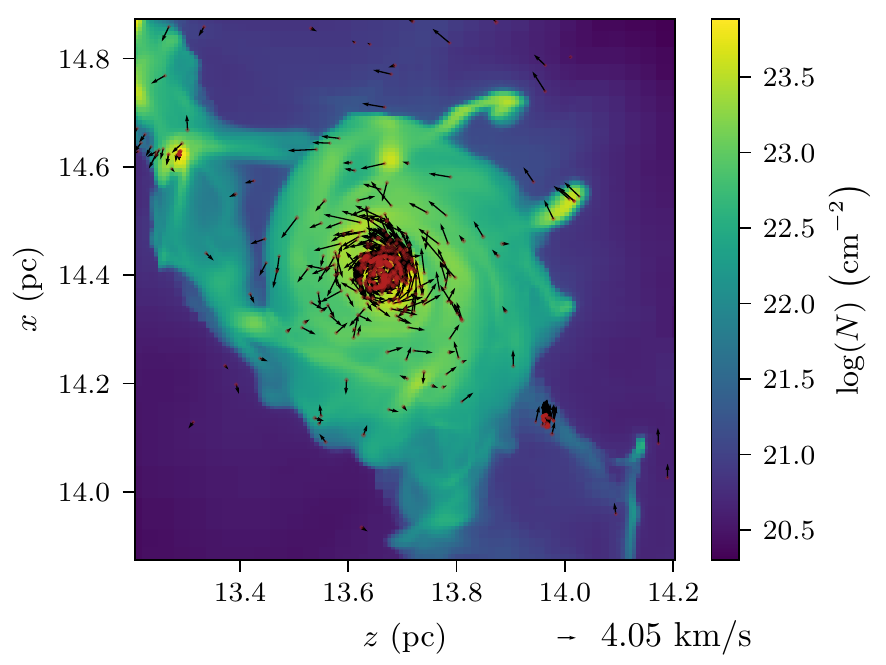}
\caption{Evolutionary sequence of the global appearance of the star cluster. Each row corresponds to a different simulation. The four simulations are visualised at three different times, \SI{2}{Myr}, \SI{2.75}{Myr} and \SI{3.5}{Myr}. From top to bottom are the simulations without feedback, with protostellar jets only, with HII regions only, and with both jets and HII regions, seen from the $y$ axis of the simulation. The colour scales are not common and depend on each map. The overplotted red circles represent the sink particles and the arrows associated represent their velocities in the plane of the visualisation.}
\label{figure: global appearance of simulations - small scale}
\end{figure*}

An evolutionary sequence of the cluster seen at small scales is visible on Fig. \ref{figure: global appearance of simulations - small scale}. The arrows attached to the sink particles represent their velocities in the plane perpendicular to the line of sight. At \SI{2}{Myr}, \snj{} and \srh{} are very similar as the expansion of HII regions has not yet occurred. At this time, the simulations with jets --- \sj{} and \sjrh{} --- are already more messy and also look both similar. As time progresses, the cluster in \snj{}, \sj{} and \sjrh{} becomes organised in the disk-like shape which seems to rotate. In \srh{}, the cluster has been completely depleted of gas, and its rotation is less obvious.

\noindent

\noindent

\noindent

\subsection{Star formation rate and efficiency}
\label{section-sub: Star formation rate and efficiency}

The first and straightforward quantity to investigate  is the total mass of the sink particles as a function of time. The slope of this curve represents the
rapidity at which the stars accrete and is often refered to as the 
SFR while the final value of the accreted mass divided by the initial 
cloud mass represents the star formation efficiency (SFE). 

\begin{figure}[hbtp]
\includegraphics[width=\linewidth]{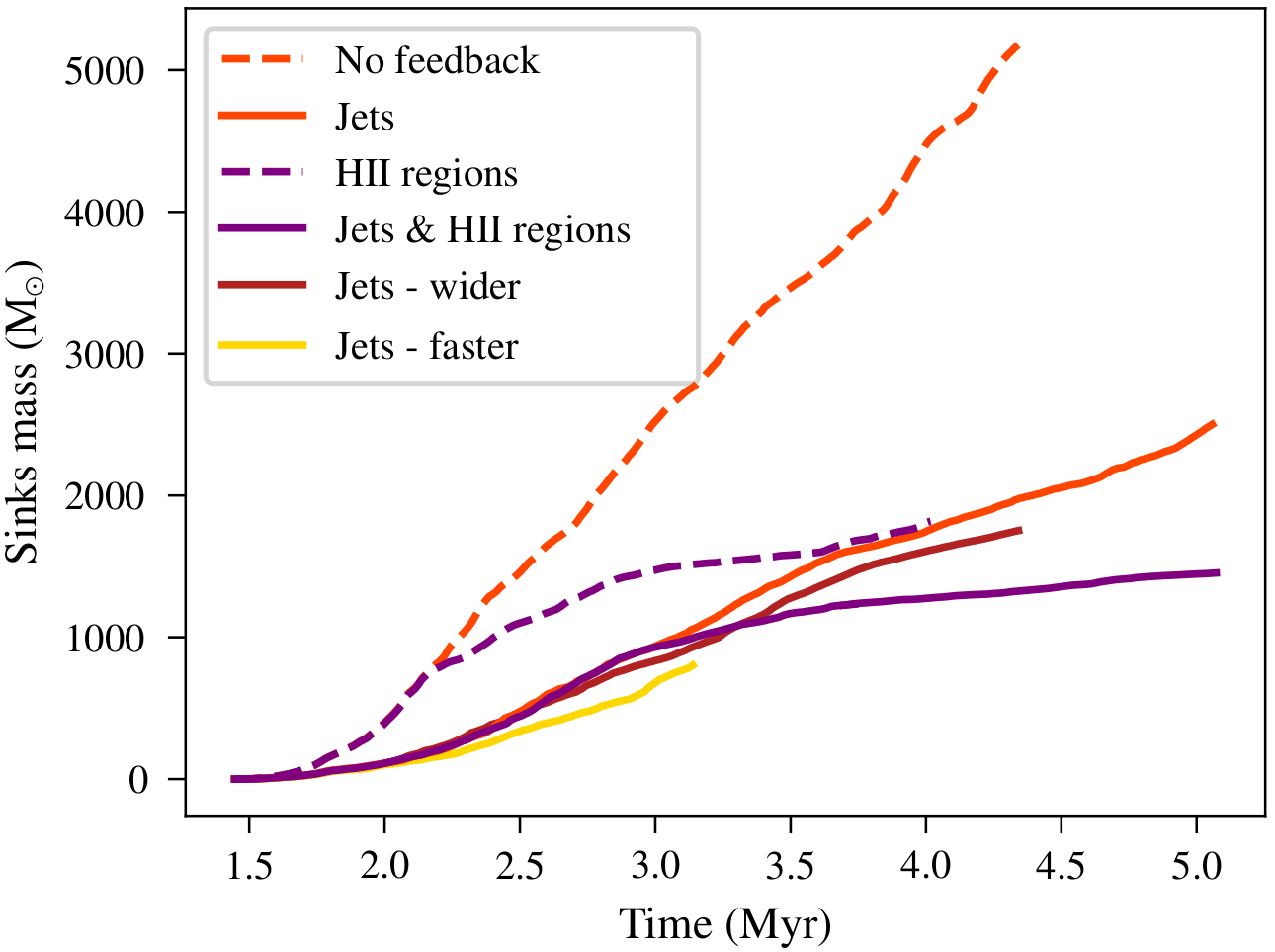}
\caption{Total mass of the sink particles in the four simulations including or not different types of feedback. The two orange (respectively purple) curves stand for the simulations without (respectively with) HII regions. The dashed (respectively solid) lines represent the simulations without (respectively with) protostellar jets.}
\label{figure: total sink mass all vs time}
\end{figure}

The evolution of this total mass over time is presented on Fig.~\ref{figure: total sink mass all vs time}. The two orange curves represent the two simulations without HII regions, while the two purple ones represent the two simulations including HII regions. The two dashed lines represent the simulations without protostellar jets, while the two solid lines represent the simulations including them. The brown solid line and the yellow solid line represent the simulations with respectively wider and faster jets.

\noindent

Let's compare the two orange curves, thus the simulations without feedback at all on one hand, and with protostellar jets only on the other hand. After \SI{2}{Myr} for the simulation without feedback, and \SI{2.6}{Myr} for the simulations with jets, the total sink mass seems to evolve almost linearly with time.
The SFR of these simulations is thus about \SI{2e-3}{\Msun . yr^{-1}} for the fisrt one, and about \SI{8e-4}{\Msun . yr^{-1}} for the second one. The difference in term of SFR when adding protostellar jets from a situation without feedback is thus a bit more than a factor two. This is consistent 
with what has been inferred previously in the litterature 
for instance by \citet{wang2010} and \citet{Federrath2014} as can be 
respectivelly seen from Fig.~1 and 2 of these papers. 

\begin{figure}[hbtp]
\includegraphics[width=\linewidth]{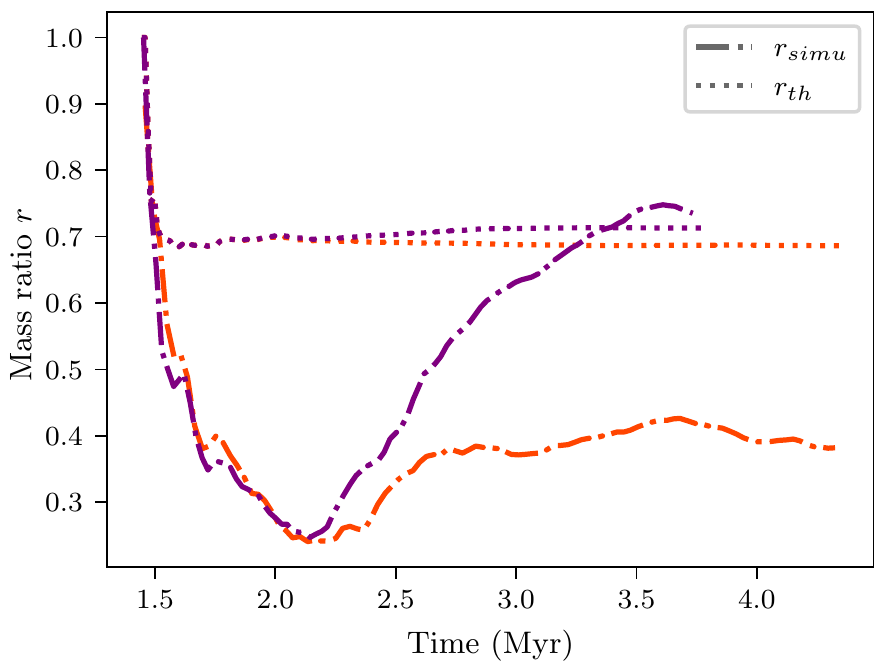}
\caption{In dashdotted lines, $M_\text{sink,\sj{}} / M_\text{sink,\snj{}}$ in orange, and $M_\text{sink,\sjrh{}} / M_\text{sink,\srh{}}$ in purple. In dotted lines, the same ratios which would be obtained if the difference of total sink mass in those simulations would only be due to the fraction of mass ejected by protostellar jets.}
\label{figure: ratio of mass all vs time}
\end{figure}

 In Fig.~\ref{figure: ratio of mass all vs time}, the orange dashdotted curve shows the ratio of the total sink mass in the simulation with protostellar jets over the total sink mass in the simulation without feedback : $M_\text{sink,\sj{}} / M_\text{sink,\snj{}}$. This curve exhibits a plateau for times larger than \SI{2.6}{Myr}, which corresponds to the linear regime of the orange curves in Fig. \ref{figure: total sink mass all vs time}. The value of this plateau lies around 0.4, corresponding to the ratio $\frac{\SI{8e-4}{\Msun.yr^{-1}}}{\SI{2e-3}{\Msun.yr^{-1}}}$ of the two SFR. The orange dotted curve shows what this ratio would be if the difference in mass would only be due to the fraction of mass ejected in protostellar jets. This difference is easy to compute as the sub-grid model used for the protostellar jets specifies that each sink particle whose mass is larger than \SI{0.15}{\Msun} expels one third of the mass it accretes. In the limit of large times, the majority of the sink particles have a mass greatly larger than the mass threshold of \SI{0.15}{\Msun}, thus the mass expelled by each sink tends toward the limit of $\frac{1}{3}$ of its accreted mass, leading to a ratio of $\frac{2}{3}$. We see that the ratio in the simulation is significantly lower than the one given by considering only ejected mass. This shows that an important effect of the protostellar jets is to diminish the accretion onto the sink particles. Thus the difference in mass is due to both the loss of the mass directly expelled by the jets and the reduction of the accretion rate onto the sink particles due to the interaction between the jets and the environment.

\noindent

The brown and yellow solid lines in Fig. \ref{figure: total sink mass all vs time} represent the two simulations with protostellar jets only but with different parameters for the jets than \sj{}. The ``jets - wider'' uses wider jets while ``jets - faster'' uses faster jets. We see that the difference between these two curves and the orange solid one standing for \sj{} is small compared to the influence of including or not the different types of feedback. All the analysis presented in the rest of this article were conducted for our six simulations. As the ``jets - wider'' and ``jets - faster'' simulations show similar results than \sj{} for all the analysis, we will then ignore these two simulations for the remaining presentation.

\begin{figure*}[hbtp]
\includegraphics[width=0.5\textwidth]{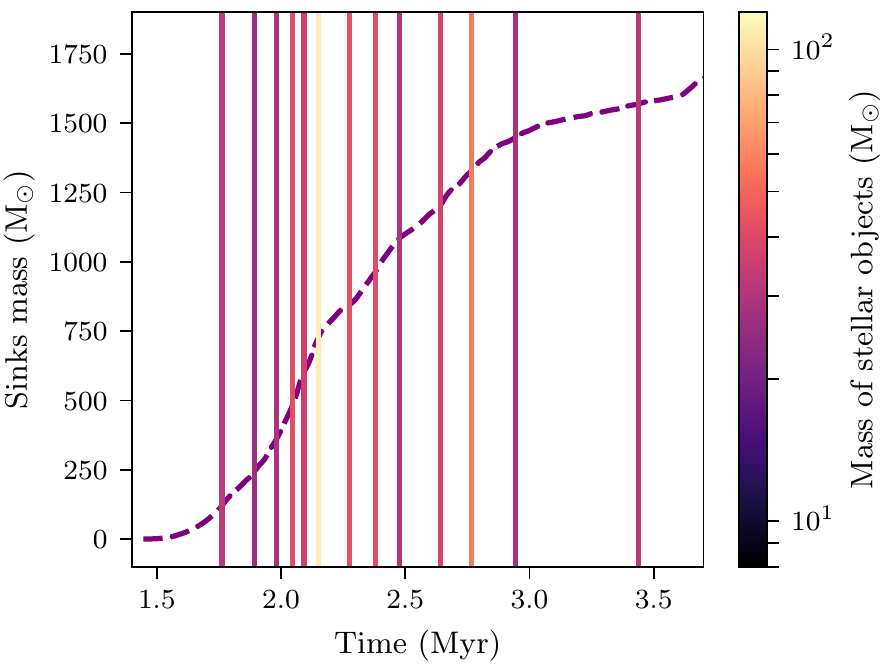}%
\includegraphics[width=0.5\textwidth]{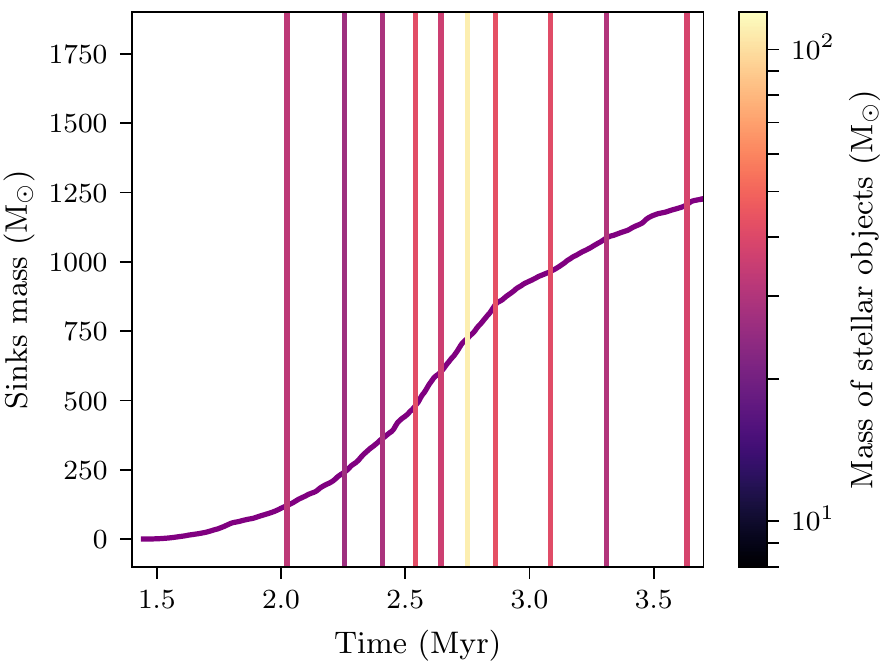}
\caption{\emph{Left:} total mass of the sink particles in the simulations including only HII regions. \emph{Right:} total mass of the sink particles in the simulations including HII regions and protostellar jets. The purple curves in the two panels are thus corresponding to the same ones in Fig. \ref{figure: total sink mass all vs time}. In the two panels, the vertical lines show the moments of creation of stellar objects, with the colours of the lines coding their mass.}
\label{figure: total sink mass for simu with HII regions}
\end{figure*}

Let's now compare the four simulations \snj{}, \sj{}, \srh{}, \sjrh{}. In  Fig. \ref{figure: total sink mass all vs time}, we see that for times lower than \SI{2.2}{Myr}, the simulations without feedback at all, and with only HII regions behave strictly the same. This reflects the late onset of HII regions. The same observation can be made comparing the simulations with protostellar jets only, and with both protostellar jets and HII regions for times lower than \SI{2.9}{Myr}, with an onset of HII regions a bit delayed due to a lower SFR in the simulations including protostellar jets. Once the HII regions begin to have an impact on its parent cloud, the behaviour of the sinks mass starts to be different as the expansion of the HII regions becomes more and more important with time and with the apparition of stellar objects. Thereby the SFR in the simulations including HII regions tends to be lower and lower with time, entirely stopping star formation within a few million years, while the SFR in the simulations without HII regions is roughly constant as the sinks mass is nearly linear with time. Figure \ref{figure: total sink mass for simu with HII regions} shows the sinks mass separately for the two simulations that include HII regions, with indicators for the moments of formation of stellar objects which give birth to HII regions. On these two panels, the mass sequence of the different stellar objects is the same as the seed chosen to initialise the random number generator is the same. This permits  close comparison, in particular  ensuring that the differences between the two simulations are not coming from the random draw on stellar objects mass \citep{geen2018}. We see that the first five stellar objects to form are not so massive, between 20 and \SI{40}{\Msun}. As the mass of those stellar objects remains limited, and since they are formed in a dense environment, the HII regions associated are not powerful enough to expand and have a significant impact on the cluster. This explains the strong similarity in the sinks mass versus time between the corresponding simulations without HII regions before \SI{2.2}{Myr} for the ones without jets, and before \SI{2.9}{Myr} for the ones with jets. This similarity is also observed qualitatively on the gas distribution. The sixth stellar object to form has a mass of \SI{102.6}{\Msun}. This rather massive stars is associated to a photon flux of about \SI{e50}{s^{-1}} and is able to form a HII region powerful enough to expand in this dense medium. In fact, we see in the two panels of Fig. \ref{figure: total sink mass for simu with HII regions} a change in behaviour just after the formation of this massive stellar objects (the vertical yellow line). Before its formation the SFR is increasing over time, and just after its formation the SFR starts decreasing over time. The stellar objects that form later have less impact than the \SI{102.6}{\Msun} one, but still contribute to the decrease in SFR.

In Fig. \ref{figure: ratio of mass all vs time}, the purple dashdotted curve shows the ratio of the total sink mass in the simulation with protostellar jets and HII regions over the total sink mass in the simulation including only HII regions: $M_\text{sink,\sjrh{}} / M_\text{sink,\srh{}}$. Between 1.6 and \SI{2.2}{Myr}, this curve is below the dotted purple curve, which represents what this ratio would be if the difference in sinks mass would only be due to the fraction of mass ejected in protostellar jets. As stated previously for the simulations without HII regions, this shows that the difference in mass is due to both the loss of the mass directly expelled by the jets and the reduction of the accretion rate onto the sink particles due to the interaction between the jets and the environment. At \SI{2.2}{Myr}, the ratio begins to increase over time. This is due to the appearance of the first HII regions which does not occur at the same time for the two simulations. In fact, between 2.2 and \SI{2.9}{Myr}, in the simulation including HII regions only, the first HII regions began to appear, while in the simulation which includes protostellar jets and HII regions, the HII regions have not expanded yet. Here we are then comparing simulations which have different behaviour in this time interval, and thus this increase in the ratio is not relevant.

\noindent

In Appendix \ref{annexe: evolution of the number of sinks} we present the evolution of the number of sinks in time, for different masses of sinks. Figure \ref{figure: number of sinks vs time} shows that for simulations with HII regions, the onset of HII regions stop most of the accretion onto the stars, while new small stars continue to form insensitively to the expansion of HII regions.

%% file: Results_gas.tex
\section{Gas density PDF and kinematic}
\label{section: article2-gas}
In this section we investigate the gas properties 
focusing on the density PDF and the Mach number 
density relation within the whole computational box. 
We complement the analysis by
studying the velocity 
dispersion of the dense structures.

\subsection{Density PDF}

\noindent
\begin{figure*}[hbtp]
\begin{overpic}[width=0.5\textwidth]{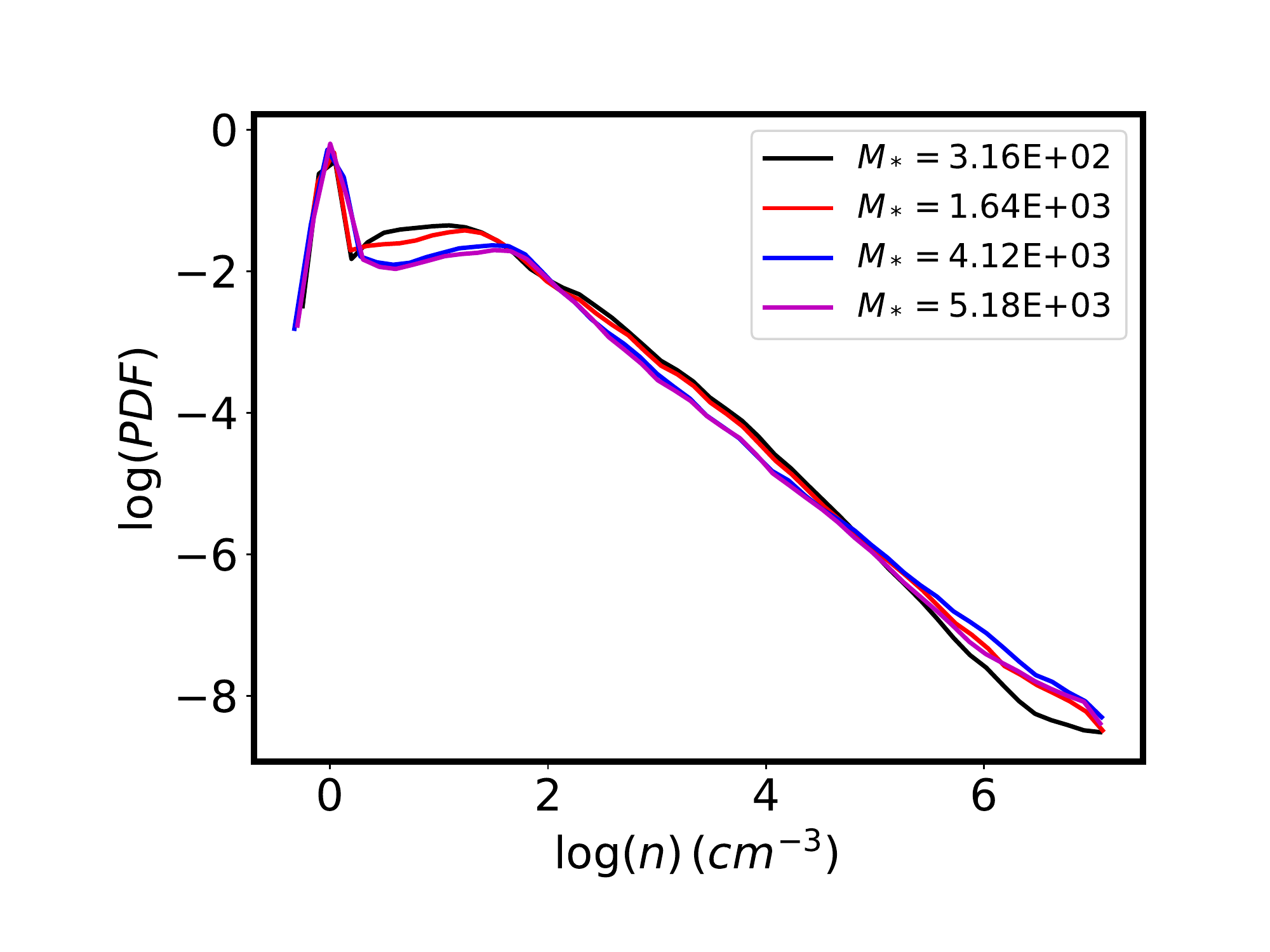} \put (50,60) {\textcolor{black}{\snj{}}} \end{overpic}%
\begin{overpic}[width=0.5\textwidth]{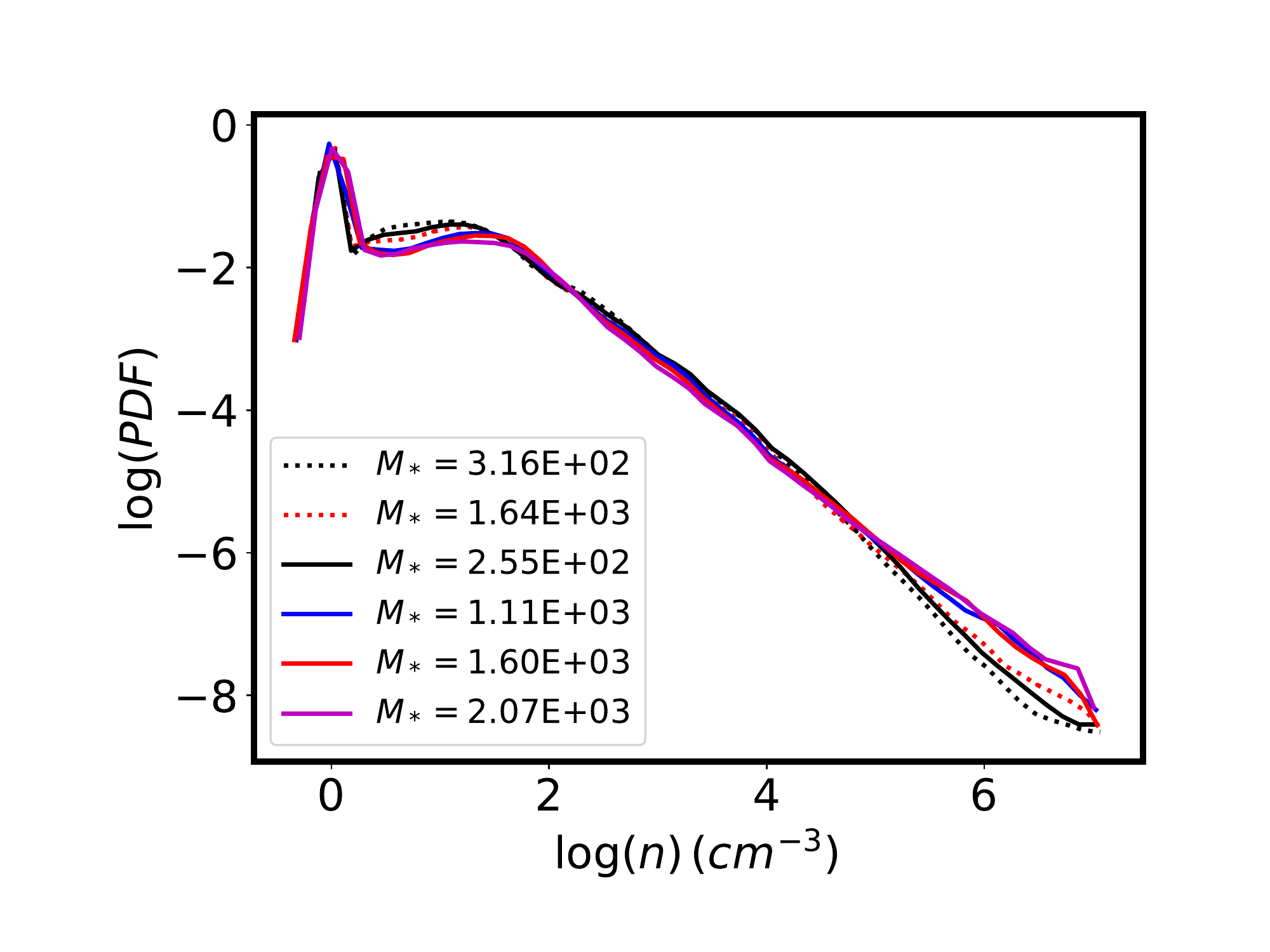} \put (50,60) {\textcolor{black}{\sj{}}} \end{overpic}
\begin{overpic}[width=0.5\textwidth]{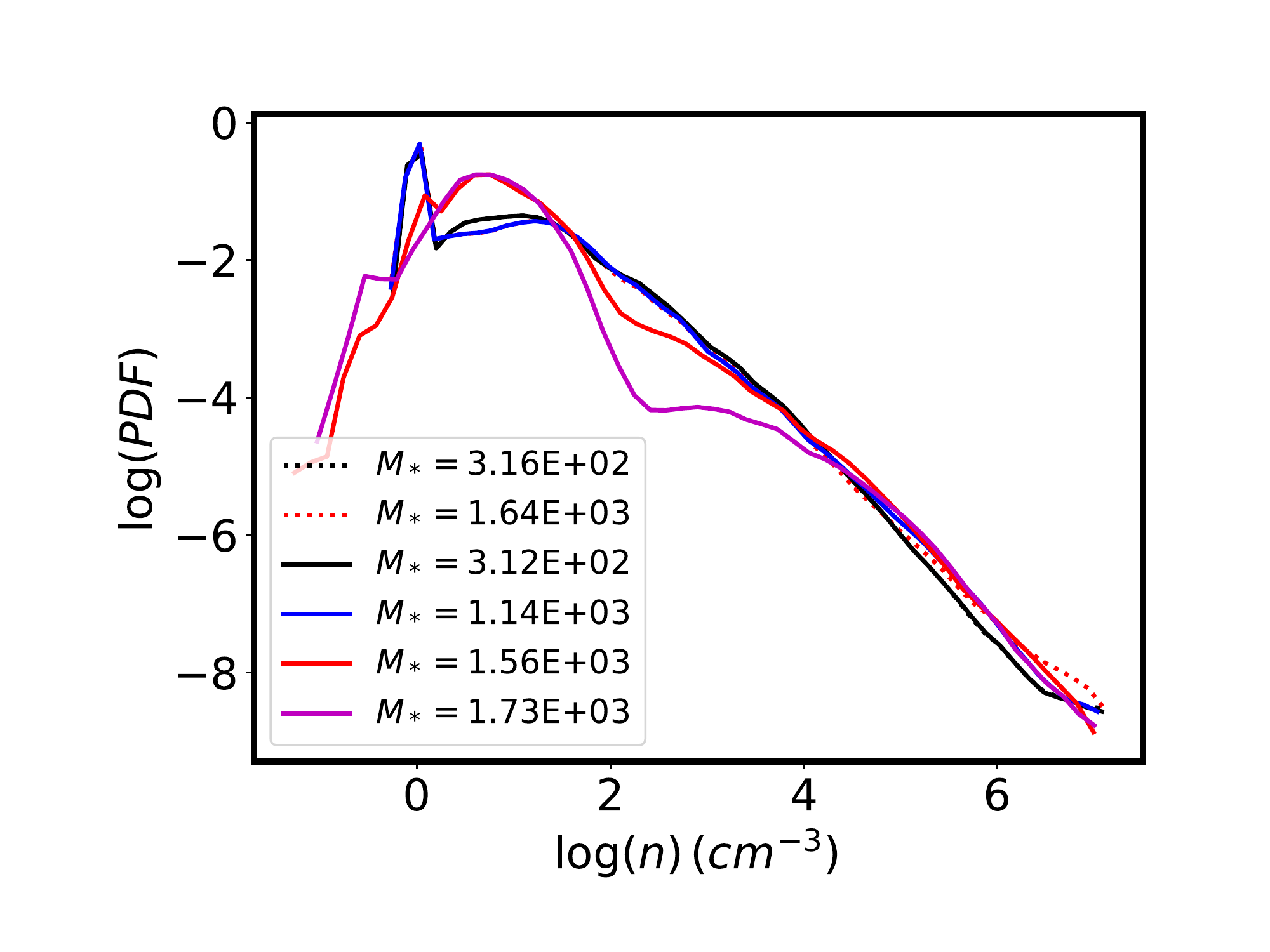} \put (50,60) {\textcolor{black}{\srh{}}} \end{overpic}%
\begin{overpic}[width=0.5\textwidth]{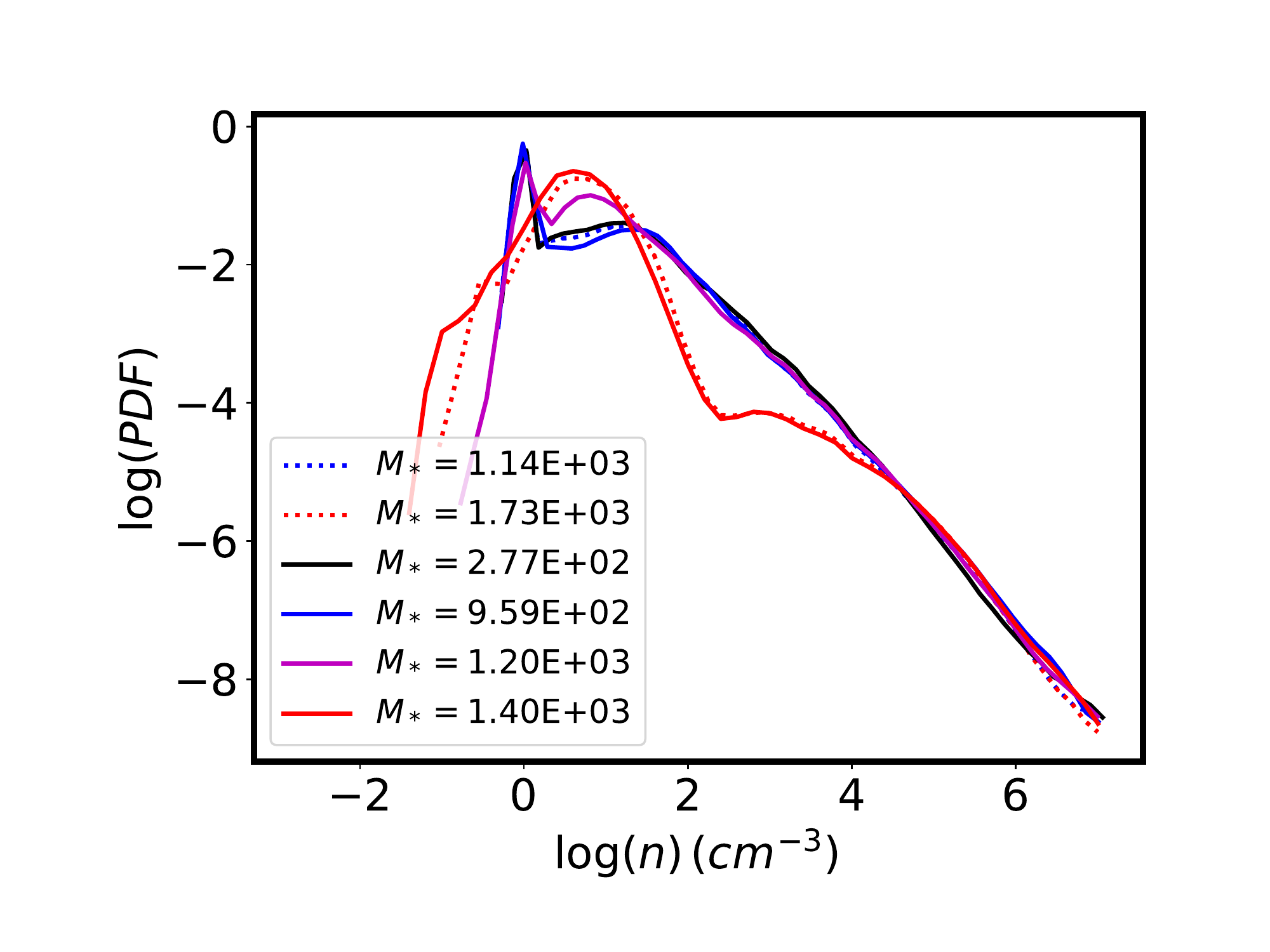} \put (50,60) {\textcolor{black}{\sjrh{}}} \end{overpic}
\caption{Density PDF for the runs 
\snj{} (top-left), \sj{} (top-right), \srh{} (bottom-left) and \sjrh{} (bottom-right) for four snapshots (solid lines).
The dotted lines visible in top-right and 
bottom left represent two snapshots of run
\snj, while the ones on bottom-right panel are from 
run \srh.
The total accreted mass is indicated as it is more
representative to make comparison between the runs.
}
\label{figure: density_pdf}
\end{figure*}

The density PDF is a fundamental quantity to investigate 
not only because it reflects the dynamical state of 
the gas but also because it is believed to 
play a fundamental role regarding the mass spectrum 
of stars \citep{padoan1997,HC08,leeh2018a}.

Figure~\ref{figure: density_pdf} displays 
the density PDF for the four runs 
\snj{}, \sj{}, \srh{} and \sjrh{} for various 
snapshots. To make comparison easier, 
we have reported in the top-right panel 
(\sj) and
bottom-left one (\srh), two snapshots of run
\snj{} (dotted lines) using the color code to 
indicate what timesteps should be most comparable. In the bottom-right one (\sjrh), two snapshots of run \srh{} (dotted lines) have been reported.

The top-left panel portrays the density PDF of run \snj.
It is a clear powerlaw of index $\simeq -1.5$ typical 
of gravitational collapse arising in turbulent flows \citep{Kritsuk11,leeh2018a}. The powerlaw is remarkably
stable and does not evolve significantly over a period
that goes from the cluster formation time up to the 
end of the simulation where about half of the gas has been accreted. 

The top-right panel reveals that the jets have no significant 
influence on the density PDF, while the bottom-left panel
shows that ionising radiation has a drastic role once
massive enough stellar objects have formed. In 
particular, the quantity of  gas at densities between \SI{e2}{} and \SI{e3}{cm^{-3}}  is diminishing by one to two orders 
of magnitude compared to the case without feedback. 
On the contrary the quantity of gas at densities around 
\SI{10}{cm^{-3}} increases by almost one order of magnitude, 
which is a clear signature of the strong photo-ionisation 
that is induced by the massive stars. 
The bottom-right panel shows that even in the 
presence of ionising radiation feedback, the jets 
have very little influence on the density PDF. 
We note from both bottom panels that the high density 
gas is not strongly affected by the ionising radiation 
which is due to the recombination of electron and 
proton being very efficient at high densities.

\subsection{Mach number density relation}

\noindent
\begin{figure*}[hbtp]
\begin{overpic}[width=0.5\textwidth]{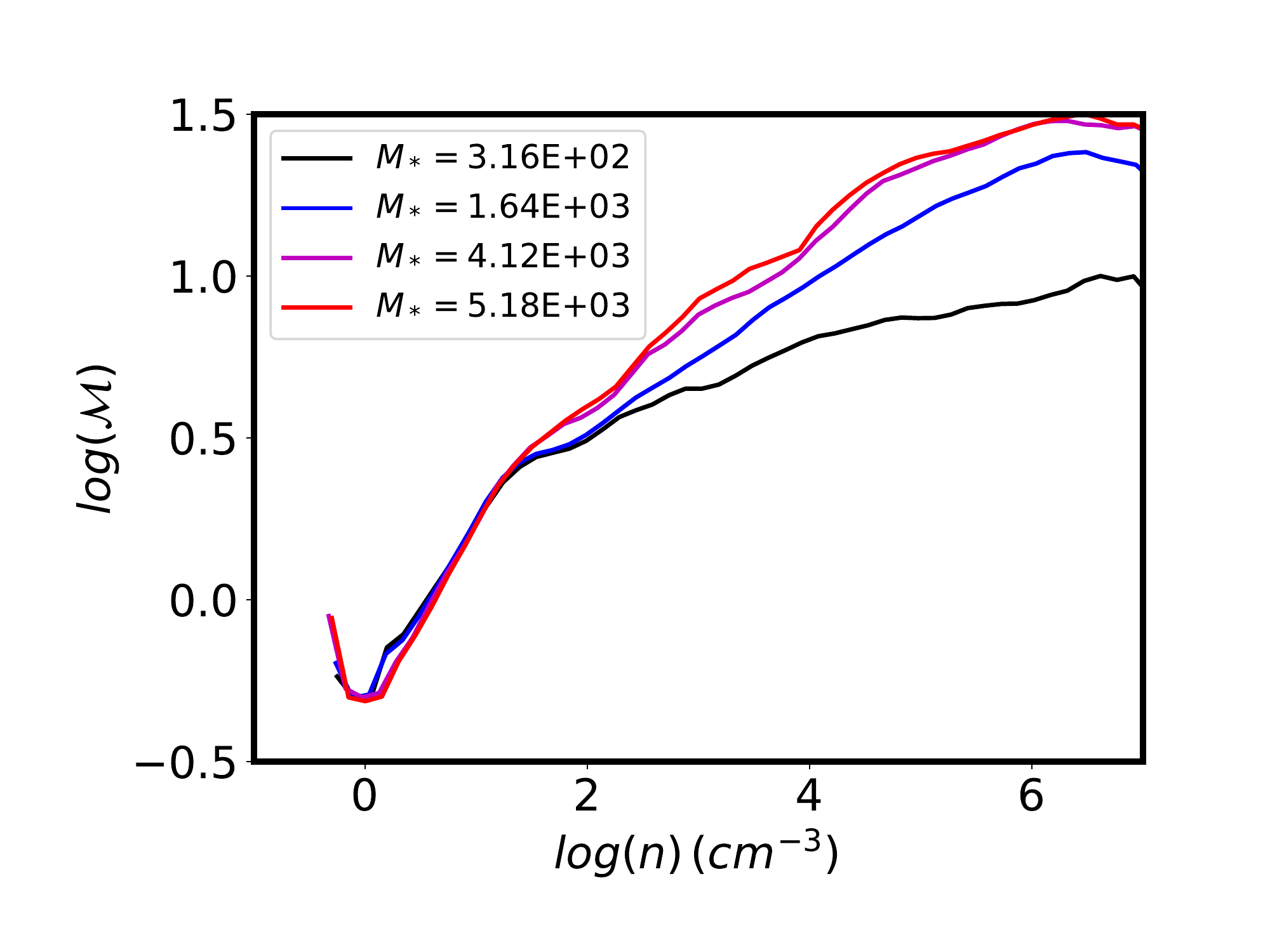} \put (40,18) {\textcolor{black}{\snj{}}} \end{overpic}%
\begin{overpic}[width=0.5\textwidth]{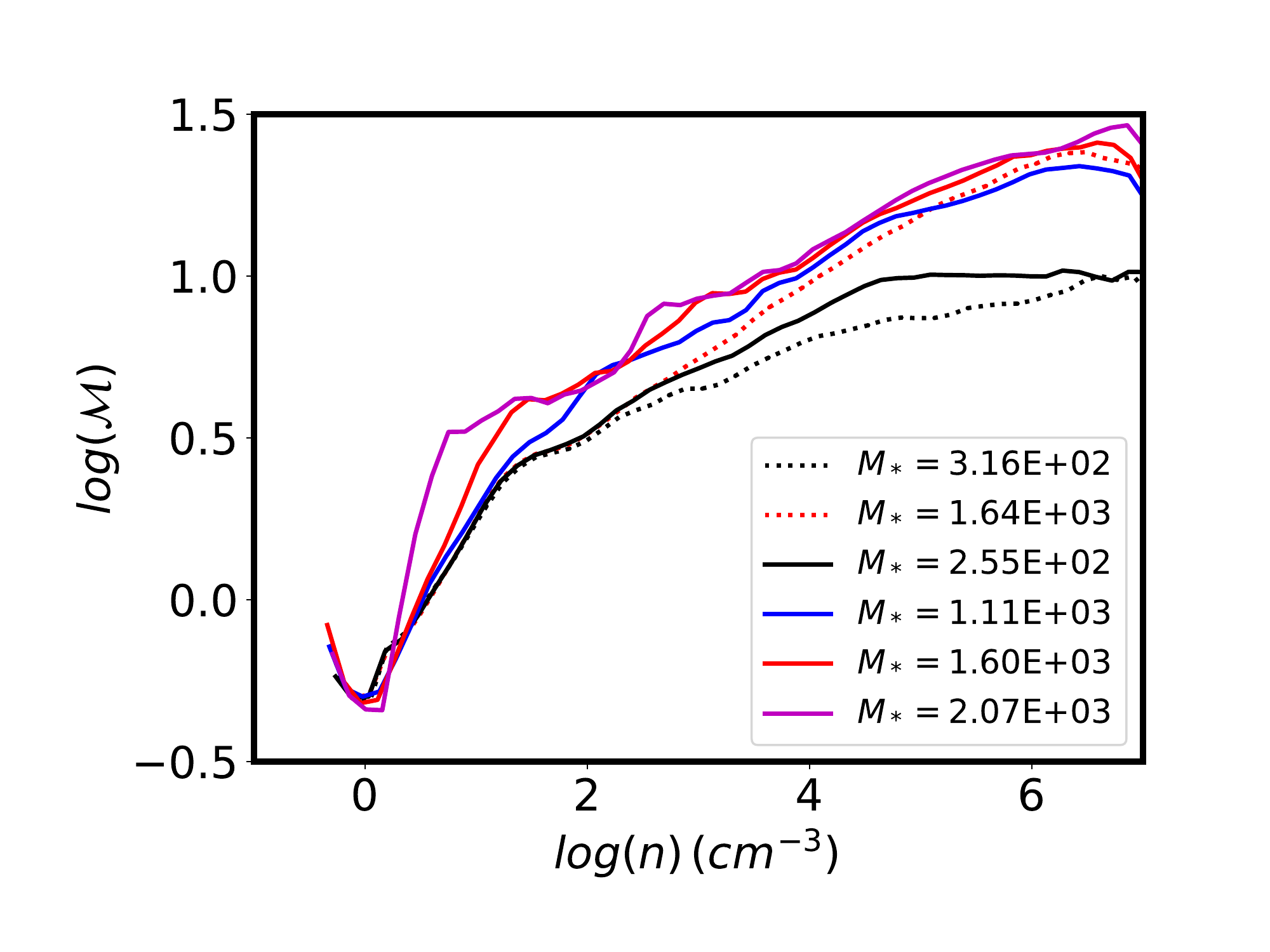} \put (40,18) {\textcolor{black}{\sj{}}} \end{overpic}
\begin{overpic}[width=0.5\textwidth]{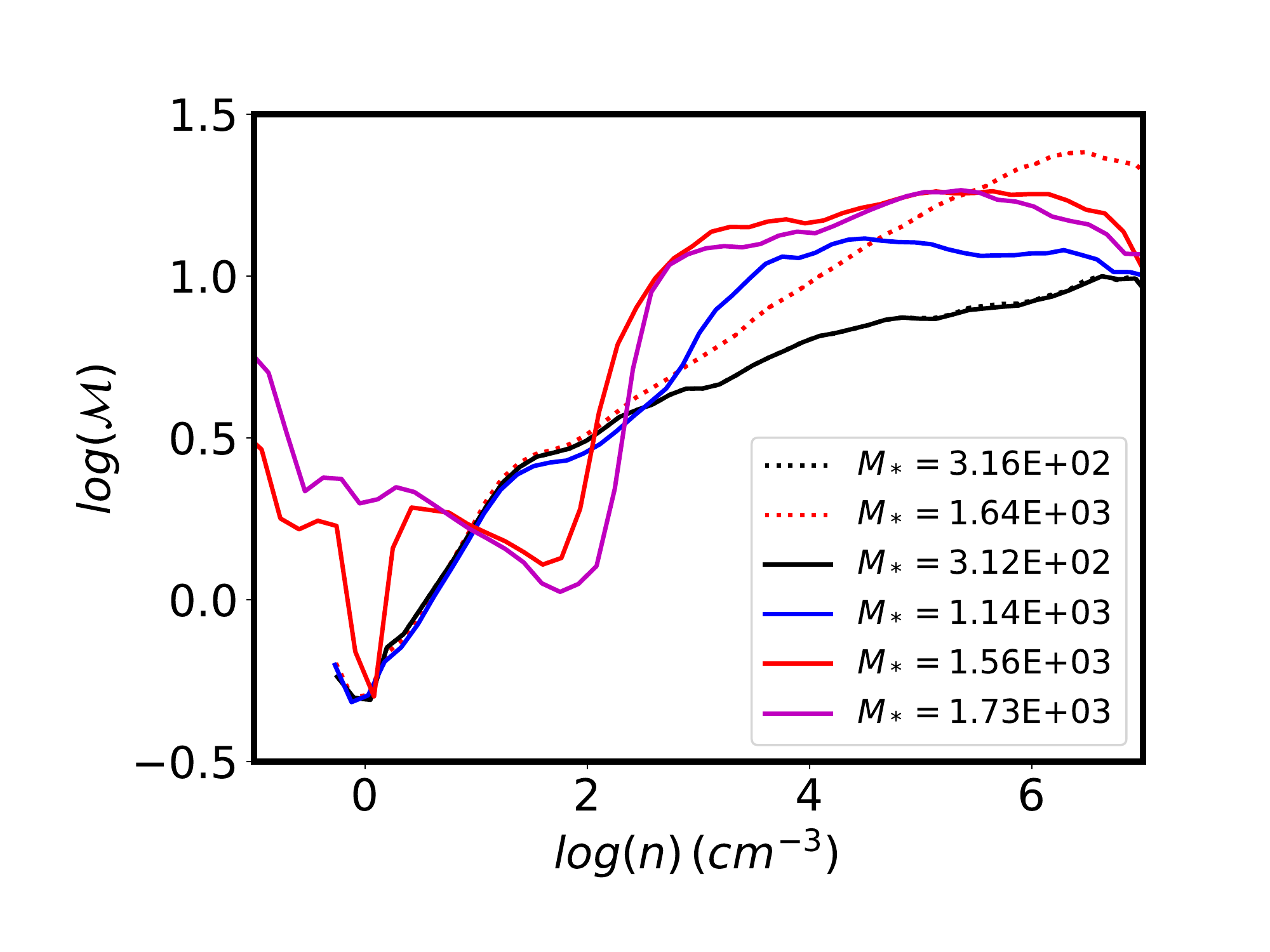} \put (40,18) {\textcolor{black}{\srh{}}} \end{overpic}%
\begin{overpic}[width=0.5\textwidth]{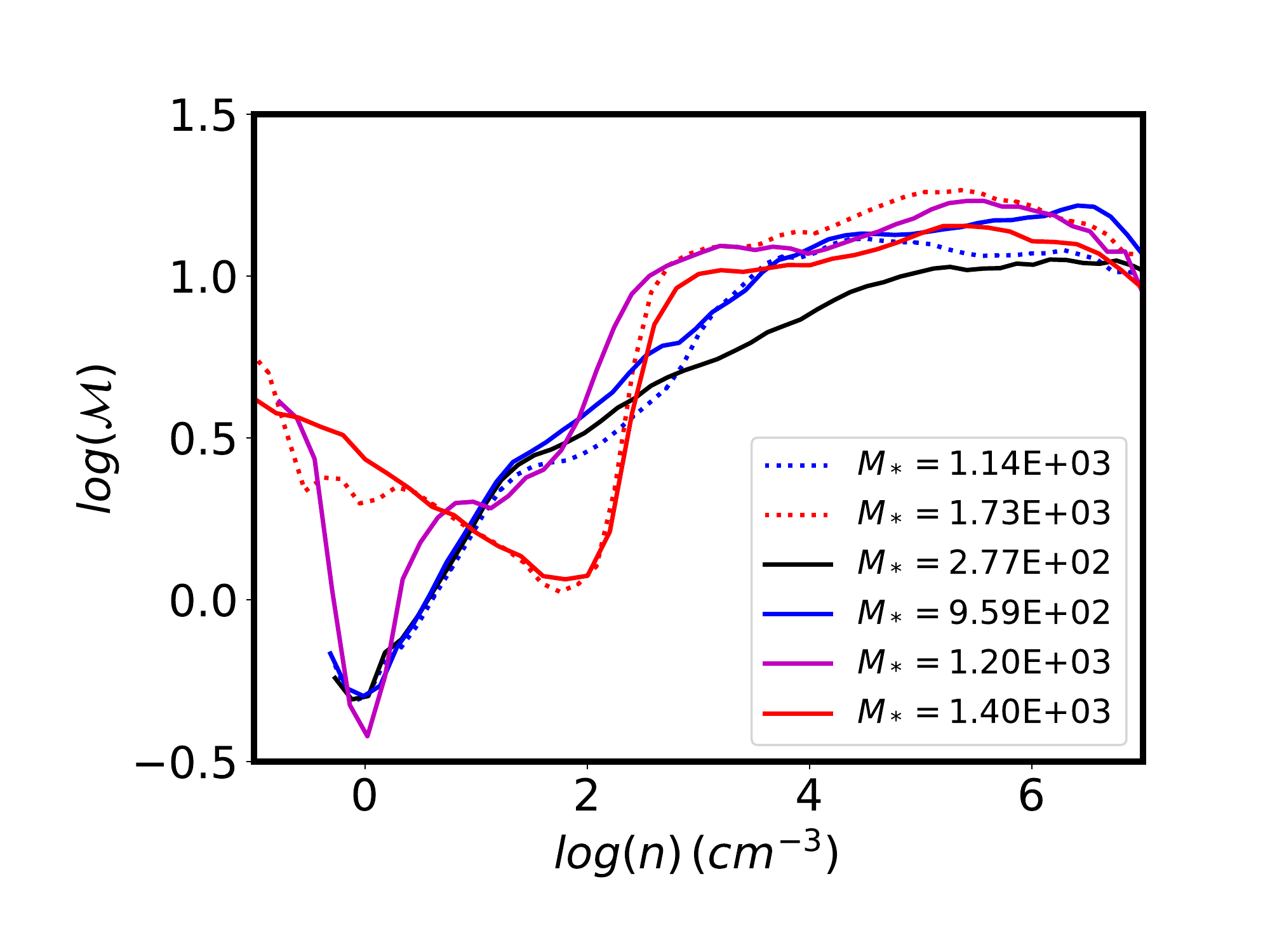} \put (40,18) {\textcolor{black}{\sjrh{}}} \end{overpic}
\caption{Same as Fig.~\ref{figure: density_pdf}
for the Mach-density relation. }
\label{figure: density_mach}
\end{figure*}

A fundamental question with the feedback processes is 
how they exactly operate on the surrounding gas. 
For instance, do the jets increase the turbulence 
and by how much are crucial issues. Various teams \citep[e.g.][]{carroll2009,wang2010,Federrath2014,offner2017,murray2018} have all concluded that jets 
do trigger some turbulence although the importance 
of the effect vary between studies. This may be a
consequence of the various setups that have been 
employed. Figure~\ref{figure: density_mach}
displays the mean Mach number as a function 
of density for the four runs using the same conventions 
and snapshots than for  Fig.~\ref{figure: density_pdf}.

The top-left panel shows that in the absence of feedback, the Mach number increases with density from about 1 to 30. 
It also increases with time for densities larger than 
$\sim \SI{e2}{cm^{-3}}$. This is obviously a consequence 
of gravity which produces both infall and virial 
motions of the order of $\sqrt{GM/R}$, where $M$ is the 
accreted mass in the cluster and $R$ the typical distance from the cluster center. 

The top-right panel reveals that the presence of jets 
increases the mean Mach number by tens of percents, 
mainly for intermediate densities of about \SI{e3}{cm^{-3}}, as visible 
when more than \SI{e3}{\Msun} have been accreted. 
We therefore conclude that in the present configuration 
in which the turbulence is fed by accretion and 
 in which the jet directions are not widely 
dispersed (see Sect.~\ref{section-sub: Alignment of the stars in the cluster}), turbulence is not strongly 
enhanced by the presence of jets. 
It is therefore likely that they mainly act by repulsing 
some of the gas that would have otherwise been accreted 
as indeed envisioned by \citet{matzner2000}. 

The importance of ionising radiation 
is visible from bottom panels. 
Clearly this feedback substancially 
modifies the mean Mach number at almost 
all densities. At about 
\SI{e3}{cm^{-3}} it is on the order of 2 which implies 
a kinematic energy roughly four times larger and therefore 
influences much the gas behaviour. Note that the drop 
seen at \SI{e2}{cm^{-3}} is due to the increase of 
the gas temperature. Let us stress that while the ionising radiation has little effect on gas temperature 
for densities above 10$^3$ cm$^{-3}$ and below $\simeq$5 cm$^{-3}$, it can significantly heat, typically up to $T=$5000-8000 K, gas of intermediate densities.
The presence of jets (bottom-right panel) further 
increases the Mach number for gas of densities \SI{e3}{cm^{-3}} (compare blue dotted line and purple solid one), which is consistent with the final SFE being 
a bit lower when both jets and ionising radiation are included.

\subsection{Dense structures internal velocity dispersion}

\noindent
\begin{figure*}[hbtp]
\begin{overpic}[width=0.5\textwidth]{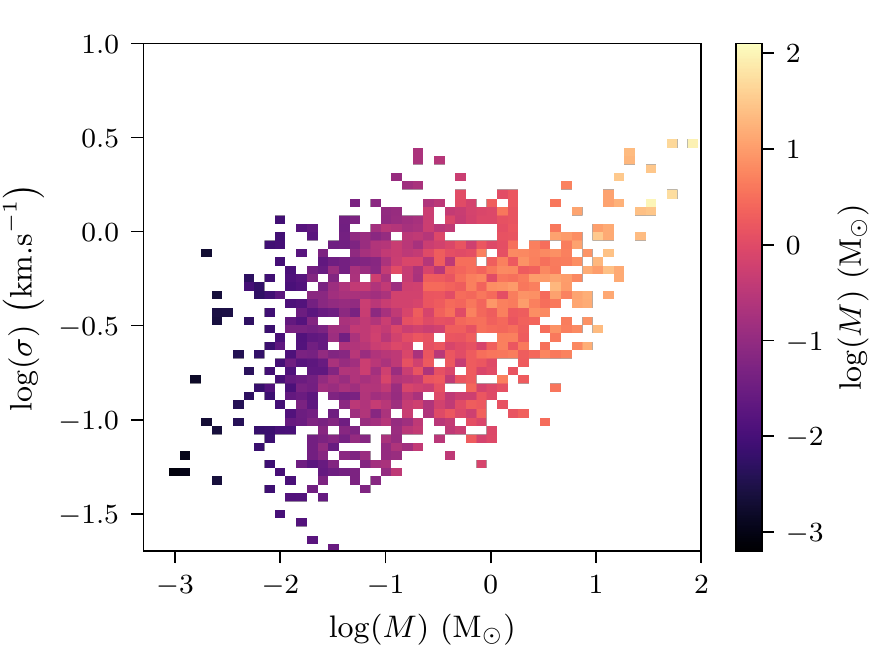} \put (58,18) {\textcolor{black}{\snj{}}} \end{overpic}%
\begin{overpic}[width=0.5\textwidth]{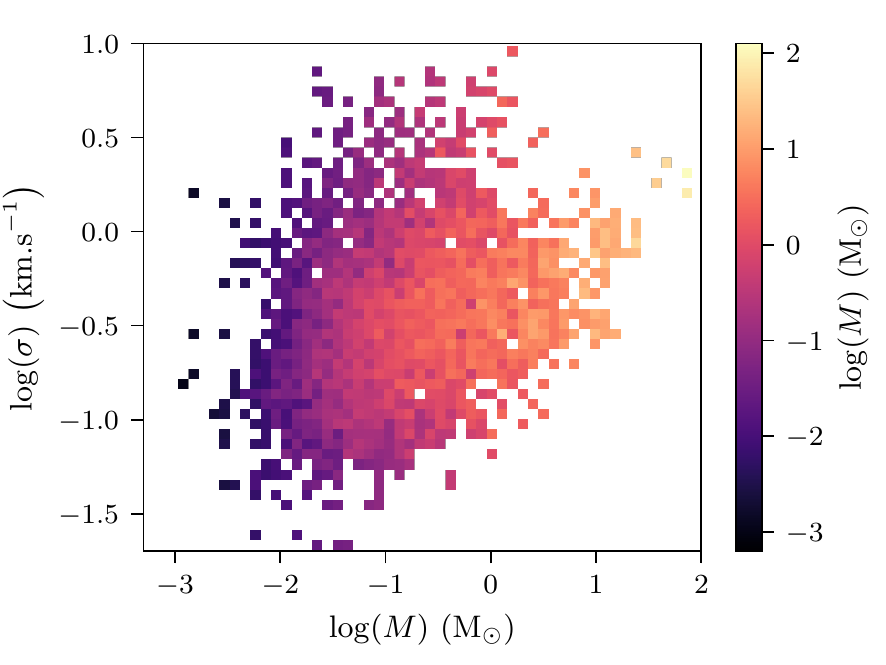} \put (58,18) {\textcolor{black}{\sj{}}} \end{overpic}
\begin{overpic}[width=0.5\textwidth]{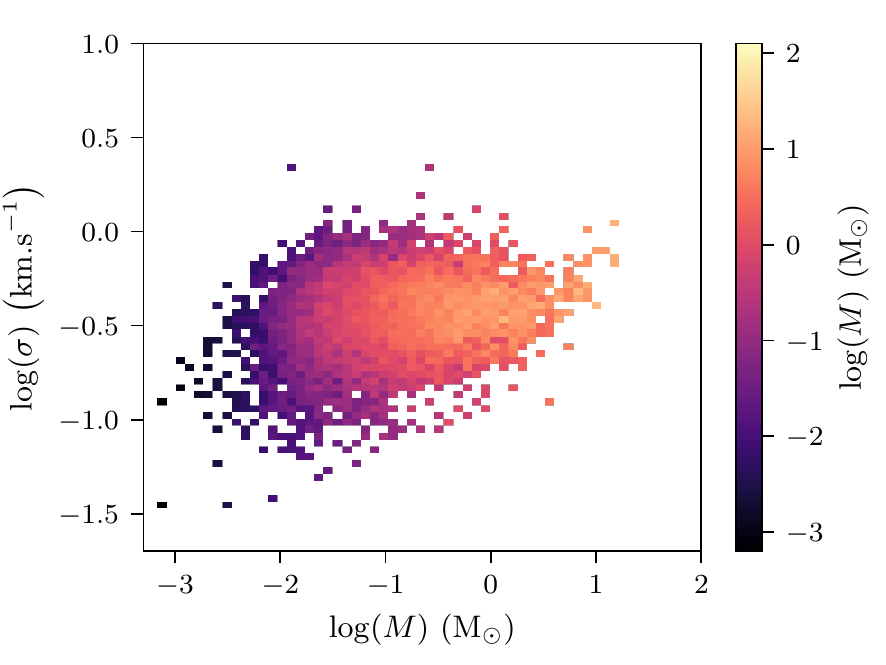} \put (58,18) {\textcolor{black}{\srh{}}} \end{overpic}%
\begin{overpic}[width=0.5\textwidth]{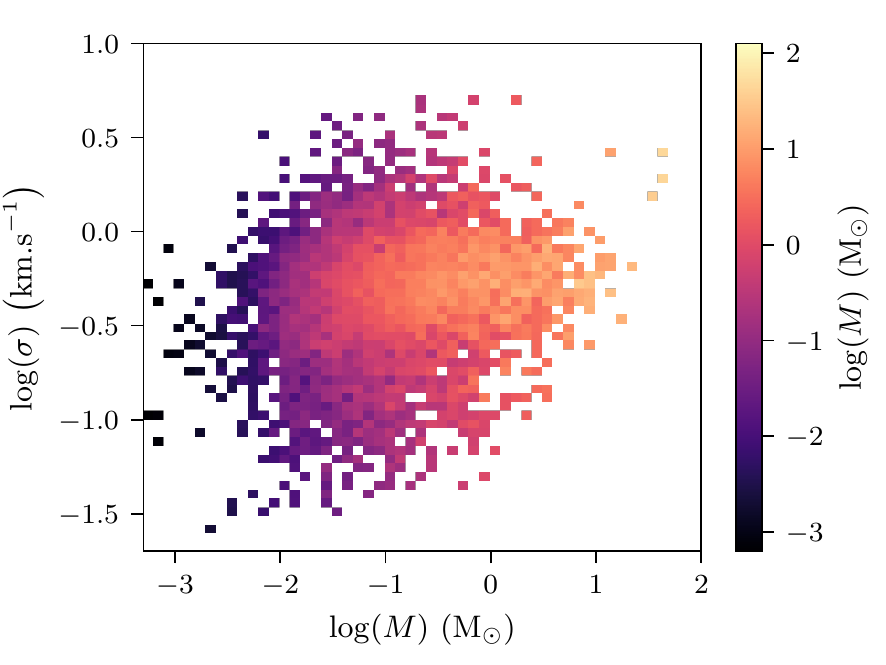} \put (58,18) {\textcolor{black}{\sjrh{}}} \end{overpic}
\caption{Bidimensional histograms showing the
velocity dispersion of dense structures for the four runs
\snj{} (top-left), \sj{} (top-right), \srh{} (bottom-left) and \sjrh{} (bottom-right) at a time of about \SI{3.2}{Myr}.}
\label{figure: vel_disp}
\end{figure*}

Finally we complement the analysis presented in the previous section, which is a global estimate on the whole simulations, by investigating the local 
velocity dispersion of dense structures. The latter 
have been identified by running the HOP 
algorithm \citep{eh1998} on gas cells 
having densities larger than \SI{e4}{cm^{-3}}.
While the mean Mach number analysis captures
the large scale motions, which dominate
in turbulent flows, the velocity dispersion of dense structures reveals the local motions in star forming clumps. 

Figure~\ref{figure: vel_disp} portrays the 
bidimensional histograms of the velocity dispersion of these structures at a time of about \SI{3.2}{Myr}. 
The top-left panel is typical of 
previous analysis reported in 
simulations without feedback \citep{frigg2018,ntormousi2019}. 
Interestingly, top-right panel shows that 
in the presence of jets, the histogram 
presents a significant population of dense structures
of gaseous masses between 0.1 and \SI{1}{\Msun} that 
have a velocity dispersion up to five times larger than 
the typical velocity dispersion $\left( \sim \SI{0.3}{km.s^{-1}} \right)$
found in the absence of jets. This indeed confirms that 
jets can locally perturb significantly the 
dense structures in which they develop. We note however that the bulk  of the dense structures, that contains 
most of the mass, remains very similar to the one
obtained in run \snj. 

The bottom panels show that ionising radiation has 
a moderate impact on the  velocity dispersion of 
dense structures. The \srh{} run presents velocity dispersion
slightly larger $\left( \text{say } \sim 50 \% \right)$ than 
in run \snj{}, while a similar, though even less pronounced, trend is observed between 
runs \sj{} and \sjrh{}. 

We conclude that altogether stellar feedback certainly
alters the velocity dispersion and motions 
in star forming clumps by increasing it by a factor of a few \citep{goldbaum2011}. 
However the nature 
of these motions cannot be simply described 
by a mere increase of turbulence. The net effect 
of the jets remains modest but they can 
have local very significant impact and this is probably why they reduce the SFR by a factor of roughly 2.
On the contrary the ionising radiation 
appears to have even less impact on the dense 
gas but on the other hand 
they lead to global expansions --- as visually obvious from 
Fig.~\ref{figure: global appearance of simulations} --- that considerably 
decreases the SFE.

%% file: Results_bound_and_rotation.tex
\section{The stellar/sink properties}
\label{section: article2-sinks properties}
We now turn to the description of the stellar properties. We will then consider the sink particles as individual stars.

\subsection{Bound and unbound stars}
\label{section-sub: Bound and unbound stars}
We saw in Fig. \ref{figure: global appearance of simulations - small scale} that a lot of stars are forming around the central, dense cluster of stars in the simulations that include HII regions. In this part we try to distinguish between the stars bound to the cluster, and the ones that are destined to leave it.

We first compute bi-dimensional histograms of the sink positions and we took the bin of highest concentration as the cluster's center. By doing it before and after the output of interest, we are also able to estimate the velocity of the cluster center. We then construct concentric spherical shells around the center. For each shell, we compute the escape velocity $V_\text{esc}$ at the shell radius $R$ 
as well as the mass, $M_\text{int}$, obtained by  
summing  all the mass of sinks and gas at radii lower than $R$:

\begin{equation}
    V_\text{esc} \left( R \right) = \sqrt{\frac{2GM_\text{int}}{R}}.
\end{equation}

\begin{figure*}[hbtp]
\begin{overpic}[width=0.5\textwidth]{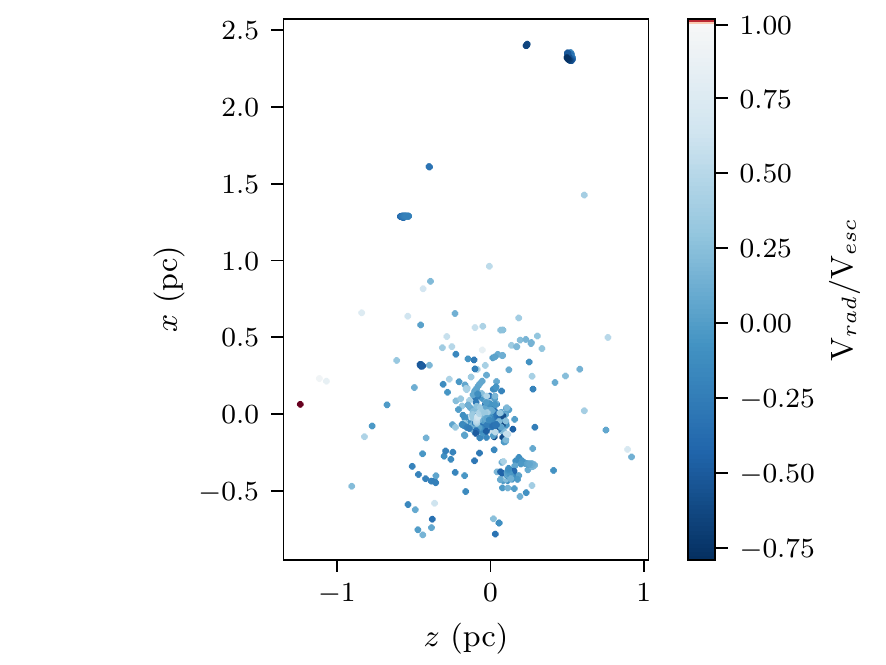} \put (35,67) {\textcolor{black}{\snj{}}} \end{overpic}%
\begin{overpic}[width=0.5\textwidth]{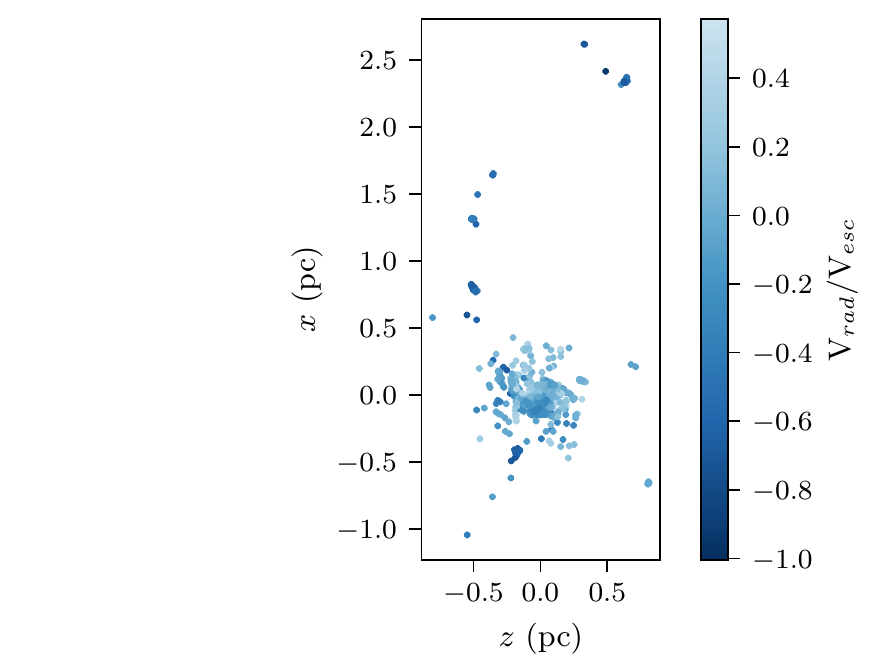} \put (50,67) {\textcolor{black}{\sj{}}} \end{overpic}
\begin{overpic}[width=0.5\textwidth]{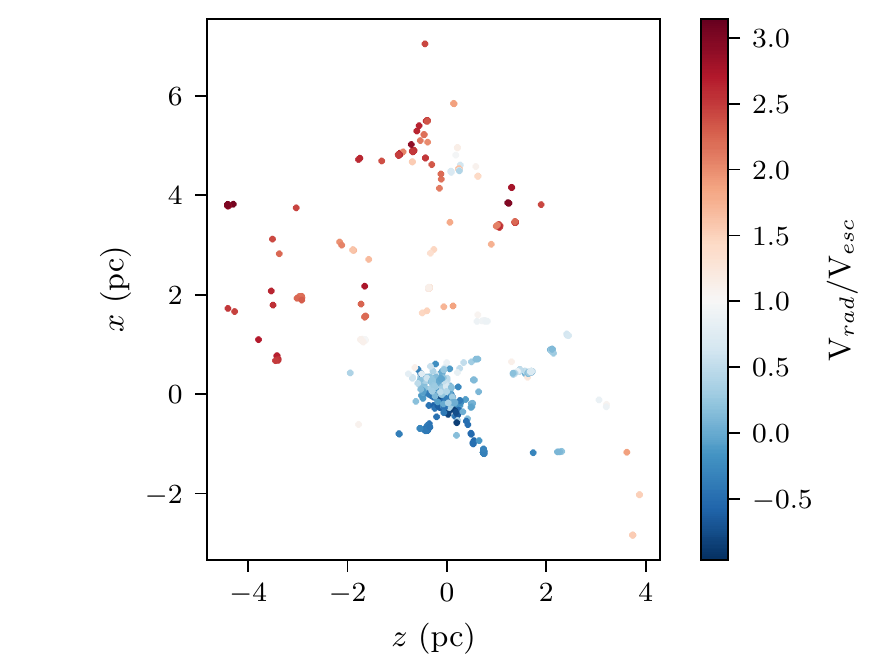} \put (26,67) {\textcolor{black}{\srh{}}} \end{overpic}%
\begin{overpic}[width=0.5\textwidth]{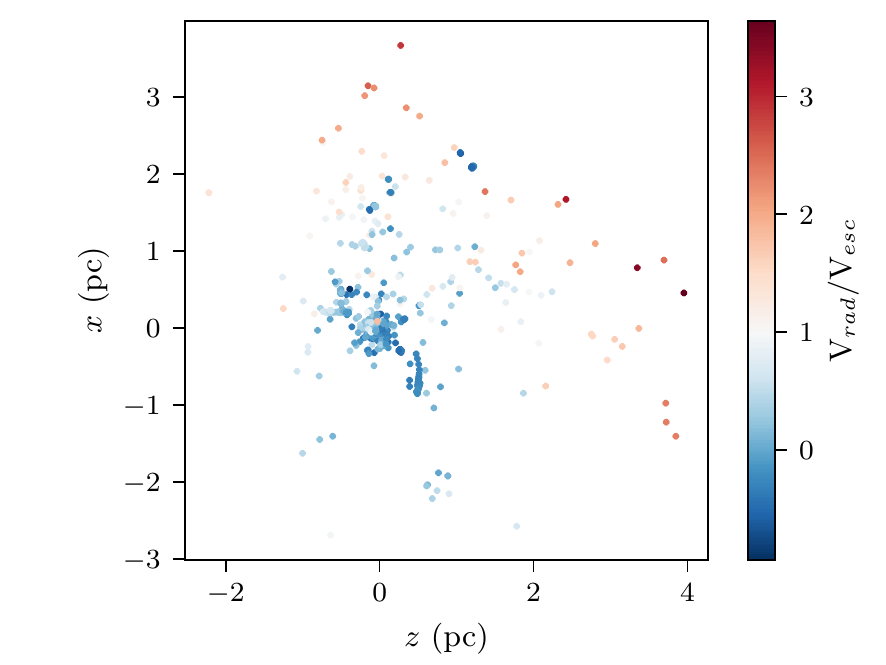} \put (24,67) {\textcolor{black}{\sjrh{}}} \end{overpic}
\caption{Distribution of stars in the four simulations at \SI{3.5}{Myr}. The colors code the ratio of the radial outward velocity of the sink particles over the escape velocity at their respective distance from the cluster's center. Particles that appear in blue are likely to be bound, whereas the ones that appear in red are likely to be unbound.}
\label{figure: bound stars}
\end{figure*}

\noindent
For each sink particle in the shell, we compute its projected velocity along the unit vector from the cluster's center to the sink particle. This velocity is noted $V_\text{rad}$ and is positive if the sink particle is going away from the center and negative if going toward the center. In Fig. \ref{figure: bound stars} we plotted all the sink particles in the four simulations with a color code indicating the ratio $\frac{V_\text{rad}}{V_\text{esc}}$. The sink particles in blue exhibit a ratio lower than 1, indicating that they should be bound at the time of interest. The ones in red exhibit on the contrary a ratio higher than 1. As they are moving away from the center at a speed higher than the escape velocity, they are destined to disperse and to leave the cluster, and we can consider these stars as unbound. We see that in the two upper panels representing simulations without HII regions (\snj{} and \sj{}), all the stars in the cluster are bound. For the two bottom panels, representing simulations with HII regions (\srh{} and \sjrh{}), the cluster consists of a bound core, surrounded by stars that are likely to disperse over time. The existence of unboud stars is likely to be due to the star formation in the material compressed by the expansion of HII regions. The escape velocity $V_\text{esc}$ is roughly constant over $R$, with a value about \SI{2.4}{km.s^{-1}}. The expansion of HII regions takes place at the sound speed in the ionized gas \citep{Geen2015}, which is about ten kilometres per second. A maximum value of about 4 for the ratio $\frac{V_\text{rad}}{V_\text{esc}}$ is coherent for stars that would have been formed in the expanding material of the HII regions.

\noindent

\subsection{Bound and unbound mass}
\label{section-sub: Bound and unbound mass}
\begin{figure}[hbtp] 
\includegraphics[width=\linewidth]{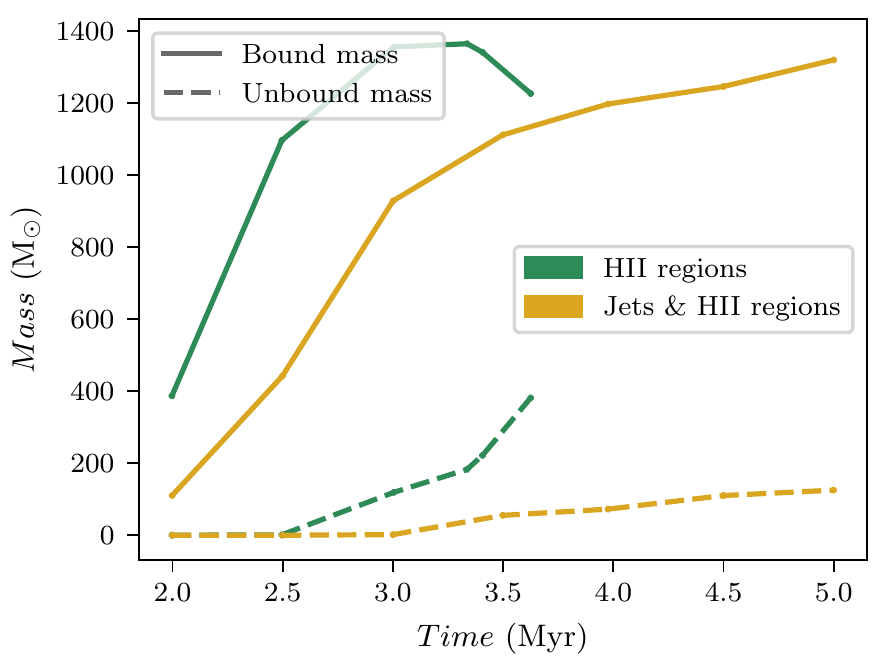}
\caption{Evolution of the total mass of bound and unbound stars. The mass of unbound stars is significantly higher in the simulation \srh{}.}
\label{figure: Bound_unbound_mass}
\end{figure}

We saw that in the two simulations that does not include HII regions, all the sinks are bound. This remains true for the entire temporal evolution. For the two simulations that include HII regions, namely \srh{} and \sjrh{}, they both exhibit unbound stars. Figure~\ref{figure: Bound_unbound_mass} shows the total amount of mass in bound and unbound stars in those two simulations. We see that in \srh{} the mass of unbound stars quickly grows to reach about a quarter of the total star mass. In \sjrh{}, the mass of unbound stars is three times lower, for the same mass of bound stars. As the evolution of the cluster in this simulation is slower, this indicates that the unbound stars form in the late stage, as HII regions are expanding in the outskirt of the cluster, triggering formation of unbound stars in these regions.

\noindent

\subsection{Rotation of the cluster}
\label{section-sub: Rotation of the cluster}

In this section we focus on the rotation of the bound part of the cluster. As in the previous section, we  identify the cluster's center as the place of highest concentration of stellar mass. We define the rotation axis of the cluster as the mean direction of the angular momentum of the bound sink particles. We define the equatorial plane as the plane perpendicular to the rotation axis and containing the cluster's center. We then construct concentric cylindrical shells around the cluster center, aligned to the rotation axis. For each sink particle we compute the azimuthal velocity. We then divide the mean of the azimuthal velocities of the sink particles by the radius of the considered shell to obtain the mean angular velocity.

\begin{figure*}[hbtp] 
\begin{overpic}[width=0.5\textwidth]{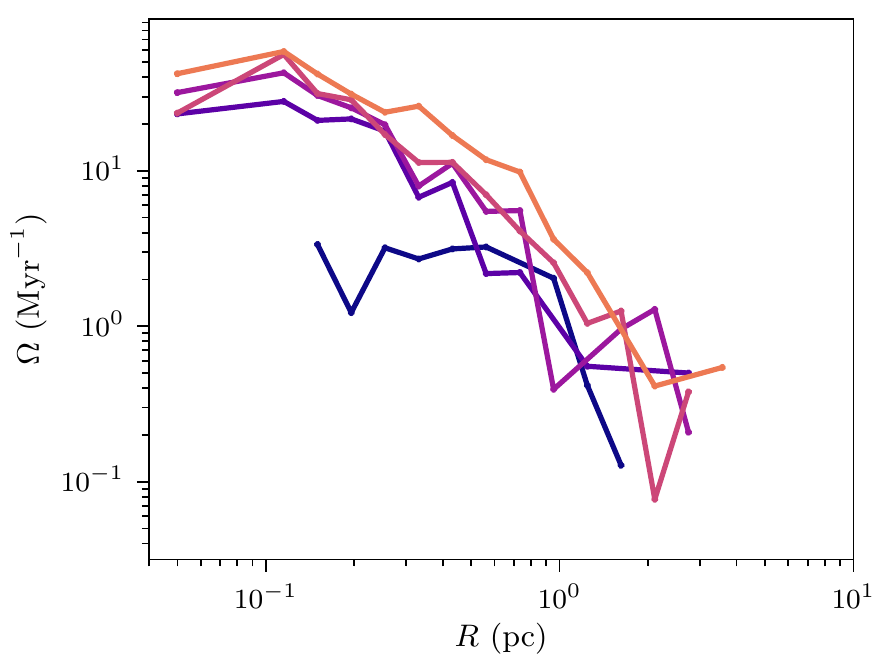} \put (85,65) {\textcolor{black}{\snj{}}} \end{overpic}%
\begin{overpic}[width=0.5\textwidth]{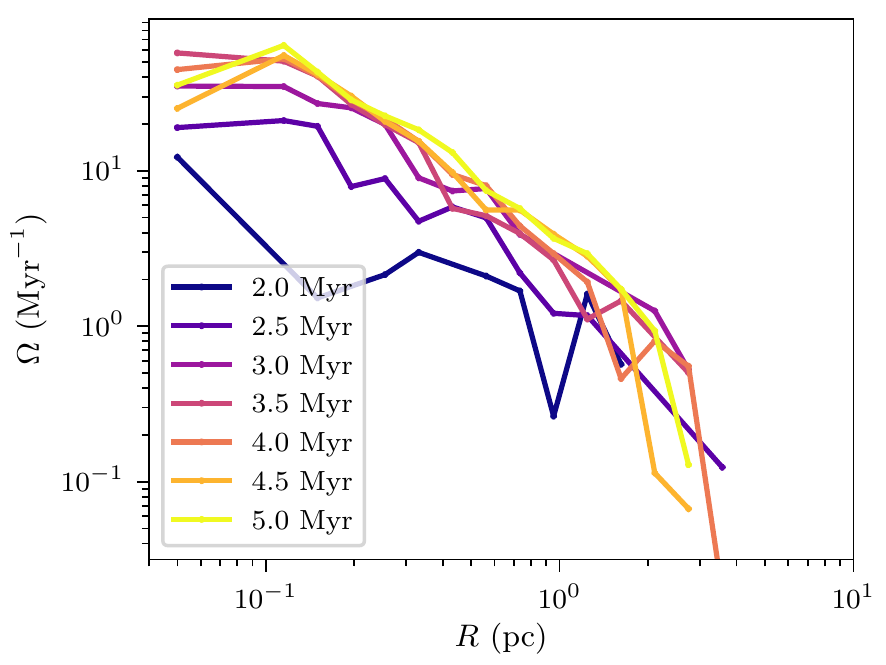} \put (82,65) {\textcolor{black}{\sj{}}} \end{overpic}
\begin{overpic}[width=0.5\textwidth]{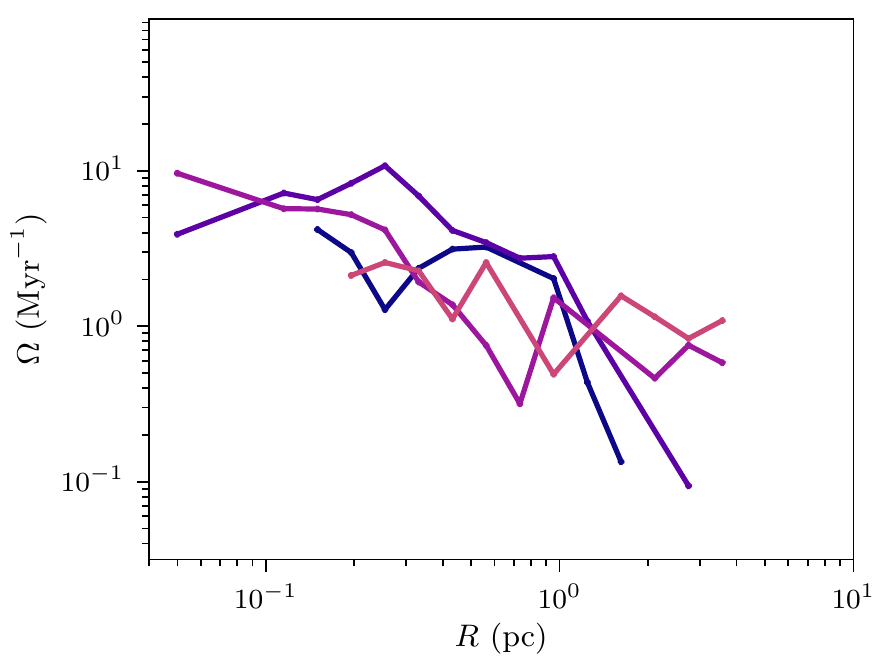} \put (80,65) {\textcolor{black}{\srh{}}} \end{overpic}%
\begin{overpic}[width=0.5\textwidth]{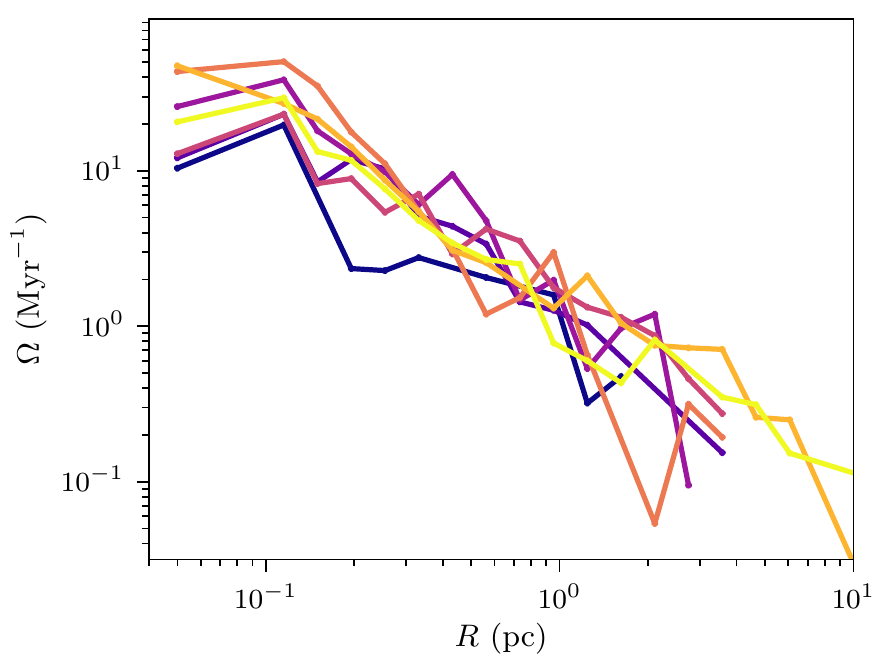} \put (75,65) {\textcolor{black}{\sjrh{}}} \end{overpic}
\caption{Angular velocity profiles of the bound part of the cluster in the four simulations, at several times. The legend is shared by the four panels. Some of the profiles are missing in the two left panels since the corresponding simulations are not advanced enough.}
\label{figure: omega cluster}
\end{figure*}

The profiles of the angular velocity is presented on Fig. \ref{figure: omega cluster}. We see that for the simulations \snj{}, \sj{} and \sjrh{}, the angular velocity profiles are well defined and stable in time, except for early times (dark blue curves) when rotation seems not to be fully developed. They roughly follow a power-law of index -1. 

The bottom left panel of Fig. \ref{figure: omega cluster} shows the angular velocity profiles for the simulation with HII regions only, \srh{}. Its velocity profiles show different behaviours compared to the ones of the other simulations. Note that the net rotation at $r \simeq \SI{1}{pc}$ is nevertheless clear though a factor 2-3 lower than for the other runs. At \SI{0.1}{pc} they are lower by a factor of 10, whereas at \SI{10}{pc} they show similar values. They thus exhibit a flatter power-law. This behaviour could be due to the very early onset of the HII regions in this simulation, as shown by Fig. \ref{figure: total sink mass all vs time} and \ref{figure: total sink mass for simu with HII regions}. This early onset could be responsible for a rapid emptying of the gas in the inner part of the cluster, disrupting the structure and preventing the stars from reaching the angular momentum corresponding to the profile observed in the other simulations. This does not happen in the simulation \sjrh{} which includes both HII regions and protostellar jets, as the jets slow down the accretion, allowing time for gas carrying large angular momentum to reach the cluster. This effect could be enhanced by the presence of unbound stars at late stages. At \SI{3.5}{Myr}, about a quarter of the mass of stars is unbound in \srh{}. These unbound stars could also play a role in perturbing the dynamics of the central bound stars.

Observationally, the situation appears to be complicated. Several authors 
\citep{henault2012,kamann2018a} report the presence of significant 
rotation on several clusters. Rotation velocities of a few \SI{}{km.s^{-1}} are 
being measured and  our simulations  qualitatively agrees with this. However 
other observations \citep{kuhn2019} do not detect significant rotation 
in a sample of young stellar clusters. In our simulations, the presence 
of rotation is a clear consequence of the large scale collapse of a gaseous 
turbulent clump. If confirmed, the absence of rotation in a fraction of 
stellar clusters will probably require alternative scenario for 
stellar cluster formation.

\noindent

%% file: Results_alignment.tex
\subsection{Alignment of the stars in the cluster}
\label{section-sub: Alignment of the stars in the cluster}

As the clusters are rotating, we study here the alignment of the angular directions of the sink particles in our four simulations.
Indeed spin alignment in two stellar clusters has been inferred 
by \citet{corsaro2017}, and by \citet{Kovacs2018} in one cluster.
We investigate whether or not a privileged direction also exists in the set of sink particles. An alignment is clearly expected in the simulations without HII regions as the gas exhibits a global rotating motion at the cluster scale. Then one may wonder what happens in the cases where HII regions are included. In particular, as the gas shows a lower global rotating motion, is the alignment still visible in the set of sink particles?

\begin{figure}[hbtp] 
\includegraphics[width=\linewidth]{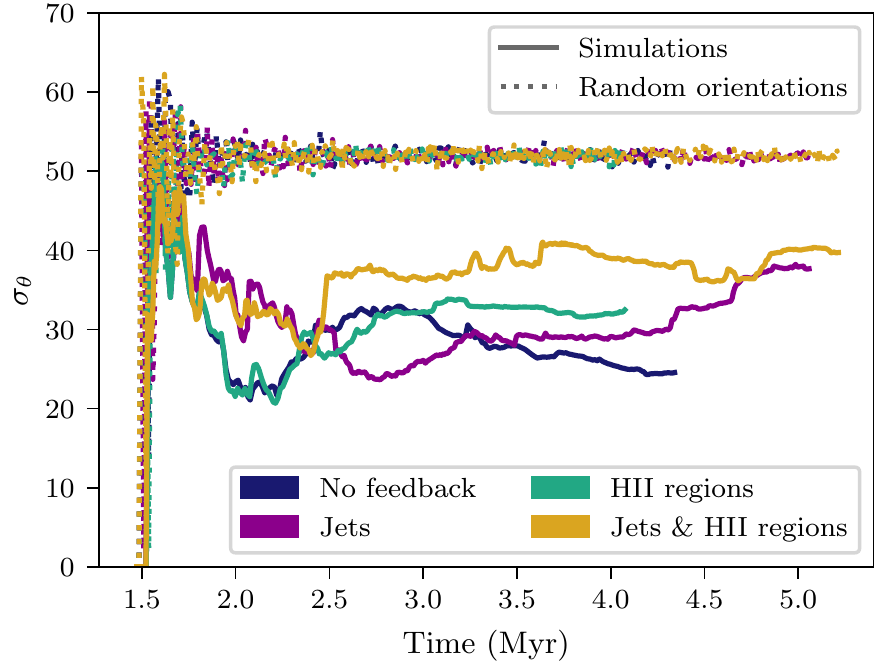}
\caption{The solid lines show the angular dispersion of the sink particles relatively to the mean angular direction, each colour corresponding to a different simulation. To consider only relevant sink particles, we selected the ones with a mass larger than \SI{0.07}{\Msun}. In comparison, the dotted lines show the angular dispersion of a set with the same cardinality than the set of sink particles at each time step, exhibiting random angular orientations. For times larger than \SI{2}{Myr}, the angular dispersion in the simulations is significantly lower than for a set of random orientations, implying that the sink particles have a preferred angular directions.}
\label{figure: dispersion_angulaire_all}
\end{figure}

\begin{figure*}[hbtp] 
\includegraphics[width=0.5\textwidth]{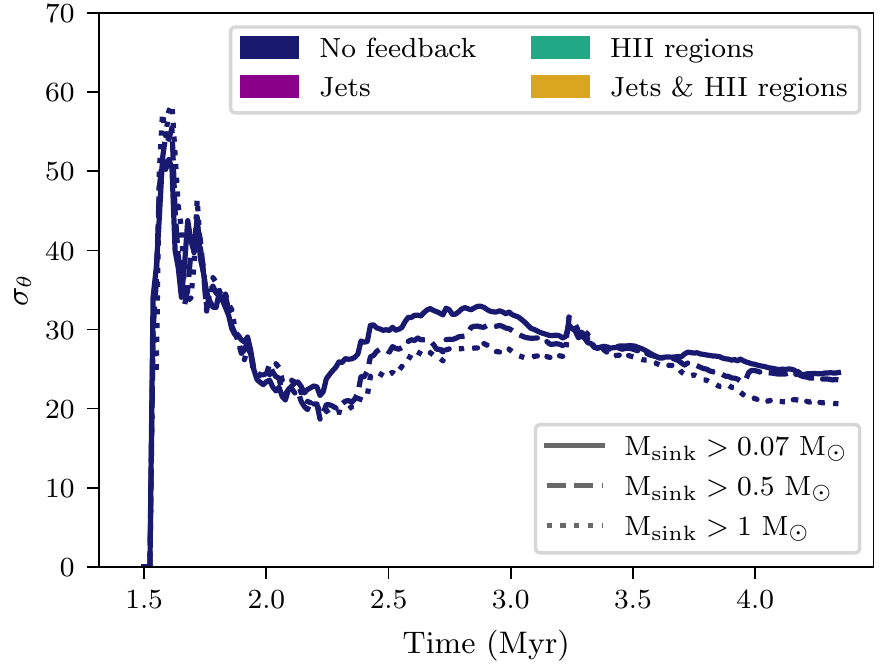}%
\includegraphics[width=0.5\textwidth]{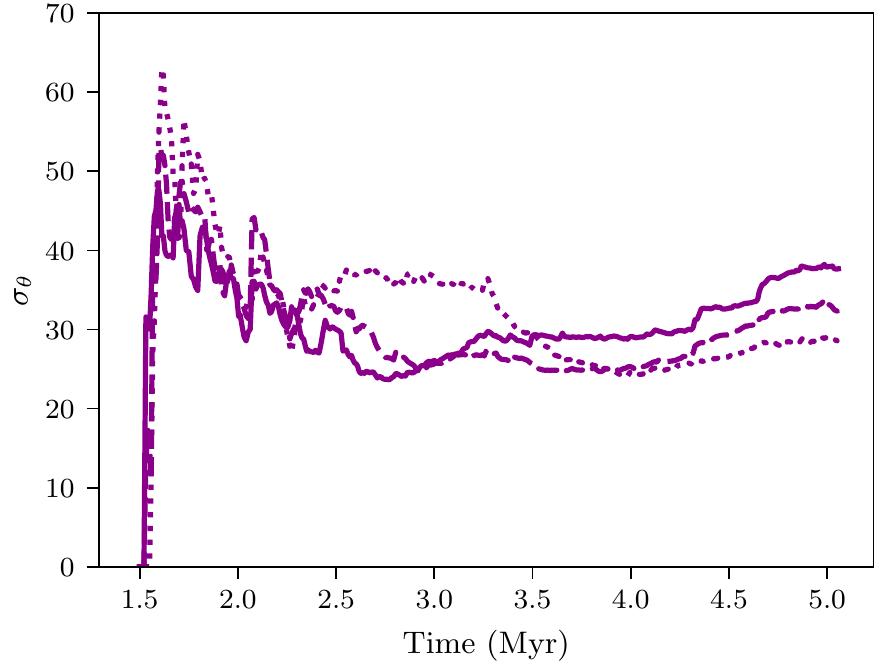}
\includegraphics[width=0.5\textwidth]{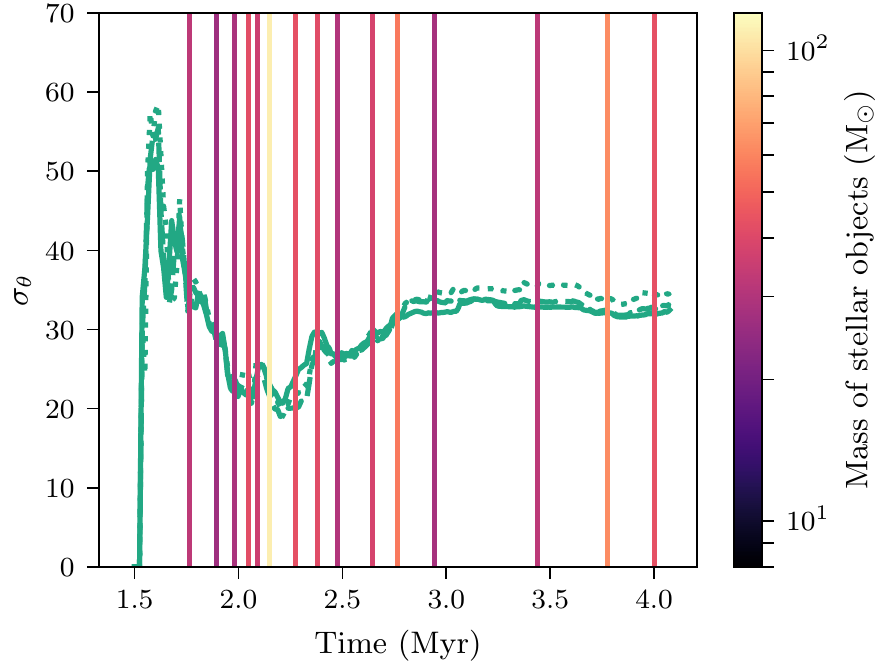}%
\includegraphics[width=0.5\textwidth]{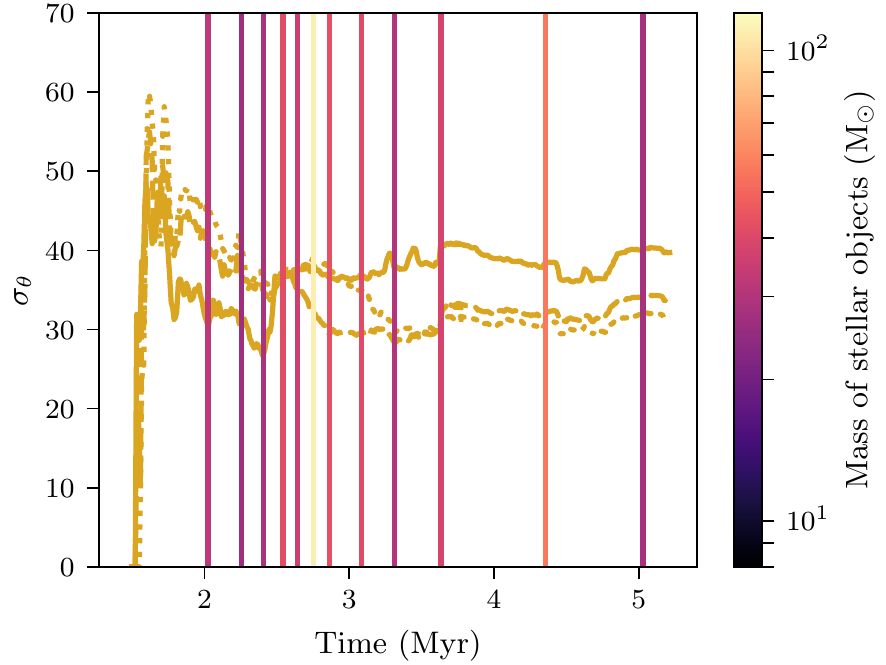}
\caption{Angular dispersion of the sink particles distribution in each simulation. From top to bottom and from left to right are the simulations without feedback, with protostellar jets only, with HII regions only, and with both jets and HII regions. The different types of lines indicate the different mass sink thresholds used to define the sets. For the two simulations with HII regions, on the second row, we indicate the moments of creation of the stellar objects with vertical lines, whose colours code their mass.}
\label{figure: dispersion_angulaire_vs_sink_seuil_mass}
\end{figure*}

To answer these questions, we computed the mean direction of the sink particles angular momentum in each simulation, at each time step. The angular momentum of a sink particle is simply that 
of all the gas it has accreted. We then get the angular dispersion of the directions around this mean value by taking the standard deviation $\sigma_\theta$ of the angles between the sink directions and the mean direction previously computed. In order to be able to tell if the sinks are aligned with the mean direction, or on the contrary randomly distributed, we computed the same quantity $\sigma_\theta$ for a set with the same cardinality than the set of sink particles at each time step, exhibiting random angular orientations. The results over time for the four simulations are presented on Fig. \ref{figure: dispersion_angulaire_all}. For times lower than \SI{1.8}{Myr}, the curves corresponding to the simulations and the ones corresponding to sets with random orientations are not distinguishable. At those times the sinks are not very massive, few in number, and the geometry of the gas flow is ill defined, which explains that we observe no particular alignment. However, after \SI{1.8}{Myr}, the angular dispersion begins to drop, stabilising at values between 20 and 40° for the four simulations. These values are lower than the ones corresponding to the sets of random orientations, which values tend to be constant and equal to about 50°. This shows that in those simulations the sink particles tend to align. This is expected from the gas geometry visible in the two first panels of Fig. \ref{figure: global appearance of simulations}, corresponding to the simulations without HII regions. In these two simulations the infalling gas forms a rotating disk-like structure around the star cluster, indicating a privileged direction as a global angular momentum emerges. In the two simulations which include HII regions, the sink particles also exhibit a preferred orientation. In Fig. \ref{figure: global appearance of simulations} and \ref{figure: global appearance of simulations - large scale} we see that the gas geometry is much more disorganised and dislocated than for the simulations without HII regions\footnote{On the last row of Fig. \ref{figure: global appearance of simulations} we see that for the simulations with jets and HII regions a small disk-like structure seems to survive at this time, while in the simulation with HII regions only, at this time the gas has been completely blown out of the cluster.}. Nevertheless, the sinks particles still exhibit a preferred orientation in these simulations which seems to survive through time as HII regions are developing. On the bottom panels of Fig. \ref{figure: dispersion_angulaire_vs_sink_seuil_mass}, the solid lines represent the same quantity than on Fig. \ref{figure: dispersion_angulaire_all}, and we indicate the creation of the stellar objects with vertical coloured lines. In this first panel standing for the simulation with HII regions only, in solid green line, we see that after the formation of the most massive stellar object at \SI{2.2}{Myr} the angular dispersion slightly increases gradually from 20° to around 33°. This increase could be due to the formation of unbound stars in the outskirt of the cluster as HII regions are expanding. Unlike central stars, unbound ones should not have a preferred direction as the dynamics of the gas is different at these places. In the second panel, representing the simulations with protostellar jets and HII regions, in solid yellow line, we see that there is no clear correlation between the slight variations of the angular dispersion and the onset of HII regions. It is thus unclear whether or not the HII regions play a role in modifying the angular dispersion here. A correlation is found only in the simulation with HII regions only. Moreover, even if the HII regions had an impact on this angular dispersion, it would be minimal as the variations in the other cases (and especially in the cases without HII regions) are of the same order of magnitude.

\noindent

\noindent

The four panels of Fig. \ref{figure: dispersion_angulaire_vs_sink_seuil_mass} represent the angular dispersion in the four simulations. The solid lines show the results for the same selection rule than in Fig. \ref{figure: dispersion_angulaire_all}, which is that only sinks with a mass higher than \SI{0.07}{\Msun} are selected. The dashed and dotted lines show results for the sets of sink particles with a mass higher than \SI{0.5}{\Msun} and \SI{1}{\Msun} respectively. We see that for the first panel representing the \snj{} simulation, the sets of higher masses exhibit slightly lower angular dispersion. This would mean that in this simulation the more massive stars are more aligned than the less massive ones. As the massive ones accrete more and more material from the oriented gas flow, they tend to align with time. When considering the less massive stars, as they are more sensitive to local slight variations in the gas flow, they exhibit a higher angular dispersion. 

In the second panel and fourth panels, representing respectively the simulation \sj{} and \sjrh{}, the sets of higher masses show higher angular dispersion during the first hundreds of kyrs of evolution, then the trend is reversing and the same behaviour than the one in \snj{} is observed. A possible explanation for this behaviour could be that at the beginning of the formation of the cluster, sinks are not particularly aligned, as showed by Fig. \ref{figure: dispersion_angulaire_all}. As time goes by, sinks are accreting matter. In \snj{}, as a preferred orientation emerged from the gas flow, it is directly reflected in the sink particles behaviour. The massive ones are accreting more matter, and then are aligned more rapidly than the lower mass ones which are more sensitive to gas properties variations. In \sj{}, the presence of jets modifies the accretion onto sink particles. The sink particles with a mass higher than \SI{0.5}{\Msun} have already started to launch protostellar jets. This modifies the direct environment of these sink particles, promoting accretion in their orthogonal plane, even if a global preferred orientation of the gas flow emerges. It is thus harder to change their initial orientation when the particles are accreting this way. The higher the mass of the particles, the more powerful the jets, and therefore the more impact they have on the direct environment of the particles. It is thus more difficult for the massive particles to align when protostellar jets are included. When considering lower mass particles with a threshold of \SI{0.07}{\Msun}, we are taking into account all the newly formed particles, which are not affected by their own protostellar jets, thus are more sensitive to the gas flow geometry, and are directly formed with a preferred orientation on average as an orientation emerges in the flow. The angular dispersion in this set is thus lower here. At \SI{3.5}{Myr} for \sj{} and \SI{2.9}{Myr} for \sjrh{}, the trend is reversing and the same behaviour than in \snj{} is observed. As the massive particles are accreting more and more, they tend to align, even if the jets slow down this alignment, and we come back to a situation where the massive stars are the most aligned whereas the newly formed stars are more sensitive to variations and exhibit higher angular dispersion.

In the third panel, representing \srh{}, the three curves present similar 
trends and it is therefore hard to interpret the slight differences insofar as they could be due to the particular realisation. Furthermore, comparing \snj{} and \srh{} in one hand -- left panels --, and \sj{} and \sjrh{} in the other hand -- right panels --, we see that HII regions only have a minor influence on the observed behaviour, while protostellar jets seem to play a more important role in the alignment of the sink particles.

%% file: Results_Q.tex
\subsection{Spatial distribution of the emerging clusters}  
\label{section-sub: Q parameter of the emerging clusters}

We now study the spatial distribution of the stars in the simulations, as traced by the sink particles. 
A first effect of feedback in the global cluster distribution can already be seen from visual inspection of previous figures in the text. 
All simulations start with a  globally similar distribution of stars, as shown in the first column of Fig. \ref{figure: global appearance of simulations - small scale}. At later stages, however, it is clear from  Fig. \ref{figure: global appearance of simulations - large scale} (first column) or Fig. \ref{figure: bound stars} (pay attention to spatial scales) that  simulations with HII regions produce distributions with larger spatial dispersion, where the stars spread over a larger fraction of the region. 

To globally characterise the  distribution of stars and its evolution with time we computed the $Q$ parameter.  This parameter, which
has been introduced and developed by \citet{Cartwright2004, Cartwright2009}, 
has been widely used to objectively evaluate the level of sub-clustering or concentration in stellar clusters and associations. The value of $Q$ is given by the ratio of the mean distance of the members and the average length of their minimum spanning tree \footnote{The minimum spanning tree of a set of points is a set of straight lines or edges connecting all the points in the sample without cycle and with minimal sum of lengths.}. 
\citet{Cartwright2004} calibrated the $Q$ parameter in synthetic clusters and found that values of $Q > 0.8$ correspond to radial concentration, $Q \simeq 0.8$ corresponds to clusters with homogeneous distribution, and $Q<0.8$ corresponds to clumpy clusters showing sub-clusterisation. To avoid over-interpretation in realistic distributions, these boundaries should not be strictly applied, and $Q$ should be used qualitatively. In appendix \ref{annexe: Q parameter} we explain the details on the calculations performed in this work.

\noindent
\begin{figure} 
\includegraphics[width=\linewidth]{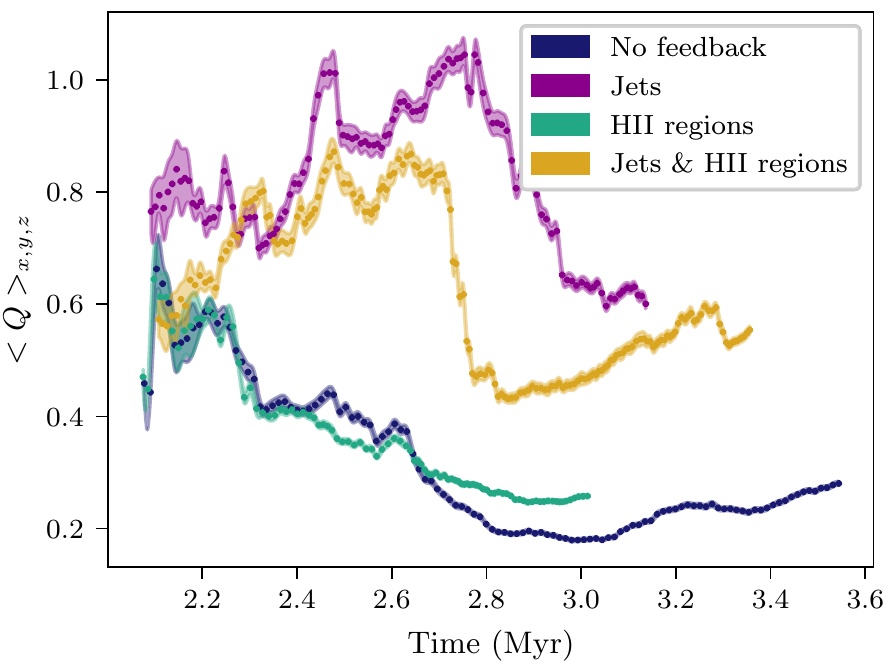}
\caption{Evolution of the $Q$ parameter with time.}
\label{figure: All_Q_stat_mean}
\end{figure}

The evolution of $Q$ with time is shown in Fig. \ref{figure: All_Q_stat_mean} where the $Q$ parameter is plotted over time for the four simulations represented by four distinct colours. The points represent the mean value of the $Q$ parameter, while the colour filled areas represent the interval of the mean value $\pm 1 \sigma$. For each simulation, we begin at a time where approximately 300 sink particles are formed,  in order to have enough sink particles in the sample.

 The first notable result is that two clear trends appear. On one hand, the two simulations that include jets -- \sj{} and \sjrh{} -- show a similar evolution over time, while in the other hand, the two simulations that do not include jets -- \snj{} and \srh{} -- also show a similar trend, different from the first one. 
 
 Both \sj{} and \sjrh{} show an increase in the $Q$ parameter between 2.1 to \SI{2.7}{Myr}, followed by a significant decrease.  The initial increase in the $Q$ parameter indicates an evolution of the initial substructure presents in \sjrh{}, and an increase of the radial concentration in \sj{} which,  indeed, starts with $Q$ values around 0.8, consistent with homogeneous distributions.
 In both cases, the decrease in $Q$ around \SI{2.7}{Myr} is associated to the growth of substructure in the outskirts of the field, far from the main cluster. These structures are small compared to the main clump of stars in the center,  but the $Q$ parameter decreases as they become dense enough to be significant. The decrease is more abrupt in the simulation with an HII region, \sjrh{}, where a subsequent slow increase is also observed.

\begin{figure} 
\includegraphics[width=\linewidth]{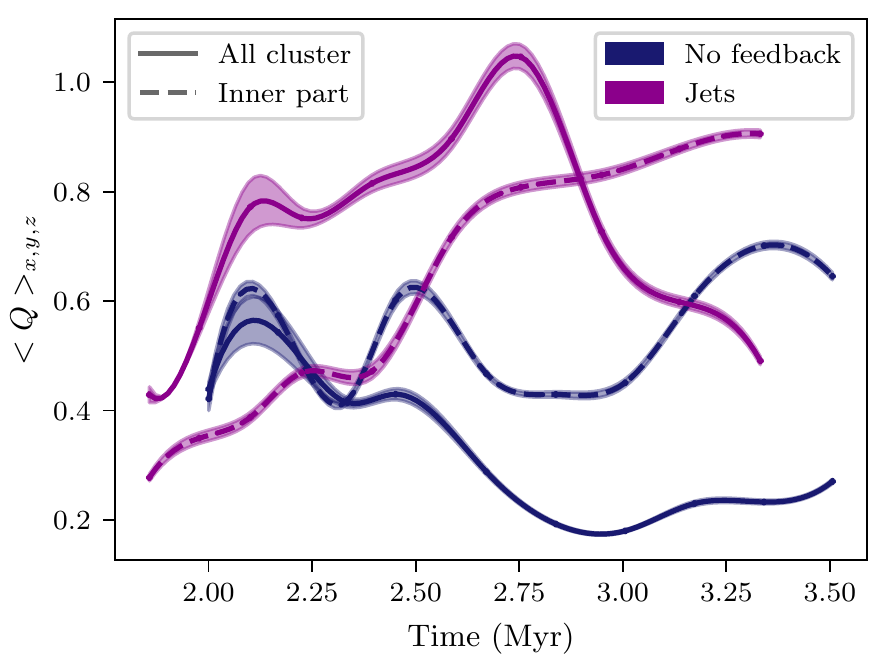}
\caption{Smoothed temporal evolution of the $Q$ parameter for the entire cluster and for the inner part only, in \snj{} and \sj{}.}
\label{figure: Q inner part}
\end{figure}

 The two simulations without jets, \snj{} and \srh{}, both show clear substructure at \SI{2.1}{Myr} with a $Q$ parameter of about 0.6,  that slowly and steadily decreases to 0.2 at \SI{2.8}{Myr}. At \SI{3}{Myr}, the $Q$ in \snj{}  slowly increases from 0.2 to 0.3 at \SI{3.6}{Myr}.  These values of $Q \lesssim 0.3$ are much  lower than the observed values for sub-clustered distributions and even that the box-fractal models in which the method was originally calibrated.  
In order to understand these very low values, we computed the $Q$ parameter of the inner part of the cluster, containing $90\%$ of the stars mass. This selection thus excludes the small substructures that grows far away from the central cluster. The $Q$ evolution is visible on Fig. \ref{figure: Q inner part} for this inner part, in the two simulations without HII regions -- \snj{} and \sj{}. We see that the inner part in \snj{} (blue dashed curve) exhibits higher and more stable values, between 0.4 and 0.7. This means that the very low values of the $Q$ parameter for the entire cluster is largely affected by the presence of sub-clusterisation in the outskirts. We also see that in the simulation \sj{} (purple dashed curve), the drop at \SI{2.7}{Myr} disappears, which confirms that this drop is associated to the growth of substructures in the outskirts of the field. Comparing the inner parts of \snj{} and \sj{}, we find that rapidly the $Q$ of \sj{} rises over the one of \snj{}. From \SI{2.5}{Myr}, \sj{} exhibits a homogeneous inner region with a $Q$ of the order of 0.8, whereas \snj{} still exhibits sub-clusterisation in its inner part, with a $Q$ between 0.45 and 0.7. 
We conclude 
that the protostellar jets seem to be efficient at preventing (or at least delaying) the apparition of substructures in the outskirts. Broadly speaking, 
the $Q$ values inferred in the presence of jets, appear in better agreement with the observations that when jets are absent.

 \noindent

Our results show that protostellar jets seem to play an important role in the structuring and distribution of stars in the cluster, while HII regions, despite producing more disperse distributions, do not seem to produce statistically significant structural changes. Looking at the results presented in sections \ref{section: article2-gas} and \ref{section: article2-sinks properties}, jets and HII regions have different impacts on the gas and the star cluster. Protostellar jets only have a minor influence on the gas appearance while having a major influence on the stars distribution in the cluster. Conversely, HII regions have a major role in shaping the gas but their impact on the distribution of stars in the cluster is limited.

%% file: Conclusion.tex
\section{Conclusions and perspectives}
\label{section: article2-conclusion}

\subsection{Conclusions}
To study the formation and evolution of stellar clusters, we performed a set a four simulations including or not the effects of HII regions and protostellar jets. We started from a turbulent, magnetised cloud of \SI{e4}{\Msun} of gas. 

The structures formed in these simulations have a very different appearance depending on whether or not HII regions are included. The expansion of the HII regions empties the central part and shreds the gas. We have shown that protostellar jets have a significant influence on the star formation rate. They slow down star formation, reducing the SFR by more than a factor of two, but do not stop star formation. The onset of HII regions also reduces the SFR but quickly leads to the dispersion of the gas in the cluster, which almost completely extinguishes the accretion onto stars and then decreasing the SFE. 

The study of the gas reveals that the jets have almost no influence on the density distribution and have a moderate influence on the Mach number. On the contrary, the HII regions strongly alter the gas at intermediate densities and strongly modify the Mach number at all densities. It is thus difficult to find evidences for the presence of jets looking only at the kinematics of the gas. Turbulence in  cluster forming clumps is primary due to gas accretion.

We then studied the emerging star clusters. In simulations without HII regions, all the stars formed are gravitationally bound. When HII regions are included, the cluster consists of a central core of bound stars surrounded by an outskirt of unbound stars. The clusters are rotating. We showed that the angular velocity profiles are stable in time. In the simulation with HII regions only, the early onset of HII regions could explain the low rotation of the cluster. We showed that the stars exhibit a preferred alignment of their own rotation axis. This alignment seems to persist through time, even in the simulation with HII regions only, where the global rotation of the cluster is less obvious. In order to characterise the spatial distribution of the formed stars, we calculated the $Q$ parameter of these clusters. Protostellar jets play a major role here. 
They seem to prevent the formation of sub-structures in the outskirt of the cluster. Simulation including protostellar jets thus showed distribution with less substructures, and the ones that still form appear in later times.

\subsection{Perspectives}
It would be interesting to further investigate the $Q$ parameter. One possibility could be to use the S2D2 algorithm developed by \citet{Gonzalez2021} to detect significant small-scale sub-structures known as \textit{Nested Elementary STructures} (NESTs). The study of these NESTs and their distribution could be a powerful way to interpret more precisely the $Q$ parameter of the emerging clusters.

One of the limitations of our study is the impossibility to run simulations with a significant statistics on the initial conditions, due to the high computational cost of this type of simulations. It would be valuable to run similar simulations with different initial conditions and different seeds for the random generator of the stellar object masses. As the expansion of HII regions is known to depend on the density of the gas where it takes place, varying the initial conditions would allow to investigate the statistics for the onset of HII regions. 

One could also try to perform similar simulations with higher resolutions. It would then be possible to study the emerging initial mass function and to determine the individual and joint influences of HII regions and protostellar jets on it. 

It would be interesting to investigate for longer times the mass fraction of bound and unbound stars and to follow those stars in time. In fact, it is known that some of the clusters are evaporating and \citet{gavagnin2017} showed that cluster survival depends on feedback strength. Leading this study in simulations with different initial conditions would probably shed lights on conditions and processes responsible for the evaporation of such stellar clusters.


%% file: Annexe_1.tex
\section{Protostellar jets modelisation and implementation}
\label{annexe: implementation jets}

In our simulations, the maximum resolution associated to the finest AMR level is about \SI{1.5e3}{au}. This prevents to observe effects emerging from physical mechanisms happening at smaller scales. In particular we do not resolve all the physics responsible for the emission of protostellar jets. The theories describing these ejections of matter from young accreting stars such as the centrifugal acceleration mechanism or the X-wind model \citep[see for example][]{Federrath2014} involve physical mechanisms taking place on scales smaller than the astronomical unit. It is thus for the moment not possible to describe consistently the emission of protostellar jets and the dynamical formation of a star cluster in a \SI{30.4}{pc} simulation box.

\subsection{The sub-grid model}
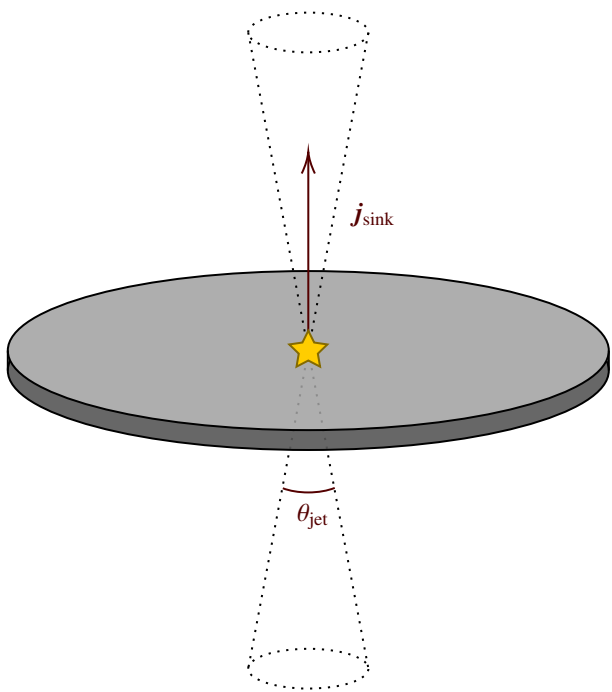
\begin{figure}[h]
\input{Jets_grometry_tikz_figure}
\caption{Illustration of the geometric properties of the protostellar jets model. The central protostar is schematised is yellow, surrounded by its accretion disk in grey. The third of the accreted mass at each time step is ejected in a right circular bicone of opening angle $\theta_\text{jet}$ and of direction the angular momentum of the sink particle ${\vec{j}_{\text{sink}}}$.}
\label{figure: illustration jets}
\end{figure}

To still be able to take into account the effect of these jets on the large scale evolution, we implement a sub-grid model based on the properties of the sinks. Once a sink grows a mass higher than \SI{0.15}{\Msun}, at each timestep it expels $1/3$ of the mass accreted during this timestep in the form of a circular biconic jet. The matter is expelled with a velocity equal to a fraction $f_v$ of the escape velocity at the surface of the star. The direction of the ejection is given by the angular momentum of the sink. Each circular cone has an opening angle of $\theta_\text{jet}$. $f_v$ and $\theta_\text{jet}$ are fixed and are the same for all the sink particles in the simulation. Figure \ref{figure: illustration jets} shows a schematic view of the jets geometry around a sink particle.

\subsection{Implementation in \ramses{}}
\label{annexe: implementation jets - implementation}
The numerical implementation of this sub-grid model of protostellar jets is made through a routine called just after the accretion onto the sinks. First of all, during a time step, the accretion of material onto each sink is processed. The mass accreted by each sink during this time step is stored in a variable which will be named here $m_{\text{acc},j}$ with the index $j$ referring to the sink number. 

Once this accreted mass is computed, the routine setting protostellar jets is called. This routine contains a first loop over the CIC particles\footnote{These CIC particles were originally introduced in \ramses{} to model dark matter. In the current version of \ramses{}, the mass of each sink particles is distributed equally on to a spherical cloud of CIC particles. The distance between each CIC particles if half the grid spacing, and the radius of the cloud of CIC particles sets the gravitational softening length. More details are given in \citet{Bleuler2014}.}. 
Doing so, it is possible to identify the gas cells around each sink particle without having to scan the entire AMR grid. During this first loop, for each sink particle, the total volume of the cells that lay in a circular bicone around the sink is computed. A second loop over the CIC particles is done to redistributed some of the accreted quantities to the gas. For each sink particle, the quantities retrieved by each gas cell are proportional to the fraction of the cell volume over the total volume of gas cells in the circular bicone. This ensures that the quantity that is redistributed in the jet launch area is homogeneous over this area. 

For each sink particle $j$ that exhibits mass greater than \SI{0.15}{\Msun}, an arbitrary fraction $1/3$ of the accreted mass $m_{\text{acc},j}$ is given back to the gas cells. This value of $1/3$ is coherent with previous analytical, numerical, and observational studies of protostellar jets. For example, theories based on centrifugal acceleration or X-wind model by \citet{Blandford1982,Pudritz1986,Shu1988,Wardle1993,Konigl2000,Pudritz2007}
and observations by 
\citet{Hartmann1995,Calvet1998,Bacciotti2002,Cabrit2007,Bacciotti2011}
 suggest a fraction of the accreted mass which is ejected in the range of 0.1 to 0.4  \citep[see page 4 of][for a review over these different studies and results]{Federrath2014}.
The arbitrary threshold of \SI{0.15}{\Msun} stands to ensure that the angular momentum of the sink particle is well defined when ejection begins. The velocity given to the ejected fraction of mass is equal to a fraction of the escape velocity at the protostar surface $v_{jet}=f_{v} \sqrt{ 2GM_{j}/R_j }$ with $M_{j}$ and $R_{j}$ being respectively the mass and radius of the protostar modelled by the sink particle. For $f_{v}$ between $0.25$ and $0.5$ this prescription gives jets velocities of the order of a hundred to a few hundreds of \SI{}{km.s^{-1}}. The matter deposited in the circular bicone is given an specific thermal energy which is the same than the one of the gas cell in which it is deposited.

Once these quantities are given back to the gas, they are deduced from the sink particles quantity: $M_j = M_j - \frac{m_{\text{acc},j}}{3}$. The linear and angular momentum, and the energy associated to each ejection is also calculated and subtracted from the linear and angular momentum of each sink particle.


%% file: Jets_grometry_tikz_figure.tex

\tikzset{every picture/.style={line width=0.75pt}} 

\begin{tikzpicture}[x=0.75pt,y=0.75pt,yscale=-1,xscale=1]

\draw [color={rgb, 255:red, 0; green, 0; blue, 0 }  ,draw opacity=1 ][fill={rgb, 255:red, 0; green, 0; blue, 0 }  ,fill opacity=0.19 ] [dash pattern={on 0.84pt off 2.51pt}]  (330,310) -- (360,470) ;
\draw [fill={rgb, 255:red, 0; green, 0; blue, 0 }  ,fill opacity=0.19 ] [dash pattern={on 0.84pt off 2.51pt}]  (330,310) -- (300,470) ;
\draw  [draw opacity=0][fill={rgb, 255:red, 103; green, 103; blue, 103 }  ,fill opacity=1 ] (476.09,310.89) .. controls (478.65,313.81) and (480,316.86) .. (480,320) .. controls (480,342.09) and (412.84,360) .. (330,360) .. controls (247.16,360) and (180,342.09) .. (180,320) .. controls (180,316.48) and (181.71,313.06) .. (184.92,309.8) -- (330,320) -- cycle ; \draw   (476.09,310.89) .. controls (478.65,313.81) and (480,316.86) .. (480,320) .. controls (480,342.09) and (412.84,360) .. (330,360) .. controls (247.16,360) and (180,342.09) .. (180,320) .. controls (180,316.48) and (181.71,313.06) .. (184.92,309.8) ;
\draw  [fill={rgb, 255:red, 174; green, 174; blue, 174 }  ,fill opacity=1 ] (180,310) .. controls (180,287.91) and (247.16,270) .. (330,270) .. controls (412.84,270) and (480,287.91) .. (480,310) .. controls (480,332.09) and (412.84,350) .. (330,350) .. controls (247.16,350) and (180,332.09) .. (180,310) -- cycle ;
\draw    (180,310) -- (180,320) ;
\draw    (480,310) -- (480,320) ;
\draw  [dash pattern={on 0.84pt off 2.51pt}] (300,150) .. controls (300,144.48) and (313.43,140) .. (330,140) .. controls (346.57,140) and (360,144.48) .. (360,150) .. controls (360,155.52) and (346.57,160) .. (330,160) .. controls (313.43,160) and (300,155.52) .. (300,150) -- cycle ;
\draw  [dash pattern={on 0.84pt off 2.51pt}] (300,470) .. controls (300,464.48) and (313.43,460) .. (330,460) .. controls (346.57,460) and (360,464.48) .. (360,470) .. controls (360,475.52) and (346.57,480) .. (330,480) .. controls (313.43,480) and (300,475.52) .. (300,470) -- cycle ;
\draw  [draw opacity=0] (343.19,379.07) .. controls (339.67,380.49) and (335.05,381.35) .. (330,381.35) .. controls (325.16,381.35) and (320.73,380.56) .. (317.27,379.25) -- (330,372.18) -- cycle ; \draw  [color={rgb, 255:red, 84; green, 0; blue, 0 }  ,draw opacity=1 ] (343.19,379.07) .. controls (339.67,380.49) and (335.05,381.35) .. (330,381.35) .. controls (325.16,381.35) and (320.73,380.56) .. (317.27,379.25) ;
\draw [color={rgb, 255:red, 84; green, 0; blue, 0 }  ,draw opacity=1 ]   (330,309.93) -- (330,212) ;
\draw [shift={(330,210)}, rotate = 450] [color={rgb, 255:red, 84; green, 0; blue, 0 }  ,draw opacity=1 ][line width=0.75]    (10.93,-3.29) .. controls (6.95,-1.4) and (3.31,-0.3) .. (0,0) .. controls (3.31,0.3) and (6.95,1.4) .. (10.93,3.29)   ;
\draw [fill={rgb, 255:red, 255; green, 0; blue, 0 }  ,fill opacity=1 ] [dash pattern={on 0.84pt off 2.51pt}]  (360,150) -- (330,310) ;
\draw [fill={rgb, 255:red, 255; green, 0; blue, 0 }  ,fill opacity=1 ] [dash pattern={on 0.84pt off 2.51pt}]  (300,150) -- (330,310) ;
\draw [color={rgb, 255:red, 0; green, 0; blue, 0 }  ,draw opacity=0.2 ][fill={rgb, 255:red, 0; green, 0; blue, 0 }  ,fill opacity=0.19 ] [dash pattern={on 0.84pt off 2.51pt}]  (330,310) -- (320.83,359.75) ;
\draw [color={rgb, 255:red, 0; green, 0; blue, 0 }  ,draw opacity=0.2 ][fill={rgb, 255:red, 0; green, 0; blue, 0 }  ,fill opacity=0.19 ] [dash pattern={on 0.84pt off 2.51pt}]  (330,310) -- (339.43,359.93) ;
\draw  [color={rgb, 255:red, 132; green, 106; blue, 0 }  ,draw opacity=1 ][fill={rgb, 255:red, 255; green, 205; blue, 0 }  ,fill opacity=1 ][line width=0.75]  (330,300) -- (332.94,305.95) -- (339.51,306.91) -- (334.76,311.55) -- (335.88,318.09) -- (330,315) -- (324.12,318.09) -- (325.24,311.55) -- (320.49,306.91) -- (327.06,305.95) -- cycle ;

\draw (323,385.4) node [anchor=north west][inner sep=0.75pt]  [color={rgb, 255:red, 84; green, 0; blue, 0 }  ,opacity=1 ]  {$\theta _\text{jet}$};
\draw (350,234.4) node [anchor=north west][inner sep=0.75pt]  [color={rgb, 255:red, 84; green, 0; blue, 0 }  ,opacity=1 ]  {${\vec{j}_{\text{sink}}}$};

\end{tikzpicture}


%% file: Annexe_2.tex
\section{Evolution of the number of sinks}
\label{annexe: evolution of the number of sinks}

\begin{figure*}
\includegraphics[width=0.5\textwidth]{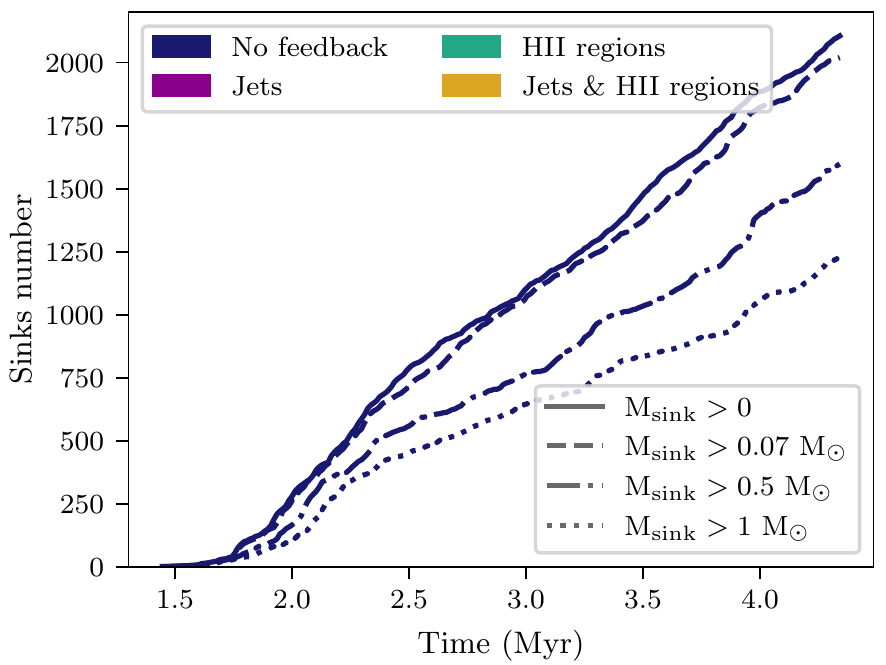}%
\includegraphics[width=0.5\textwidth]{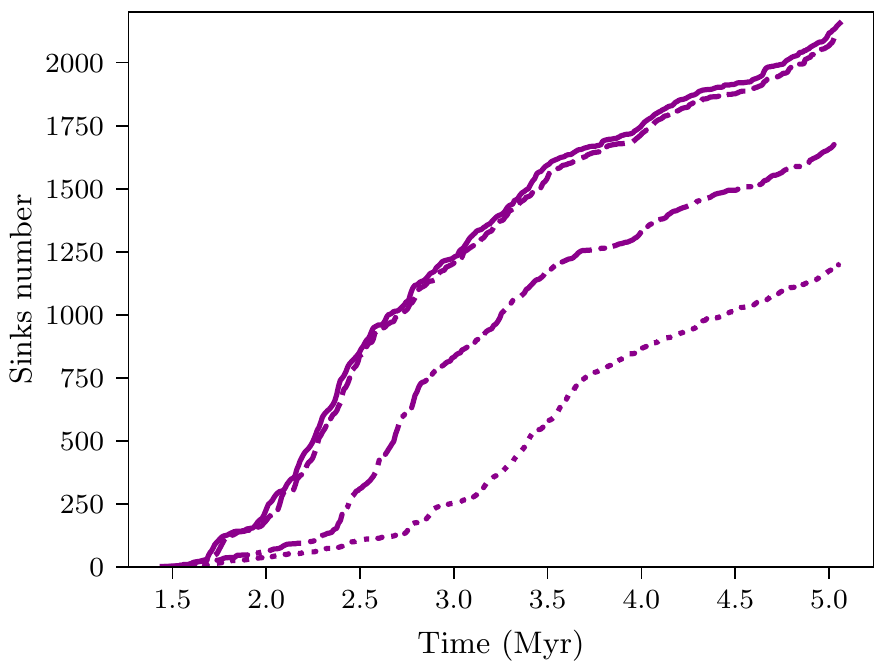}
\includegraphics[width=0.5\textwidth]{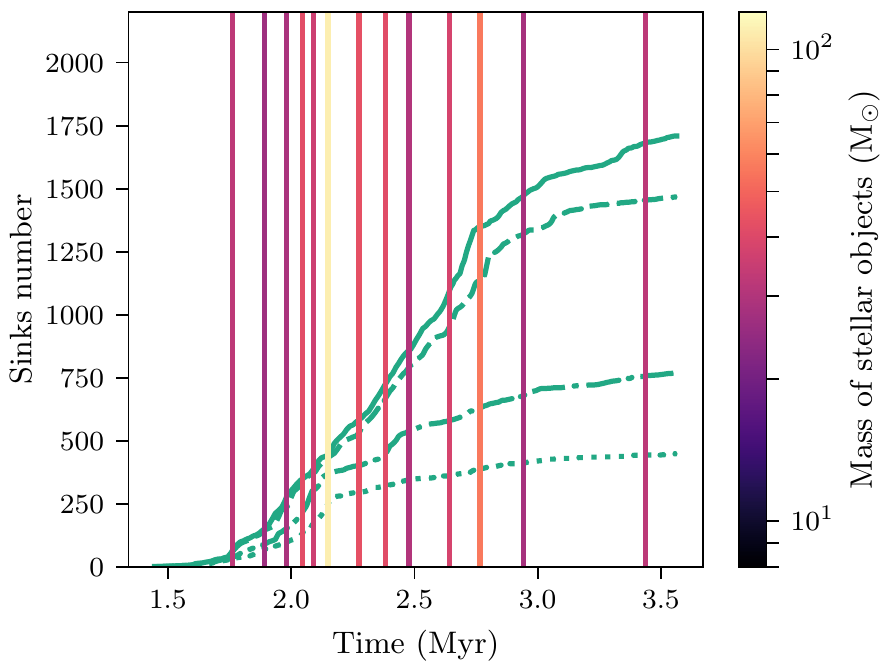}%
\includegraphics[width=0.5\textwidth]{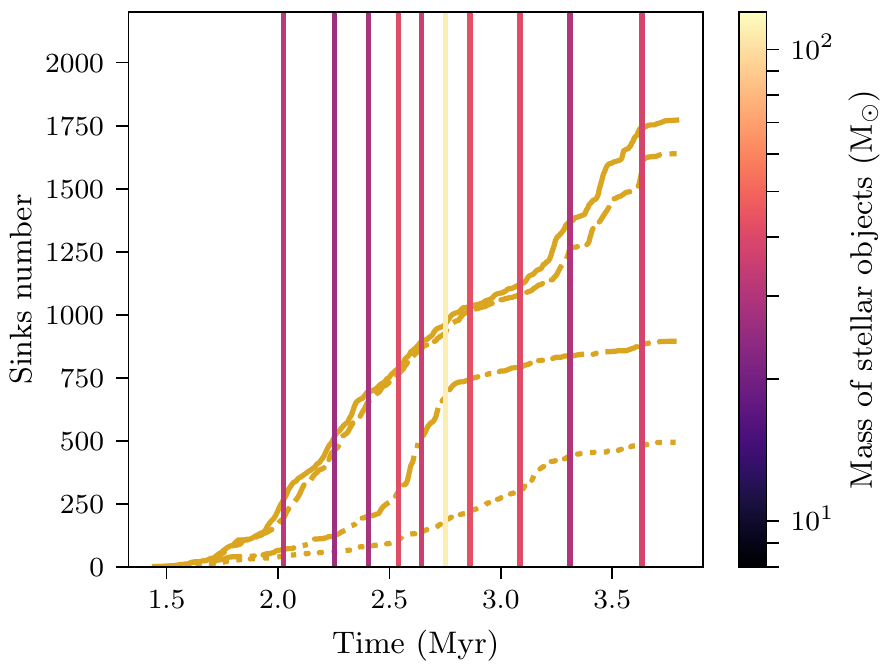}
\caption{Evolution of the number of sinks during the temporal evolution. From top to bottom and from left to right, the simulation \snj{}, \sj{}, \srh{}, and \sjrh{}. For the simulations with HII regions, vertical lines indicate the formation of stellar objects, with the colours coding for their mass. In the four panels, the solid lines indicate the total number of sinks, the other types of line (dashed, dash-dotted, and dotted respectively) indicate the number of sinks with a mass lower than a threshold (0.07, 0.5 and \SI{1}{\Msun} respectively).}
\label{figure: number of sinks vs time}
\end{figure*}

The evolution of the number of sinks with time is presented in Fig. \ref{figure: number of sinks vs time}. The four panels represent the four simulations. The y-axes have the same extent, but the x-axes extent depends on each panel. For the two simulations including HII regions, vertical lines indicate the formation of stellar objects.


We see that for the simulation including HII regions -- \srh{} and \sjrh{} -- the evolution of the number of sinks without using a mass threshold or using one of \SI{0.07}{\Msun} exhibits low influence of stellar objects formation. However, when we look at the sinks number using a mass threshold of \SI{0.5}{\Msun}, we see that just after the apparition of the most massive stellar object (yellow vertical line), the increase in the number of stars shows a drastic drop. This is also the case for the stars with a mass higher than \SI{1}{\Msun} in \srh{}, but not in \sjrh{} where this change in behaviour is seen at \SI{3.3}{Myr}. This results show that stars continue to form in simulations with HII regions, but it is essentially low mass stars. HII regions seems to be efficient at cutting accretion onto the less massive stars as they form but do not accrete enough to grow higher masses when the HII regions are developing.

\noindent

%% file: Annexe_3.tex
\section{$Q$ parameter}
\label{annexe: Q parameter}
In this appendix we describe the specific calculations of $Q$ performed in this work, along with their motivation. We refer the reader to  Appendix B in \citet{Gonzalez2021} for a recent review on the $Q$ parameter method and its alternatives. 

The $Q$ parameter is defined by:

\begin{equation}
    Q = \frac{\bar{l}_\text{MST}}{\bar{s}}
\end{equation}

\noindent
where $\bar{s}$ is the normalized mean distance between stars, and $\bar{l}_\text{MST}$ is the normalized mean edge length of the minimum spanning tree (MST) \citep[see e.g.][]{Boruvka1926,Kruskal1956,Prim1957,Gower1969}. A spanning tree is a set of straight lines connecting all the points of the sample without cycle. The MST is the unique spanning tree whose sum of edges is minimal. To allow for the comparison of regions of different size, the values of $\bar{s}$ and $\bar{l}_\text{MST}$ are normalised. Following \citet{Cartwright2004}, $\bar{s}$ should be normalised with a radius representative of the size of the cluster, and  $\bar{l}_\text{MST}$ with a factor that includes the area subtended by the MST and the minimum number of points. We use consistent normalisation area $A$ for the MST and radius $R$ for the cluster, that follow the relationship $A=\pi R^2$. 

Despite giving the boundary $Q=0.8$, there is some dispersion in the $Q$ values obtained in \citet{Cartwright2004} that can be associated to sampling effects and the different random realisations of each distribution.  This dispersion implies that even synthetic clusters with low degree of substructure or concentration are not statistically distinguishable from homogeneous.

As the $Q$ parameter has been introduced to describe observed clusters, $\bar{s}$ and $\bar{l}_\text{MST}$ were calibrated by \citet{Cartwright2004} using the projected position of the stars in the plane of the sky for a set of synthetic clusters, with radial concentration, homogeneous distribution, or fractal sub-clusterisation.  Despite the fact that in our simulations we have access to the 3D positions of the stars, we perform the calculations in projection, for consistency and to allow for comparisons with previous results. We compensate for the potential effects of projection onto a specific plane by calculating three 2D projections of each snapshot (using the $x$, $y$, and $z$ axis of the simulation as lines of sight) and averaging the obtained value of $Q$ of the three projections, $Q = \frac{1}{3} \left( Q_x+Q_y+Q_z \right)$ ,with its corresponding standard deviation $\sigma = \frac{1}{3} \sqrt{\sigma_x^2+\sigma_y^2+\sigma_z^2}$.  We note that in any case, the projection effects are not severe, with globally similar results in the three projections. 

The values of the $Q$ parameter can be sensitive to the presence of outliers, so we apply a resampling strategy to mitigate their effect. For each snapshot we randomly choose a subsample with 90\% of the stars, compute the $Q$ parameter on this set, and repeat the process 100 times. This gives a distribution of the values of $Q$ within the sample, with its associated mean value and standard deviation. 